\documentclass[11pt,preprint]{aastex}
\usepackage{emulateapj5,graphics,psfig} 

\submitted{Received 2013 Oct.\ 15; Accepted: 2013 Dec.\ 11}

\def\gs{\mathrel{\raise0.35ex\hbox{$\scriptstyle >$}\kern-0.6em
\lower0.40ex\hbox{{$\scriptstyle \sim$}}}}
\def\ls{\mathrel{\raise0.35ex\hbox{$\scriptstyle <$}\kern-0.6em
\lower0.40ex\hbox{{$\scriptstyle \sim$}}}}

\newenvironment{inlinefigure}{%
\def\@captype{figure}%
\noindent\begin{minipage}{0.999\linewidth}\small}
{\end{minipage}\smallskip}
\makeatother

\lefthead{Smail et al.}
\righthead{S2CLS: Ultraluminous star-forming galaxies in a $z=$\,1.6 cluster}

\begin{document}

\title{The SCUBA-2 Cosmology Legacy Survey:\\ Ultraluminous star-forming galaxies in a \lowercase{{\Large $z$}}\,$=$\,1.6 cluster}

\author{
Ian Smail,$\!$\altaffilmark{1} J.\,E.\ Geach,$\!$\altaffilmark{2} A.\,M.\ Swinbank,$\!$\altaffilmark{1} K.\ Tadaki,$\!$\altaffilmark{3}  
V.\ Arumugam,$\!$\altaffilmark{4}\\ W.\ Hartley,$\!$\altaffilmark{5,6}
O.\ Almaini,$\!$\altaffilmark{6} M.\,N.\ Bremer,$\!$\altaffilmark{7} E.\ Chapin,$\!$\altaffilmark{8} S.\,C.\ Chapman,$\!$\altaffilmark{9}\\ 
A.\,L.\,R.\ Danielson,$\!$\altaffilmark{1} A.\,C.\ Edge,$\!$\altaffilmark{1} D.\ Scott,$\!$\altaffilmark{10} C.\,J.\ Simpson,$\!$\altaffilmark{11} J.\,M.\ Simpson,$\!$\altaffilmark{1}\\
C.\ Conselice,$\!$\altaffilmark{6} J.\,S.\ Dunlop,$\!$\altaffilmark{4} R.\,J.\ Ivison,$\!$\altaffilmark{4,12} A.\ Karim,$\!$\altaffilmark{13}
T.\ Kodama,$\!$\altaffilmark{14} A.\ Mortlock,$\!$\altaffilmark{6}\\  
E.\,I. Robson,$\!$\altaffilmark{12} I.\ Roseboom,$\!$\altaffilmark{4} A.\,P.\ Thomson,$\!$\altaffilmark{1} 
P.\,P.\ van der Werf\altaffilmark{15} \& T.\,M.\,A.\ Webb\altaffilmark{16}
}

\altaffiltext{1}{Institute for Computational Cosmology, Durham University, South Road, Durham DH1 3LE, UK}
\altaffiltext{2}{Centre for Astrophysics Research, Science and Technology Research Institute, University of Hertfordshire, Hatfield AL10 9AB, UK}
\altaffiltext{3}{National Astronomical Observatory of Japan, 2-21-1 Osawa, Mitaka, Tokyo, Japan}
\altaffiltext{4}{Institute for Astronomy, Royal Observatory, University of Edinburgh, Blackford Hill, Edinburgh EH9 3HJ, UK}
\altaffiltext{5}{Institute for Astronomy, ETH Zurich, Wolfgang-Pauli-Strasse 27, CH-8093 Zurich, Switzerland}
\altaffiltext{6}{School of Physics and Astronomy, University of Nottingham, University Park, Nottingham NG9 2RD, UK}
\altaffiltext{7}{H.\,H.\ Wills Physics Laboratory, Tyndall Avenue, Bristol BS8 1TL UK}
\altaffiltext{8}{XMM SOC, ESAC, Apartado 78, 28691 Villanueva de la Canada, Madrid, Spain}
\altaffiltext{9}{Department of Physics and Atmospheric Science, Dalhousie University, 6310 Coburg Road, Halifax, Nova Scotia B3H 4R2, Canada}
\altaffiltext{10}{Department of Physics and Astronomy, University of British Columbia, 6224 Agricultural Road, Vancouver, BC V6T 1Z1, Canada}
\altaffiltext{11}{Astrophysics Research Institute, Liverpool John Moores University, Liverpool Science Park, 146 Brownlow Hill, Liverpool L3 5RF, UK}
\altaffiltext{12}{UK Astronomy Technology Centre, Royal Observatory, Blackford Hill, Edinburgh EH9 3HJ, UK}
\altaffiltext{13}{Argelander-Institute of Astronomy, Bonn University, Auf dem H\"ugel 71, 53121, Bonn, Germany}
\altaffiltext{14}{National Astronomical Observatory of Japan, 2-21-1 Osawa, Mitaka, Tokyo 181-8588, Japan}
\altaffiltext{15}{Leiden Observatory, Leiden University, PO Box 9513, 2300 RA Leiden, the Netherlands}
\altaffiltext{16}{Department of Physics, Ernest Rutherford Building, 3600 rue University, McGill University, Montreal, QC H3A 2T8, Canada}

\setcounter{footnote}{17}

\begin{abstract}
We analyse new SCUBA-2 submillimeter and archival SPIRE far-infrared imaging of a $z=$\,1.62 cluster, Cl\,0218.3$-$0510, which lies in the UKIDSS/UDS field of the SCUBA-2 Cosmology Legacy Survey.  Combining these tracers of obscured star formation activity with the extensive photometric and spectroscopic information available for this field, we identify 31 far-infrared/submillimeter-detected probable cluster members with bolometric luminosities $\gs $\,10$^{12}$\,L$_\odot$ and show that by virtue of their dust content and activity, these represent some of the reddest and brightest galaxies in this structure.  We exploit Cycle-1 ALMA submillimeter continuum imaging which covers one of these sources to confirm the identification of a SCUBA-2-detected ultraluminous star-forming galaxy in this structure.  Integrating the total star-formation activity in the central region of the structure, we estimate that it is an order of magnitude higher (in a mass-normalised sense) than clusters at $z\sim $\,0.5--1.  However, we also find that the most active cluster members do not reside in the densest regions of the structure, which instead host a population of passive and massive, red galaxies.  We suggest that while the passive and active populations have comparable near-infrared luminosities at $z=$\,1.6, $M_H\sim -$23, the  subsequent stronger fading of the more active galaxies means that they will evolve into passive systems at the present-day which are less luminous than the descendants of those galaxies which were already passive at $z\sim$\,1.6 ($M_H\sim -$20.5 and $M_H\sim -$21.5 respectively at $z\sim$\,0).  We conclude that the massive galaxy population in the dense cores of present-day clusters were already in place at $z=$\,1.6 and that in Cl\,0218.3$-$0510 we are seeing continuing infall of less extreme, but still ultraluminous, star-forming galaxies onto a pre-existing structure.
\end{abstract}


\section{Introduction}

%
%
\setcounter{figure}{0}
\begin{figure*}[tbh]
\centerline{\psfig{file=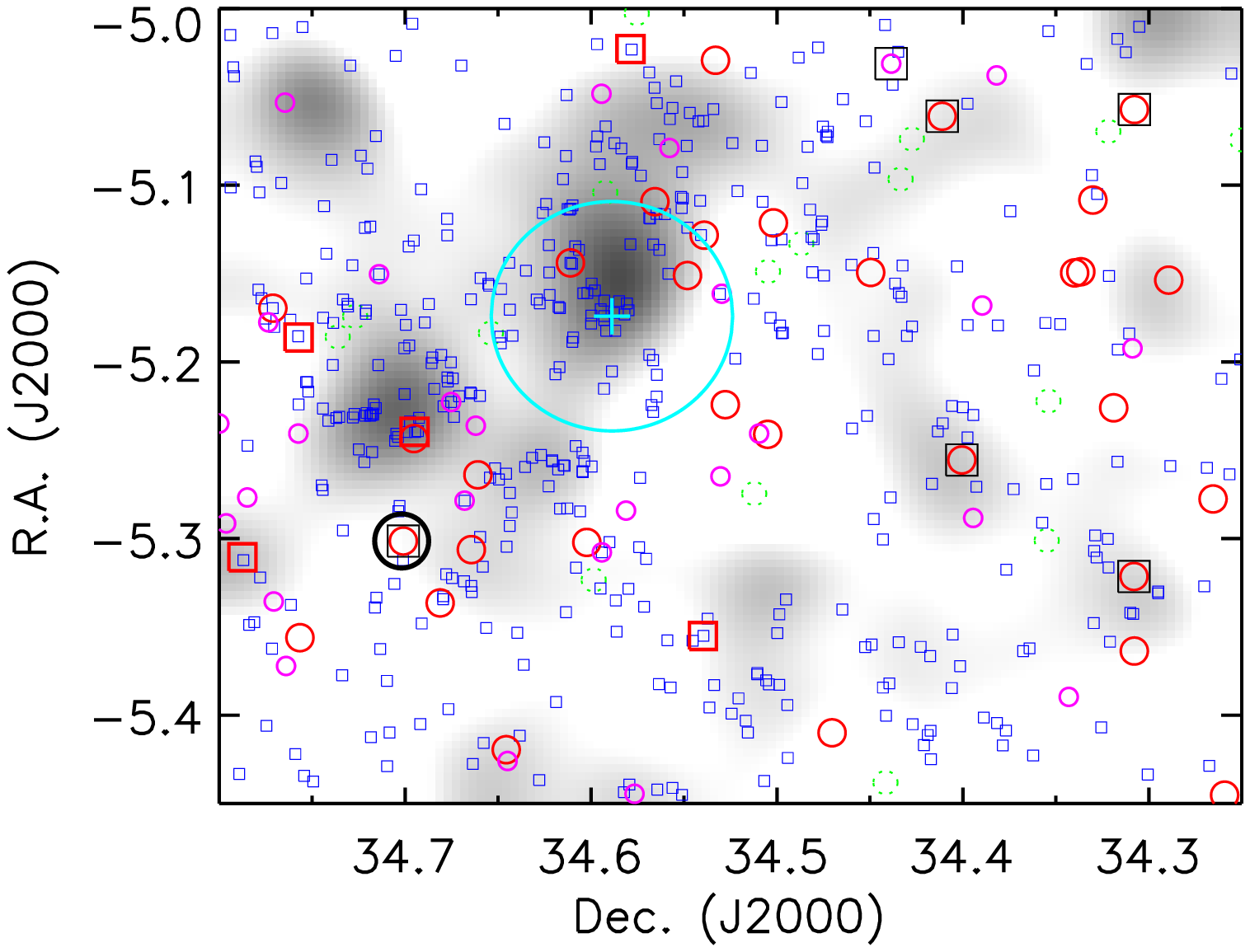,width=4.in,angle=90}\hspace*{-1cm}\psfig{file=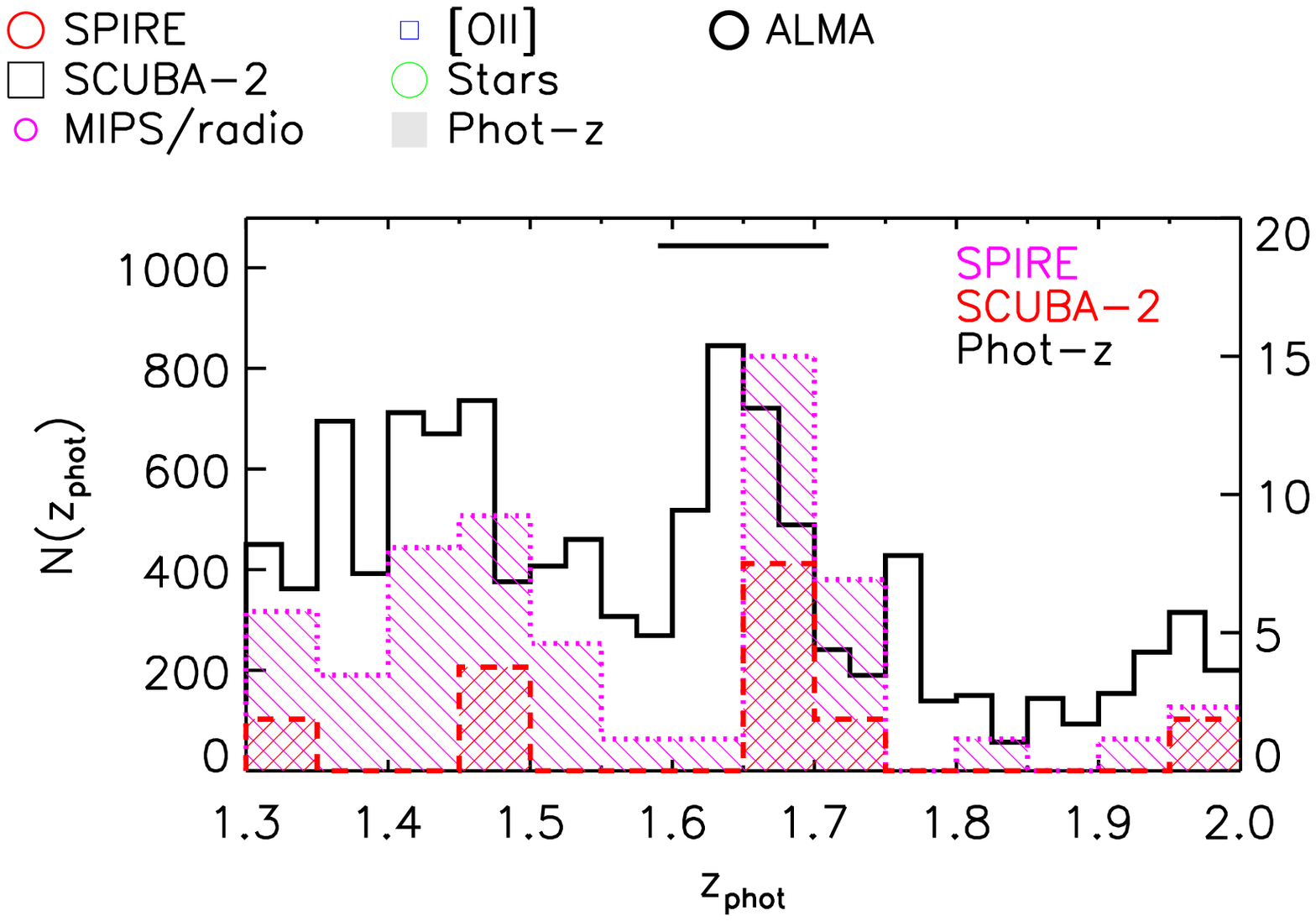,width=4.0in,angle=90}}
\caption{\small 
{\it (left)} The spatial distribution of members and probable members of the structure associated
with the $z=$\,1.6 cluster Cl\,0218.3$-$0510 in the UKIDSS/UDS field.   The area shown corresponds to the region surveyed by Tadaki et al.\ (2012) using Suprimecam on Subaru and a narrow-band filter with a wavelength corresponding to redshifted [O{\sc ii}] at $z=1.62$.  Their photometrically-selected [O{\sc ii}]-emitting cluster members are plotted and we also show the other active cluster populations, including MIPS and radio sources, the far-infrared-detected examples of this population, the SCUBA-2-detected subset of these and the  potential [O{\sc ii}]-companions to five submm sources.  We highlight the candidate SCUBA-2-detected source whose identification as a member has been confirmed by ALMA continuum observations (Simpson et al., in prep).  The background grayscale is the smoothed density distribution (smoothed with a Gaussian kernel with a {\sc fwhm} of 1.7\,Mpc) of photometrically-identified cluster members from Hartley et al.\ (2013).  In addition we show $B<13.5$ stars in the field which are sufficiently bright that they impede the detection of faint galaxies in their vicinity. Finally the  cross marks the cluster position from Papovich et al.\ (2010) with the large circle around it showing the 2\,Mpc radius region used in the analysis of the integrated star-formation rate in \S3.2. {\it (right)} A comparison of the photometric redshift distributions for the full field shown on the left.  We show the distributions of all galaxies in the photometric redshift catalog (see \S2.4), the SPIRE-detected sources (for those  detected at 250, 350 and 500\,$\mu$m) and those sources detected by SCUBA-2, as well as the photometric-redshift range used to select members (horizontal bar).  All three populations show a spike at $z\sim$\,1.65 corresponding to the cluster and we can use the average source density across the whole redshift range to assess the contamination due to unrelated interlopers of 40--60 per cent.
}
\end{figure*}

Luminous ellipticals in the central regions of rich clusters are the most massive galaxies ($\geq $\,10$^{11}$\,M$_\odot$) at the present-day.  The stars in these galaxies are metal-rich, old and surprisingly uniform:    spectral analysis of  typical $L^\ast$ ellipticals indicates luminosity-weighted ages of $\sim $\,8--11\,Gyrs and hints that those at larger radii in clusters may be slightly younger  (e.g.\ Nelan et al.\ 2005; Smith et al.\ 2012).  This homogeneity in their stellar populations means that these galaxies show a ``red sequence'' -- a tight correlation between luminosity and colour  which can be traced in clusters out to high redshifts (e.g.\ Ellis et al.\ 1997). The existence of a red sequence of passive galaxies in a cluster signals that most of these galaxies  were formed at least $\sim $\,1--2\,Gyrs prior to the epoch at which the cluster is observed, although the presence of dust reddened, star-forming galaxies can confuse this interpretation.

Red galaxy sequences have been used to identify rare, massive clusters  out to $z\sim $\,1.5  (e.g.\ Gladders \& Yee 2005; Papovich et al.\ 2010; Demarco et al.\ 2010; Muzzin et al.\ 2013).  However, in most clusters at $z>$\,1 they are sparsely populated (especially at  fainter luminosities, e.g.\ Blakeslee et al.\ 2003). This suggests that $z\sim $\,1 is the epoch where galaxies in high-density environments were evolving onto these passive sequences.   The striking decline in the galaxy population on the passive red sequence    at $z> $\,1 implies a  corresponding   increase in the population of massive, star-forming galaxies in clusters at these epochs.  Moreover, the metal-rich stars in local ellipticals suggest that this early phase of star formation is dust-enshrouded, and so   best identified in the mid-/far-infrared.

%
%
\setcounter{figure}{1}
\begin{figure*}[tbh]
\centerline{
\psfig{file=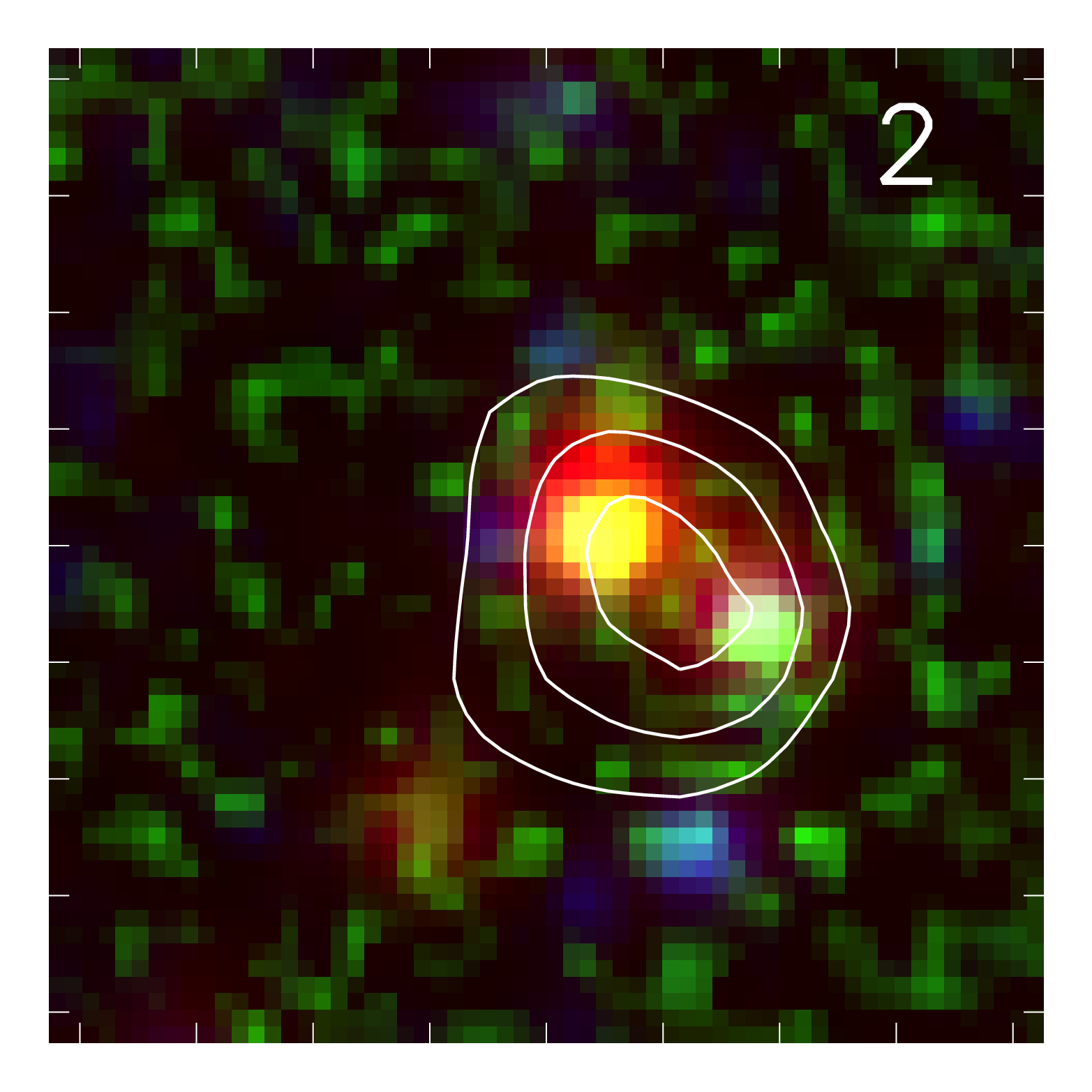,width=0.8in,angle=0}
\psfig{file=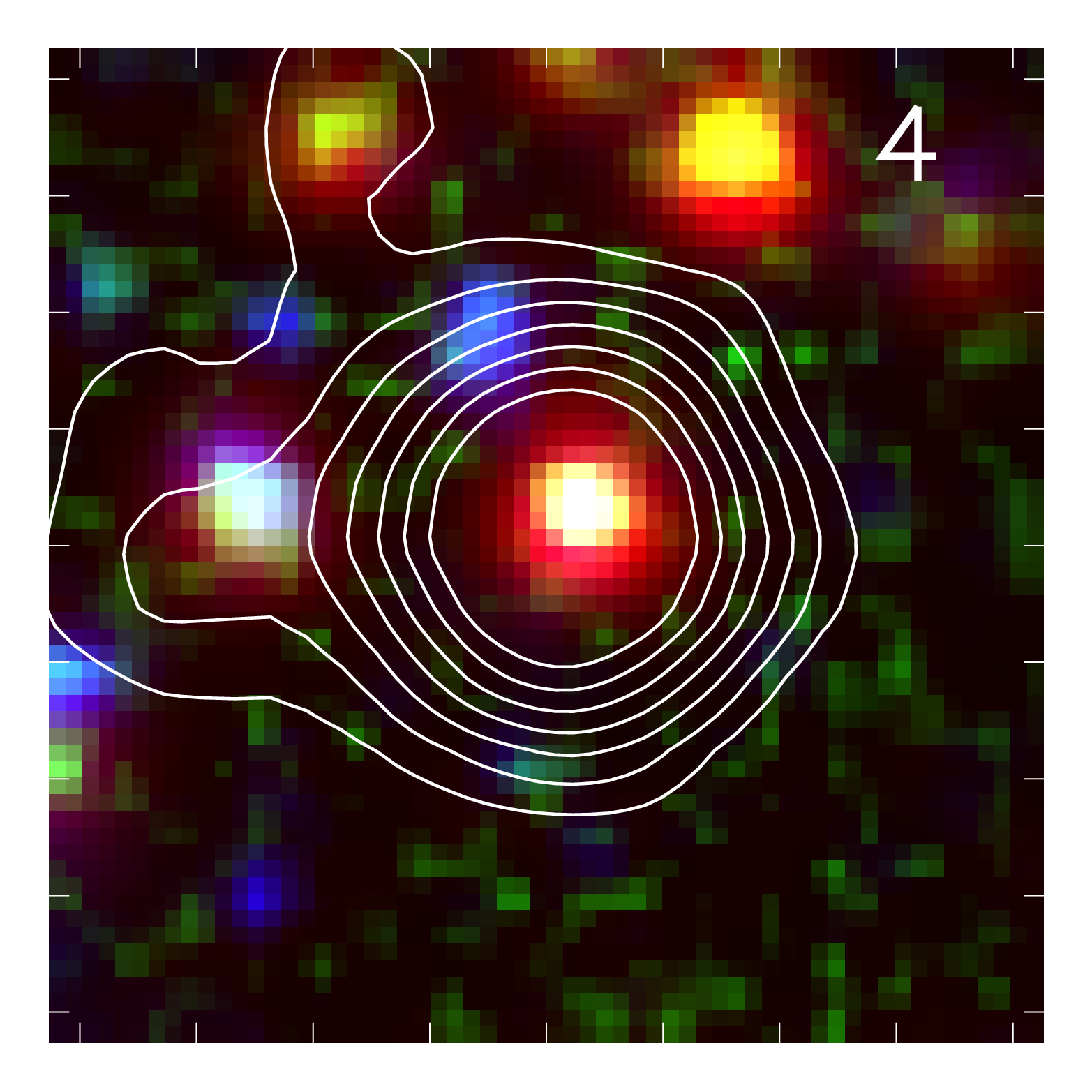,width=0.8in,angle=0}
\psfig{file=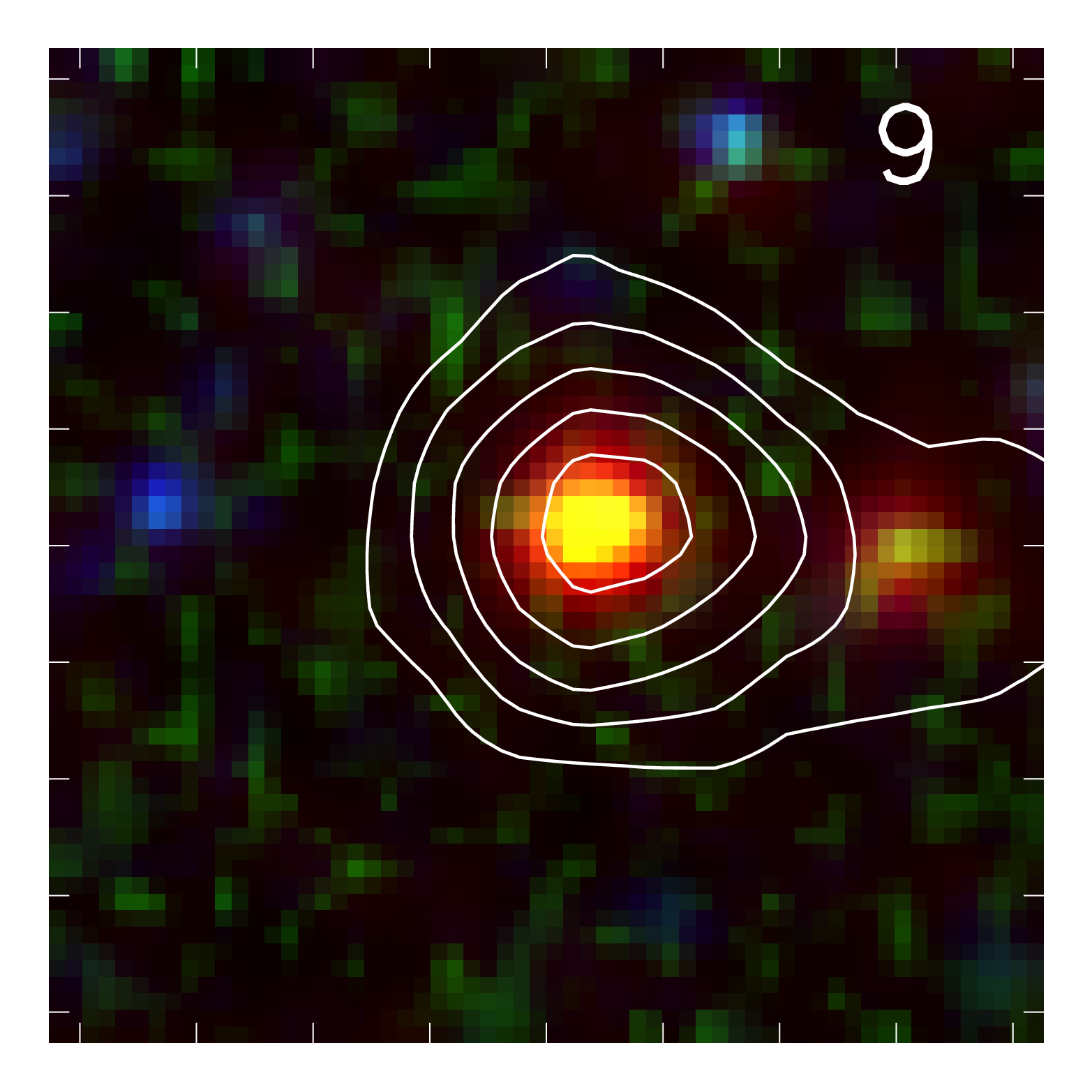,width=0.8in,angle=0}
\psfig{file=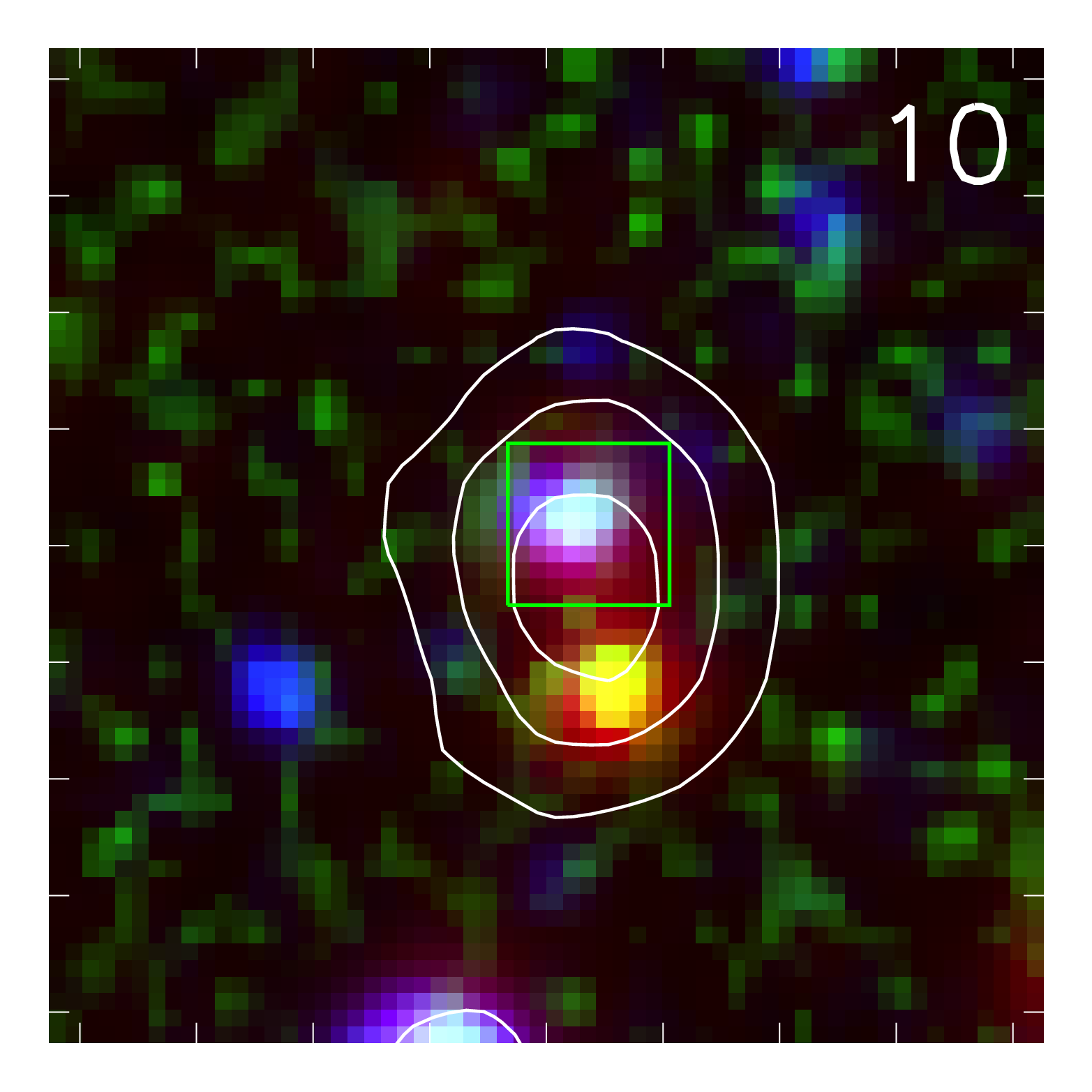,width=0.8in,angle=0}
\psfig{file=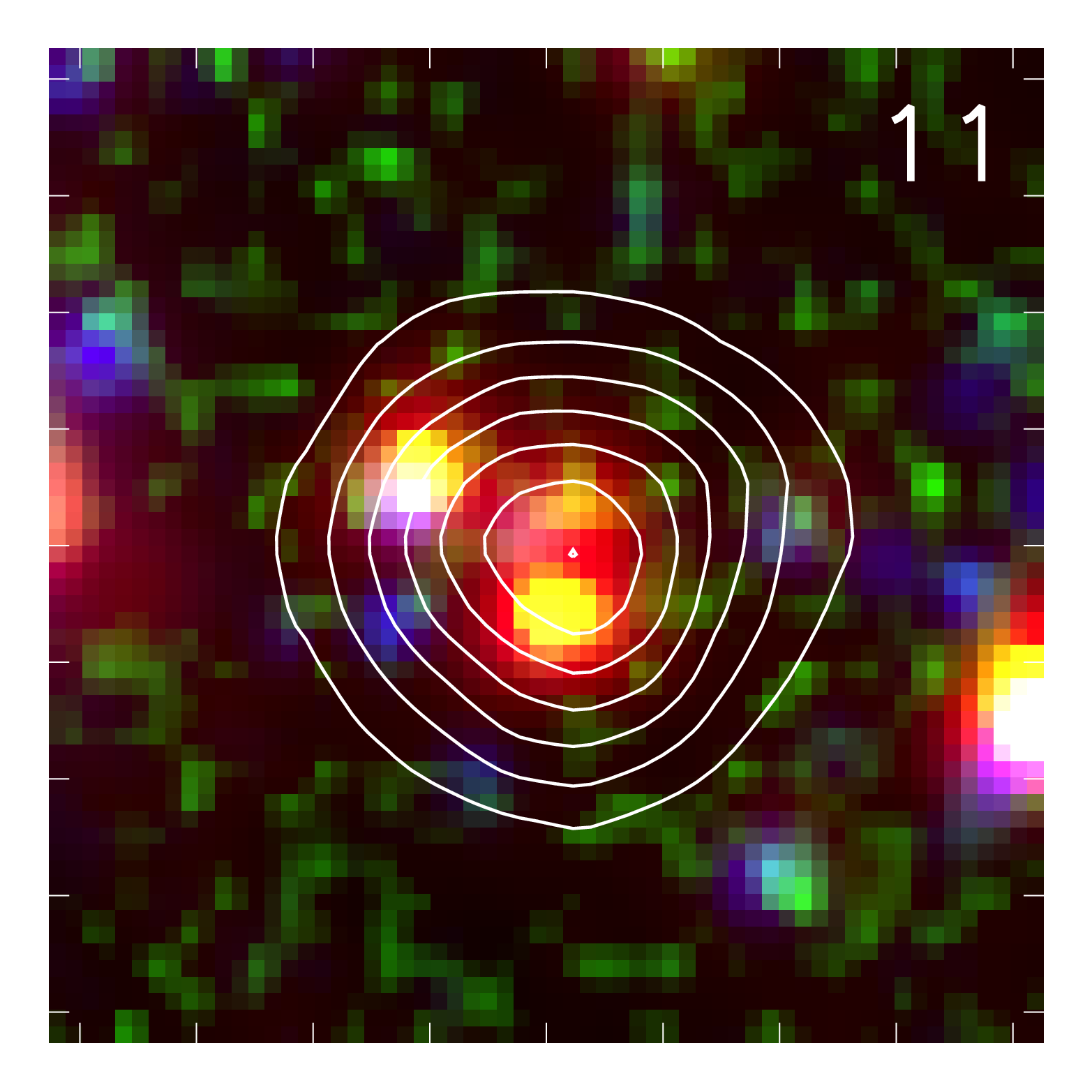,width=0.8in,angle=0}
\psfig{file=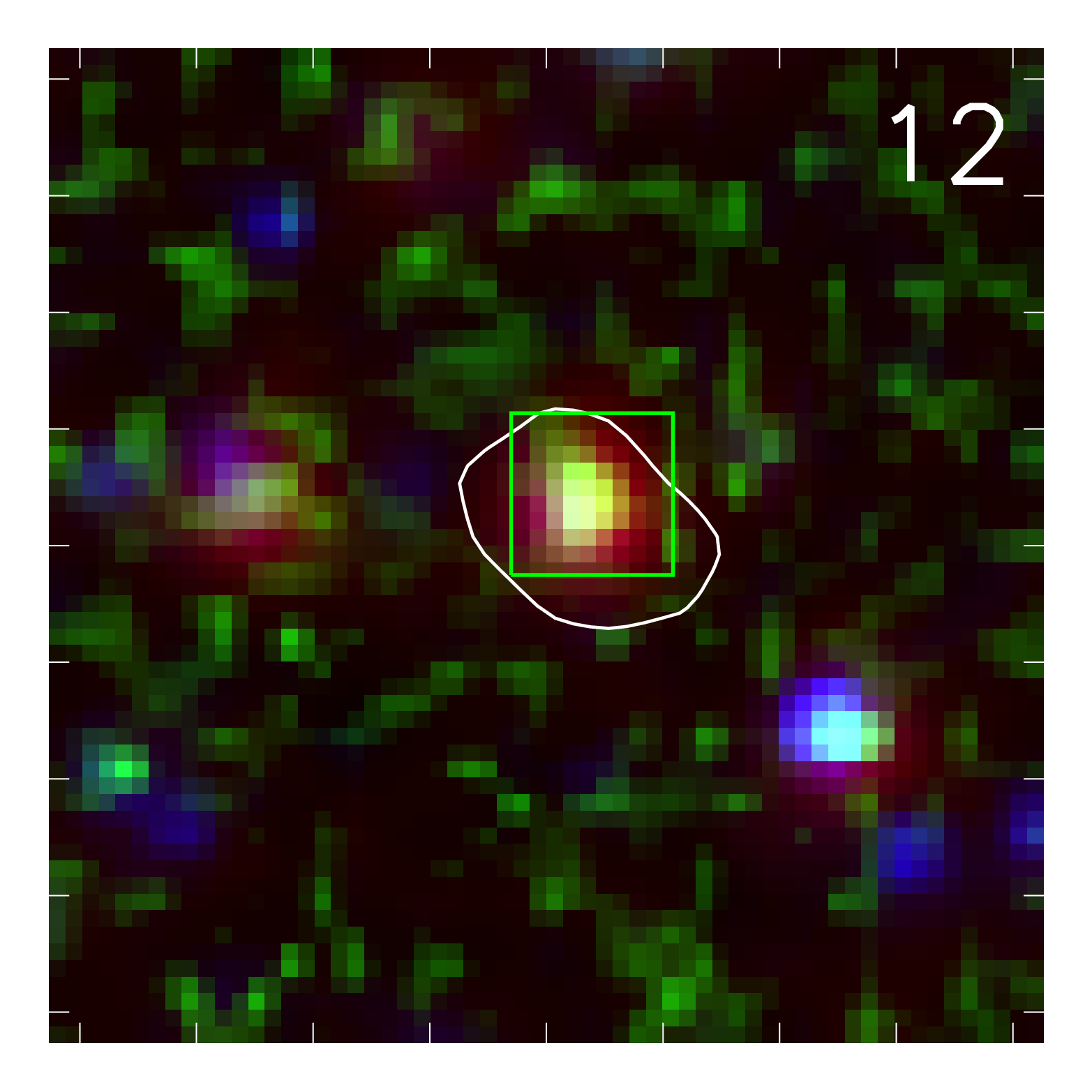,width=0.8in,angle=0}
\psfig{file=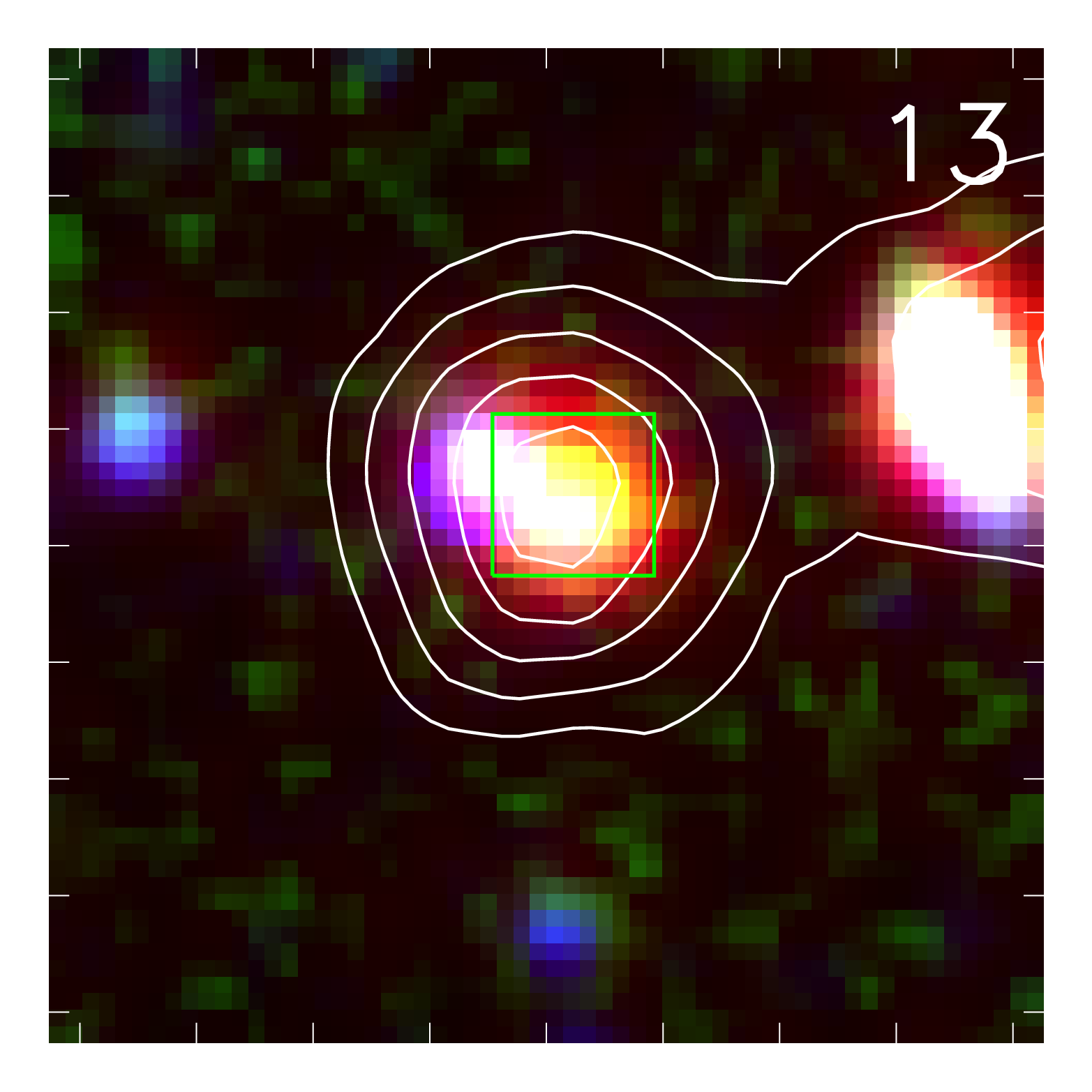,width=0.8in,angle=0}
\psfig{file=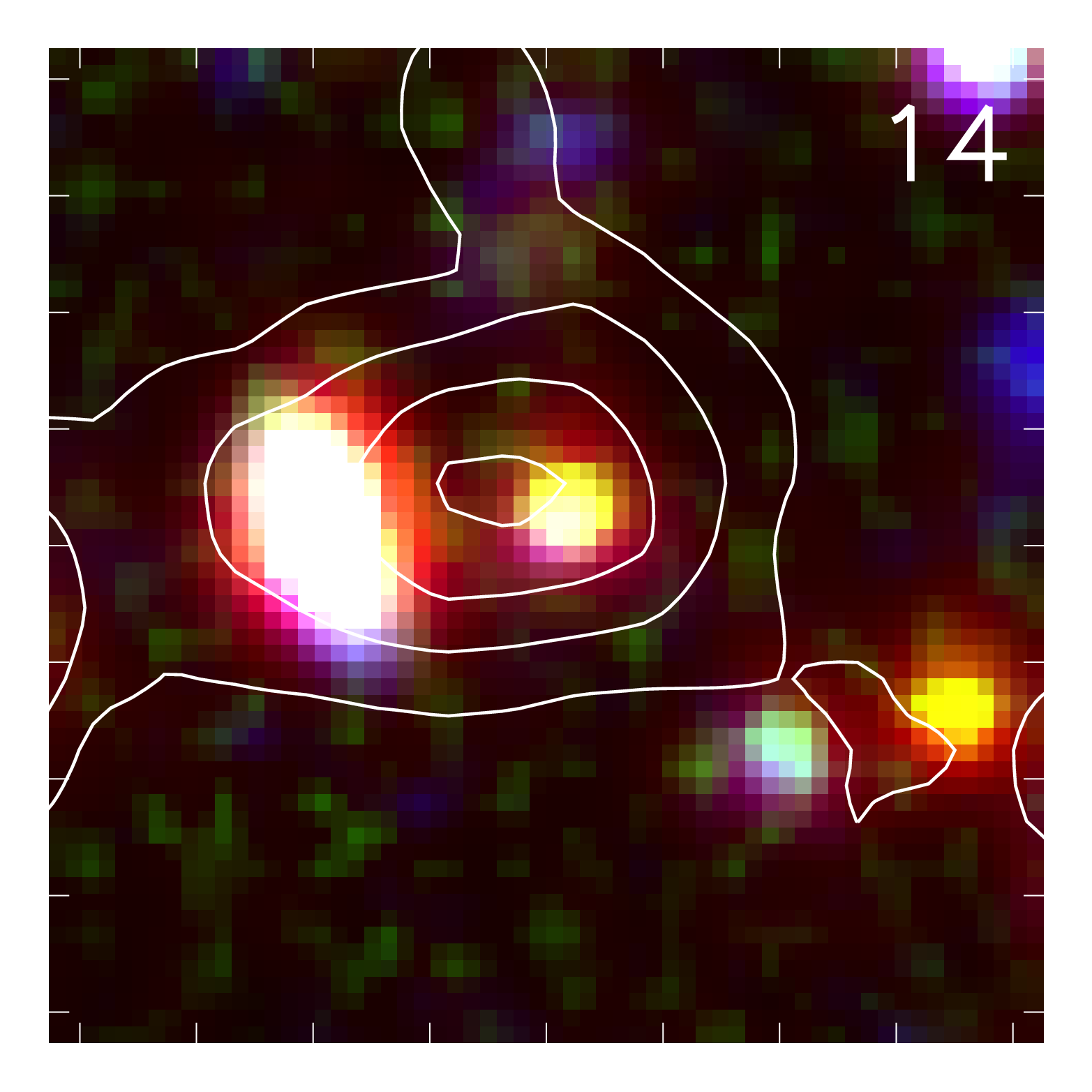,width=0.8in,angle=0}
}

\centerline{
\psfig{file=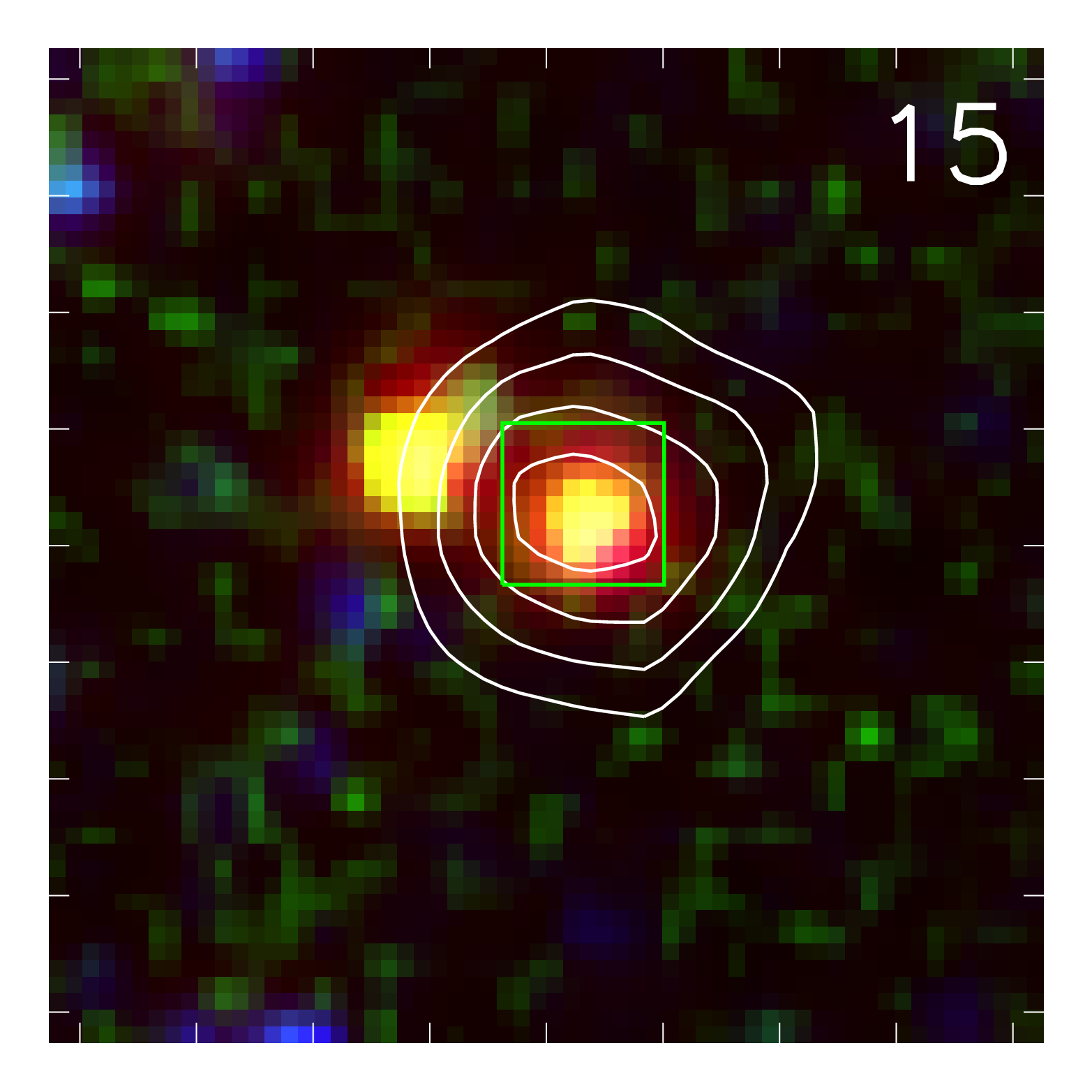,width=0.8in,angle=0}
\psfig{file=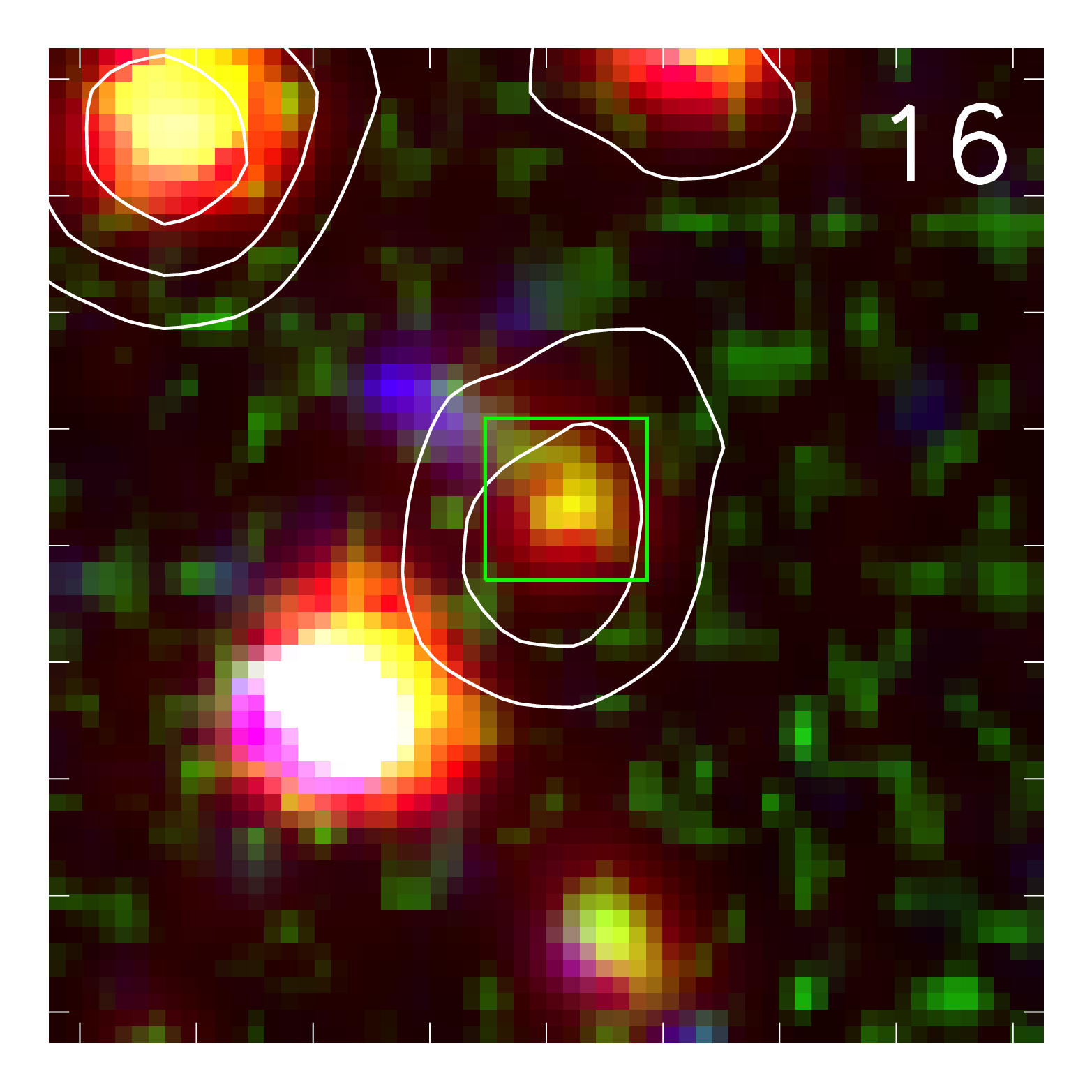,width=0.8in,angle=0}
\psfig{file=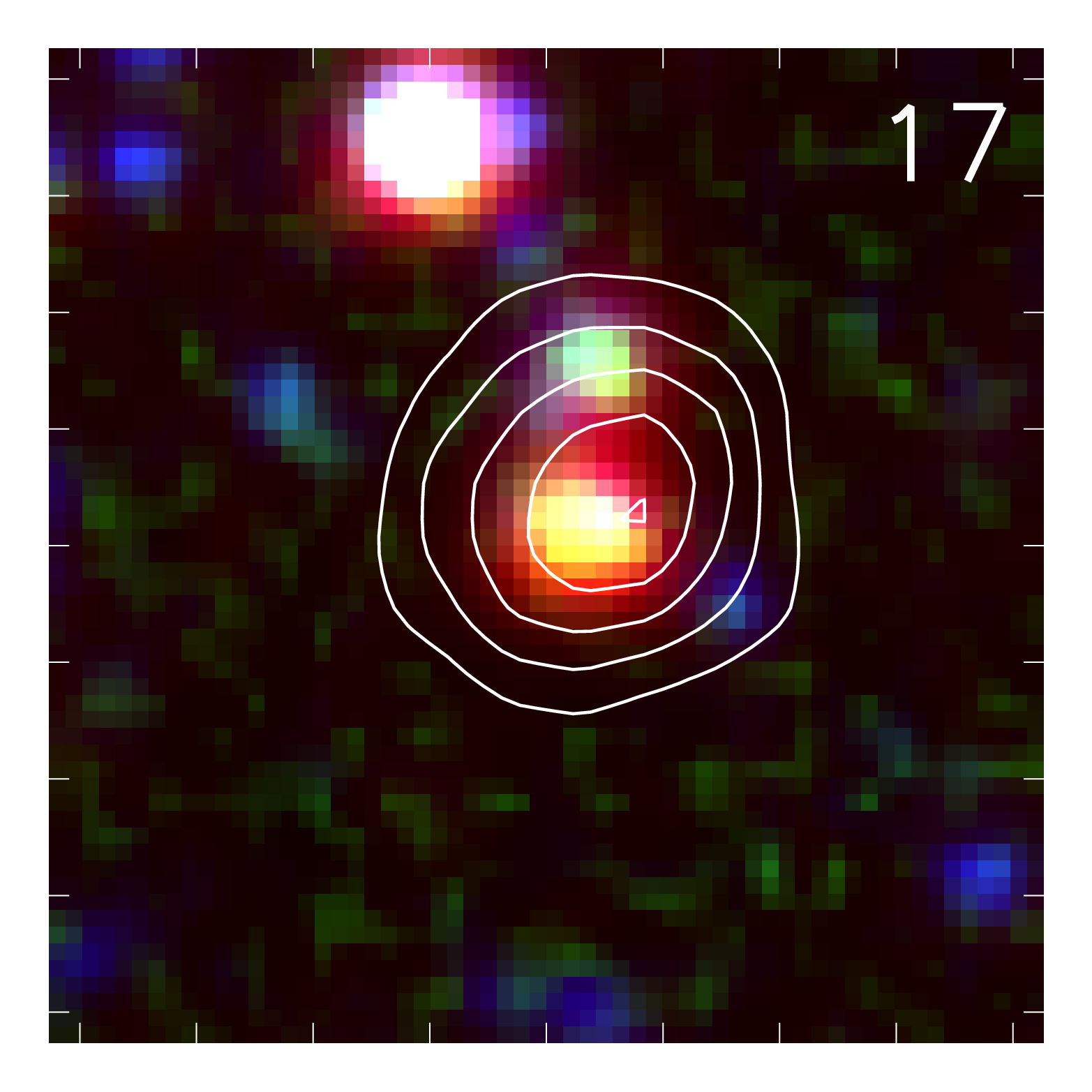,width=0.8in,angle=0}
\psfig{file=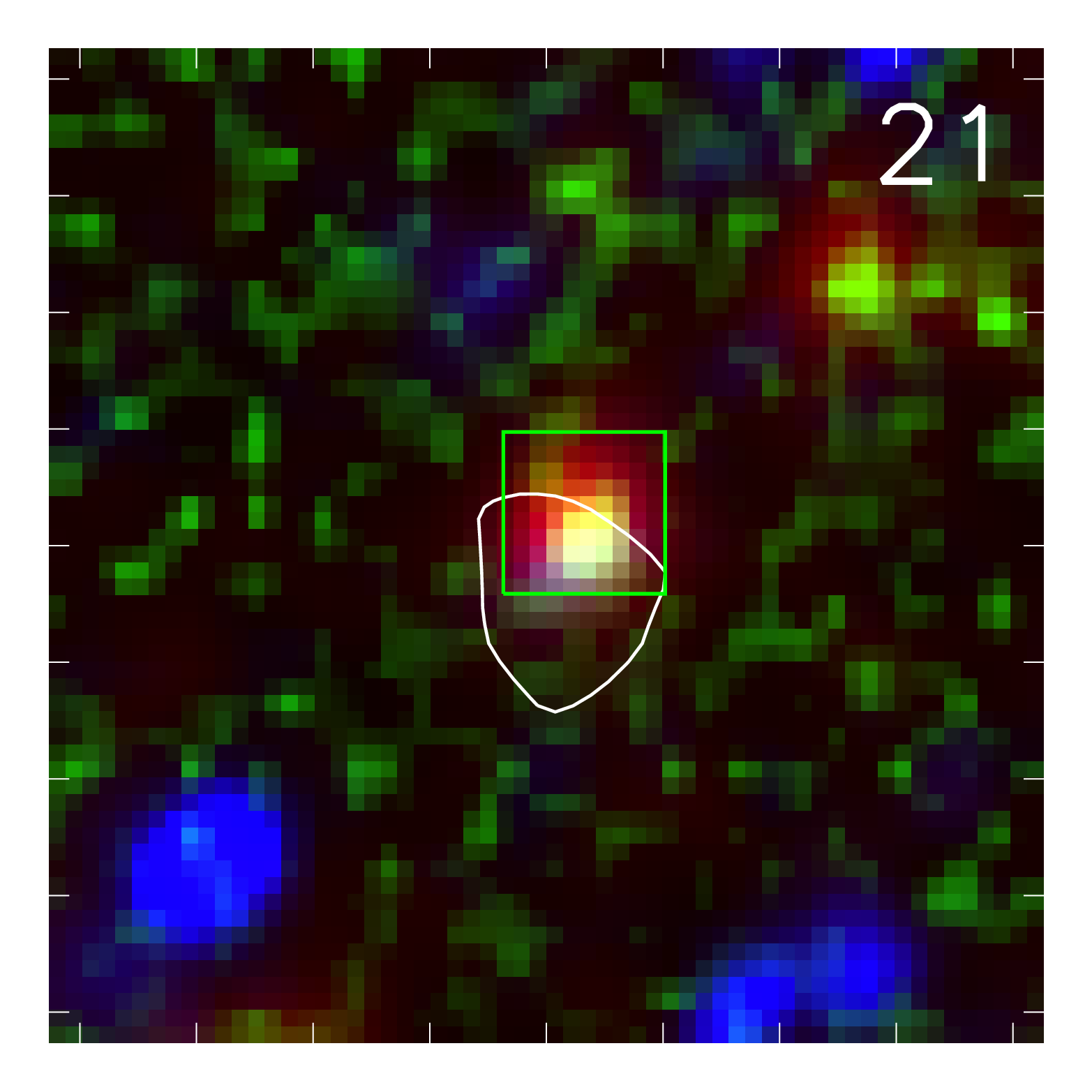,width=0.8in,angle=0}
\psfig{file=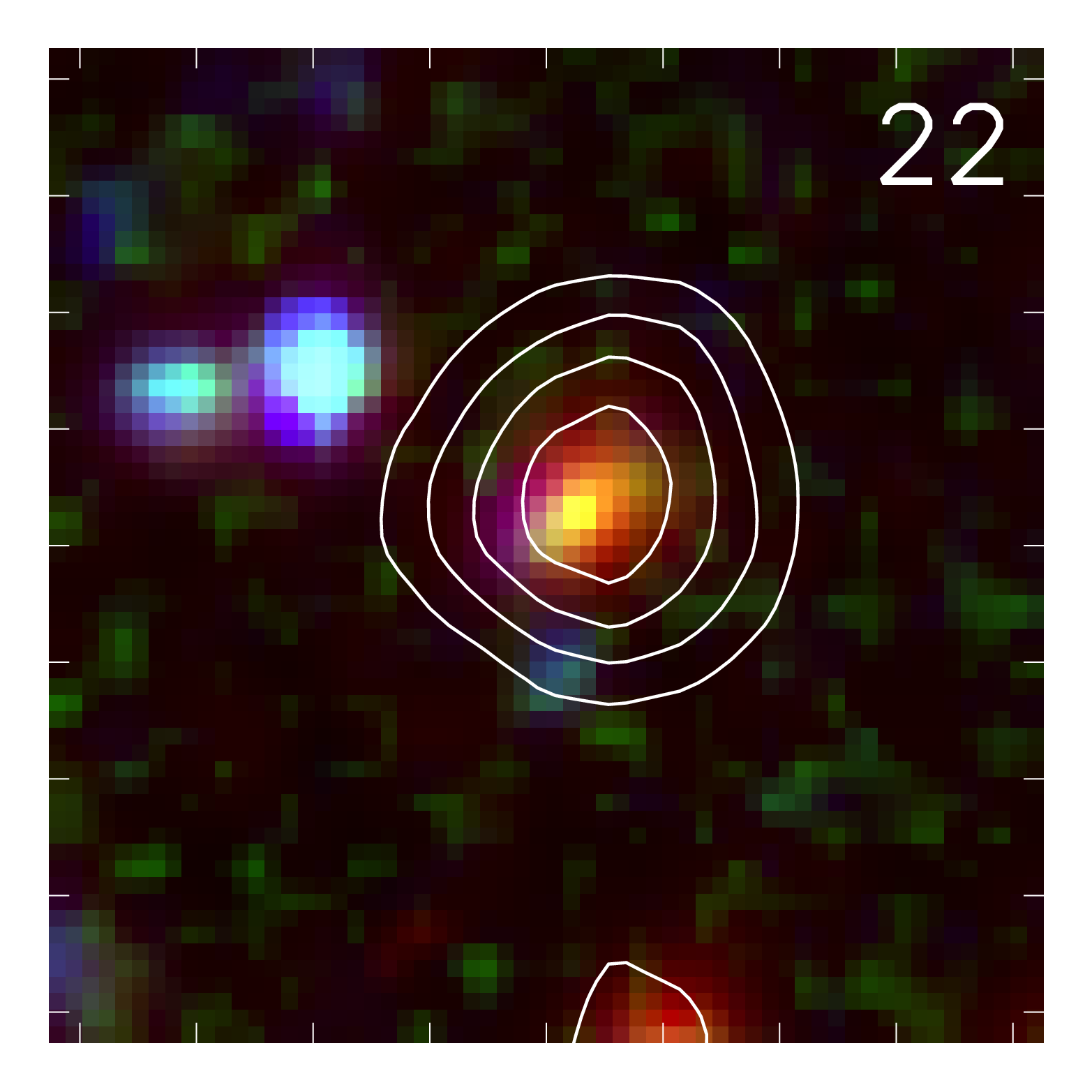,width=0.8in,angle=0}
\psfig{file=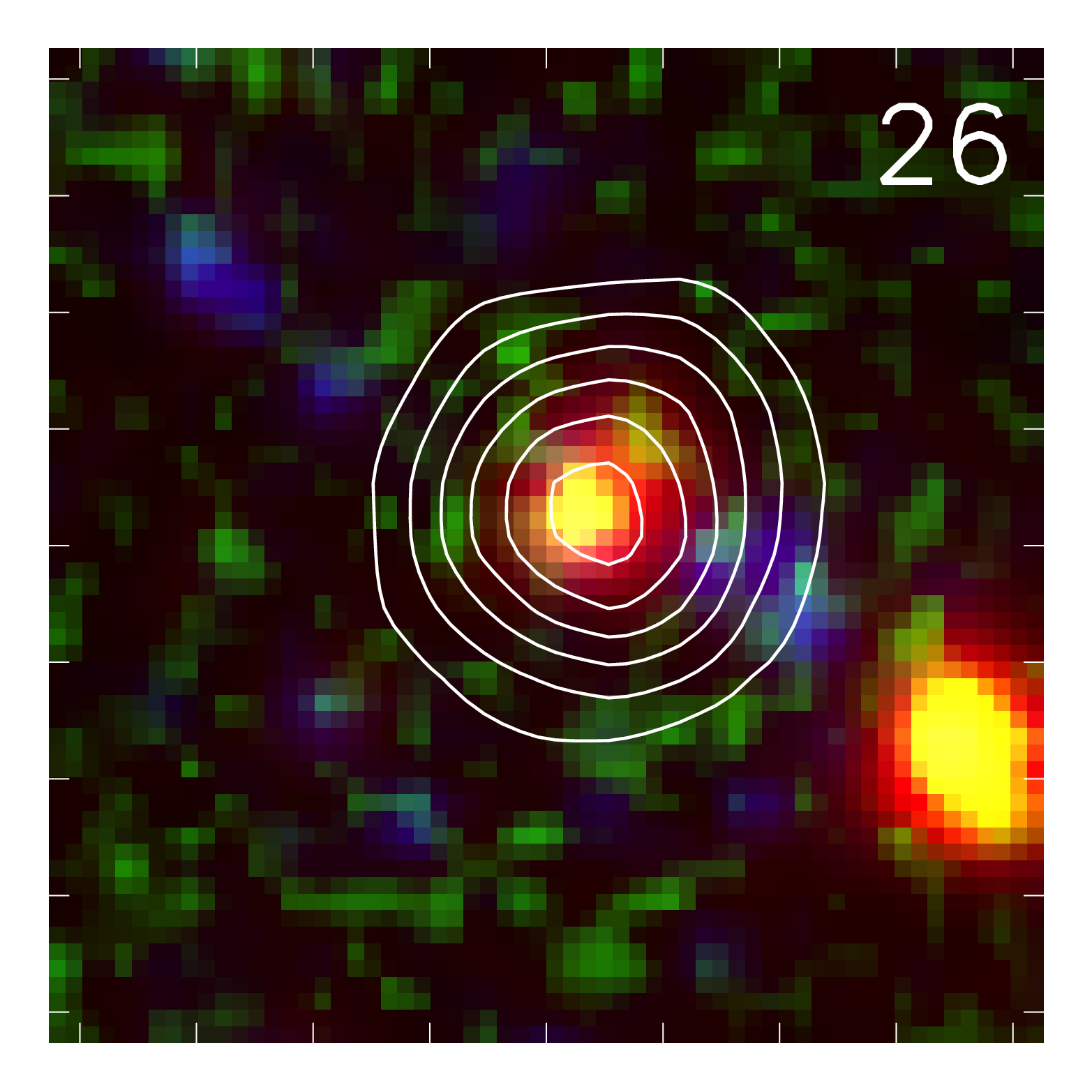,width=0.8in,angle=0}
\psfig{file=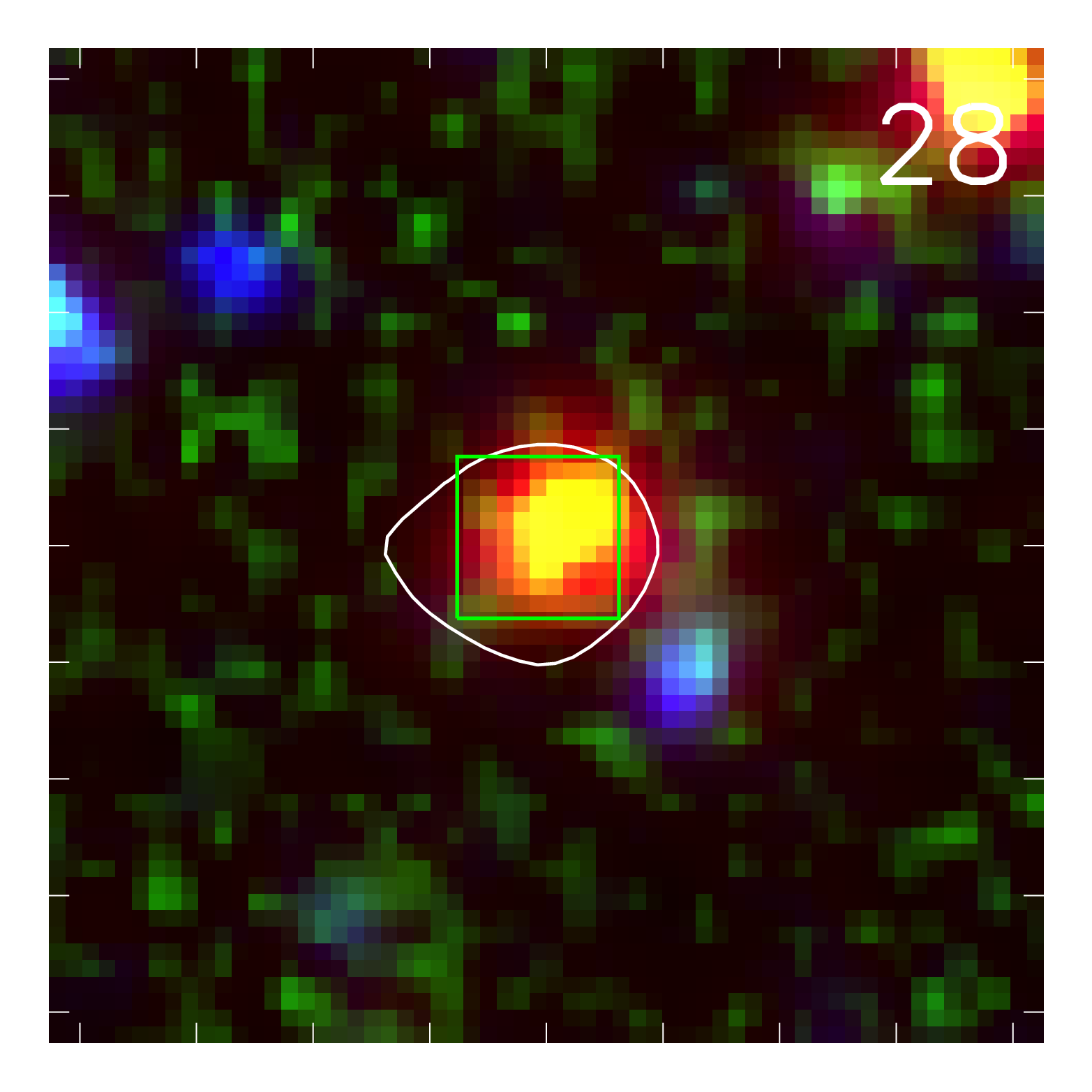,width=0.8in,angle=0}
\psfig{file=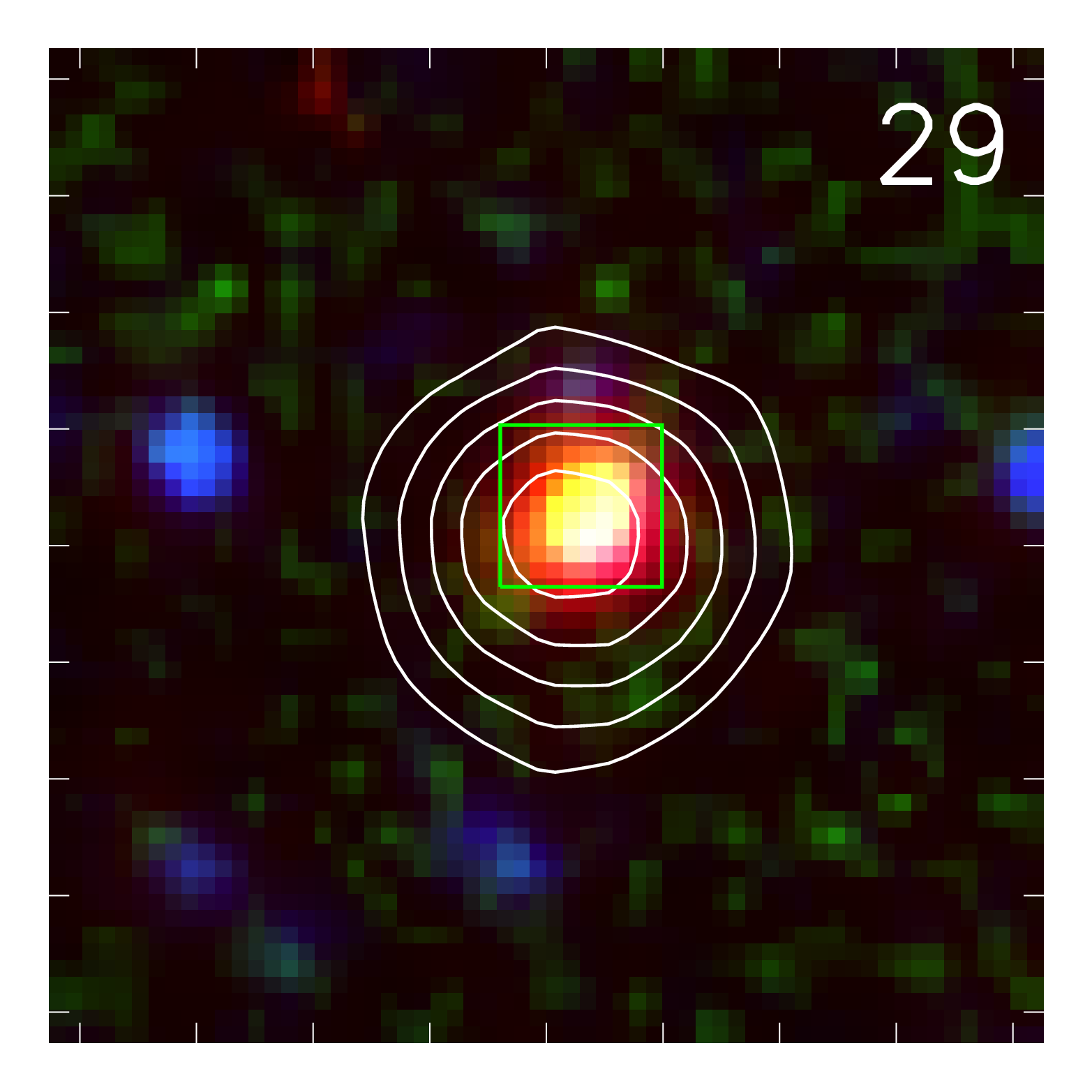,width=0.8in,angle=0}
}

\centerline{
\psfig{file=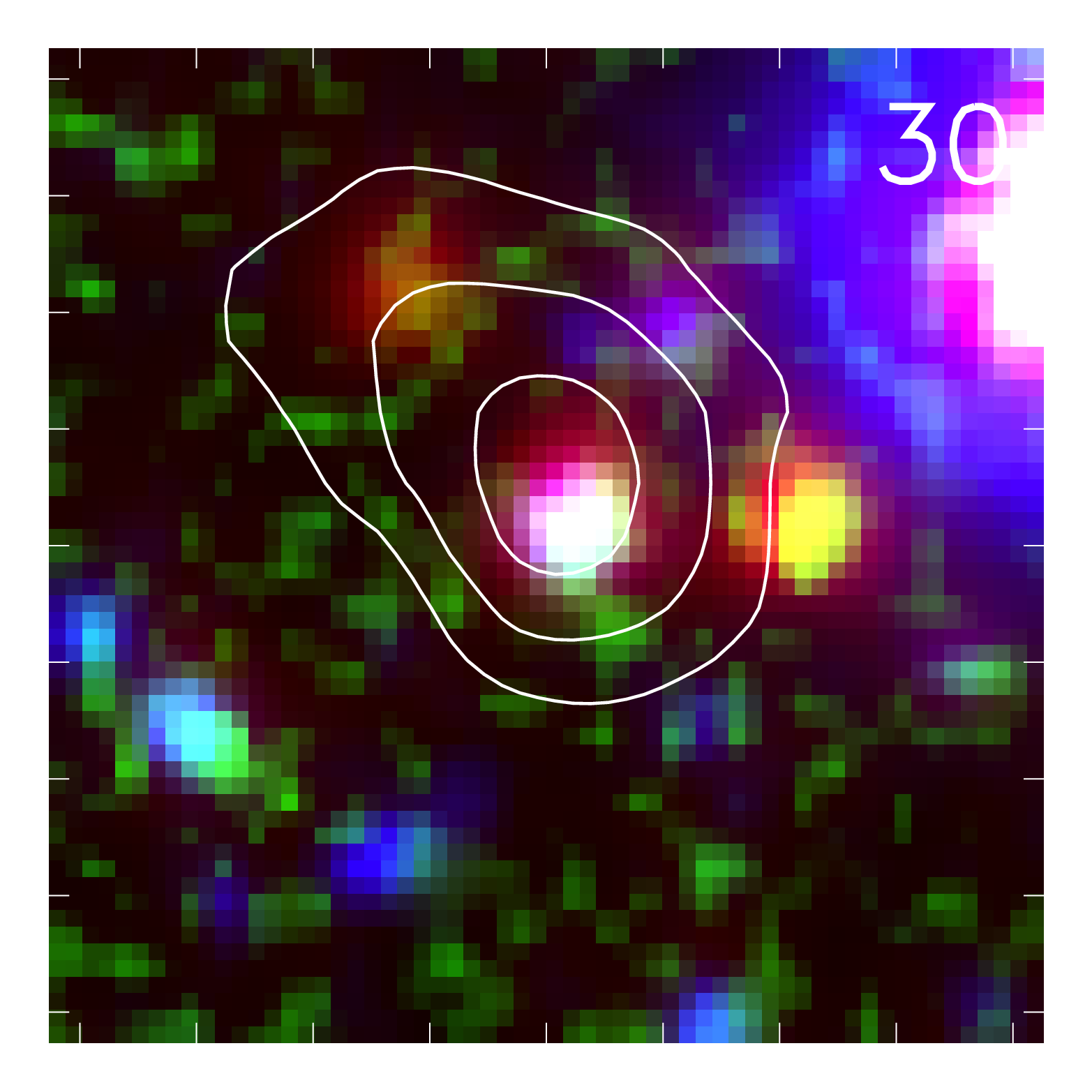,width=0.8in,angle=0}
\psfig{file=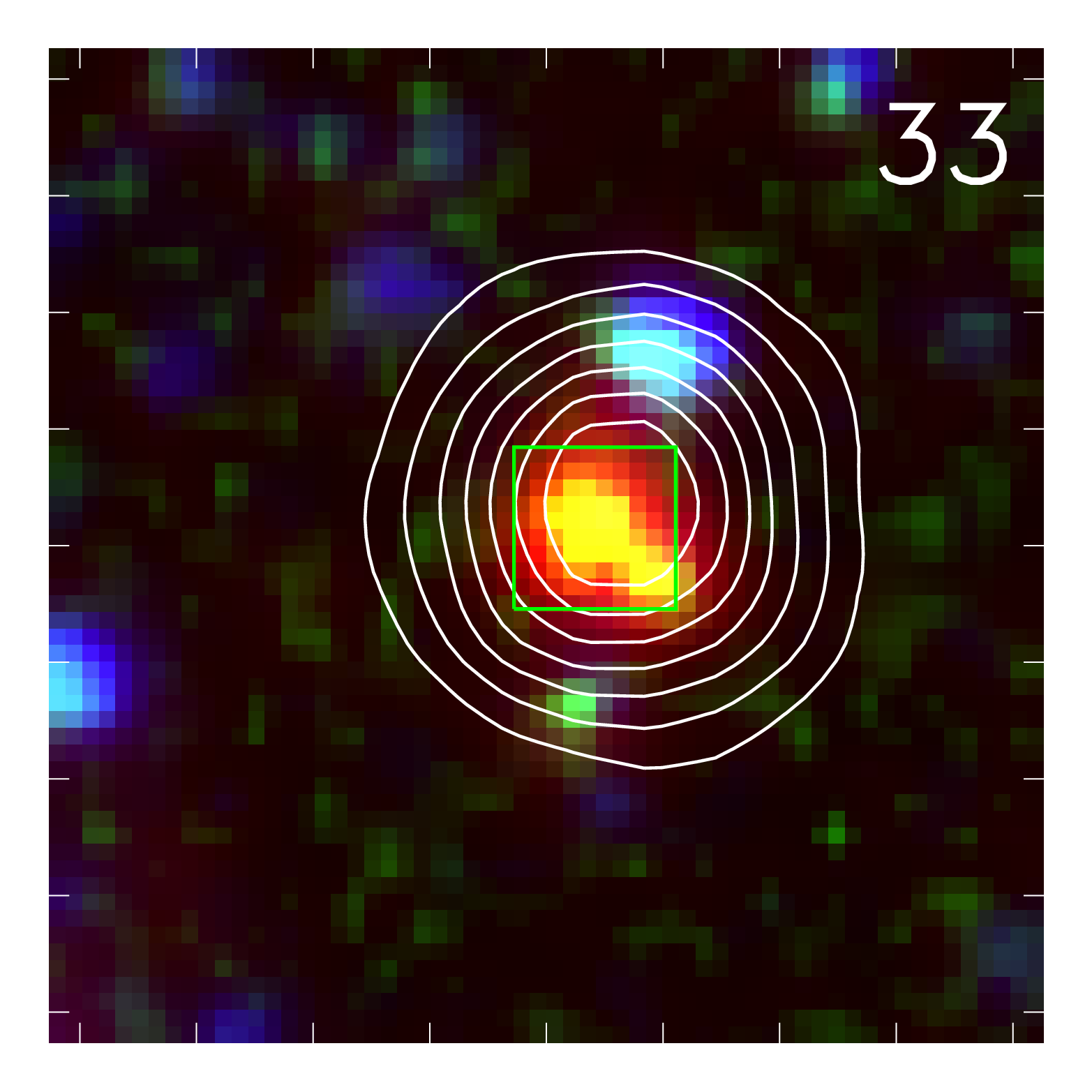,width=0.8in,angle=0}
\psfig{file=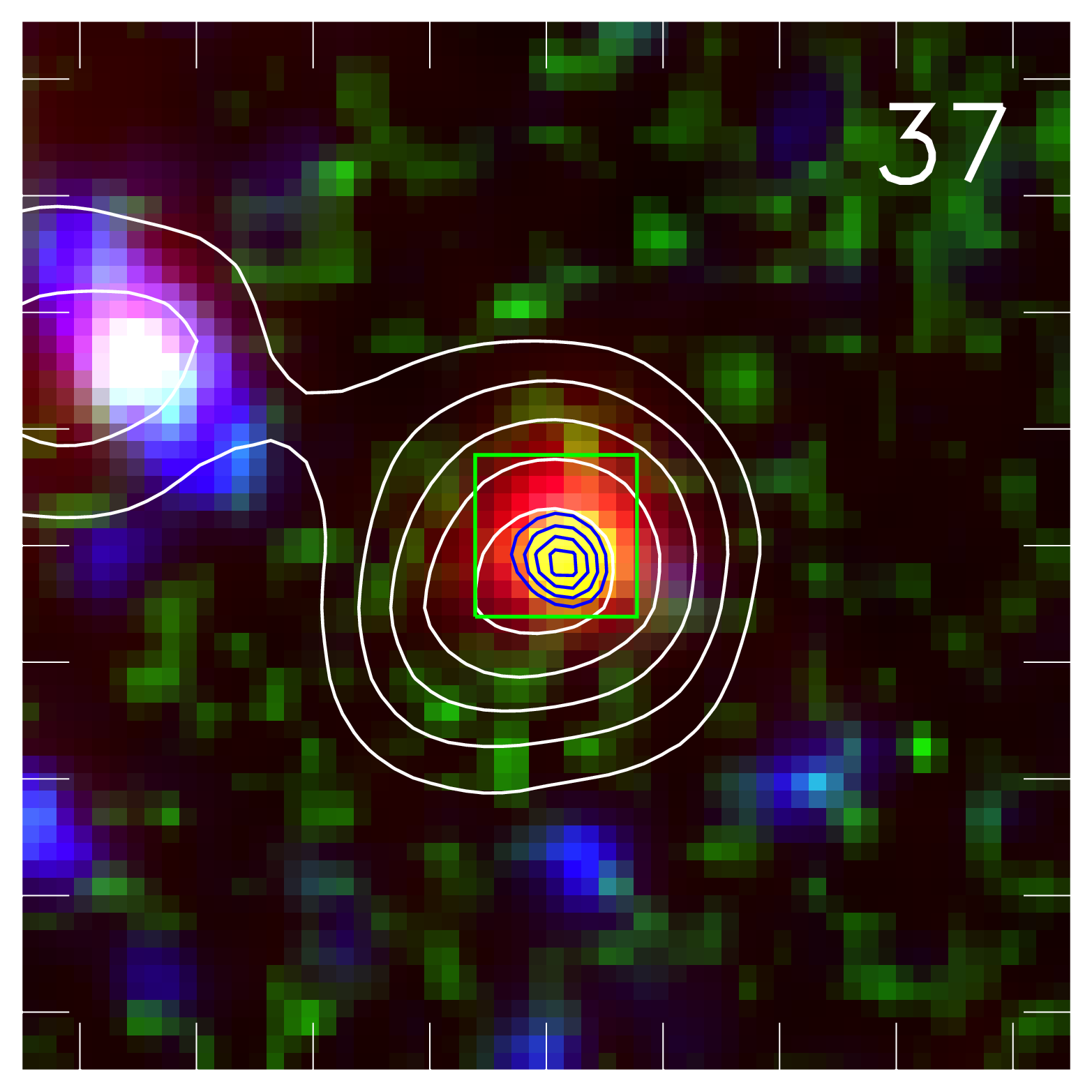,width=0.8in,angle=0}
\psfig{file=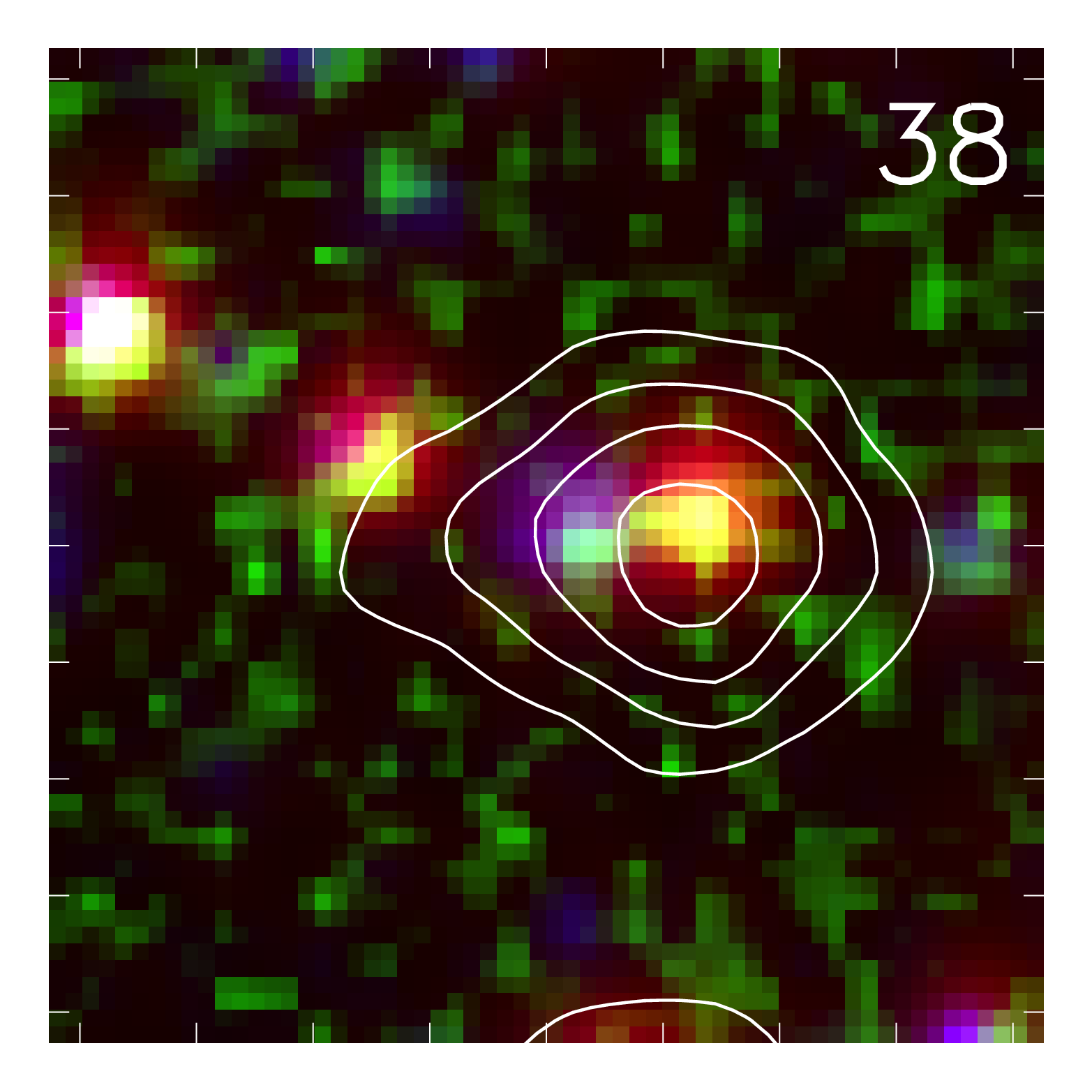,width=0.8in,angle=0}
\psfig{file=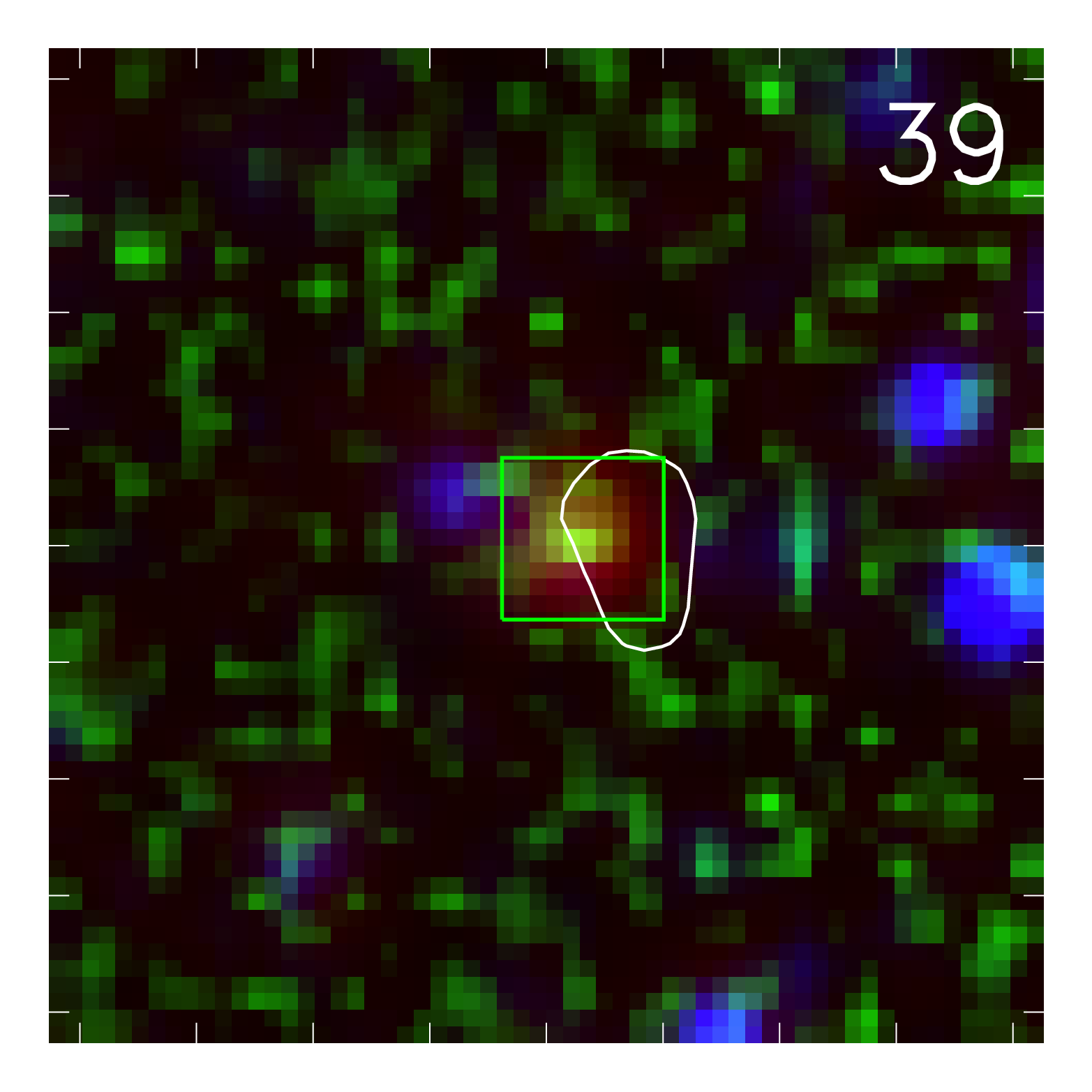,width=0.8in,angle=0}
\psfig{file=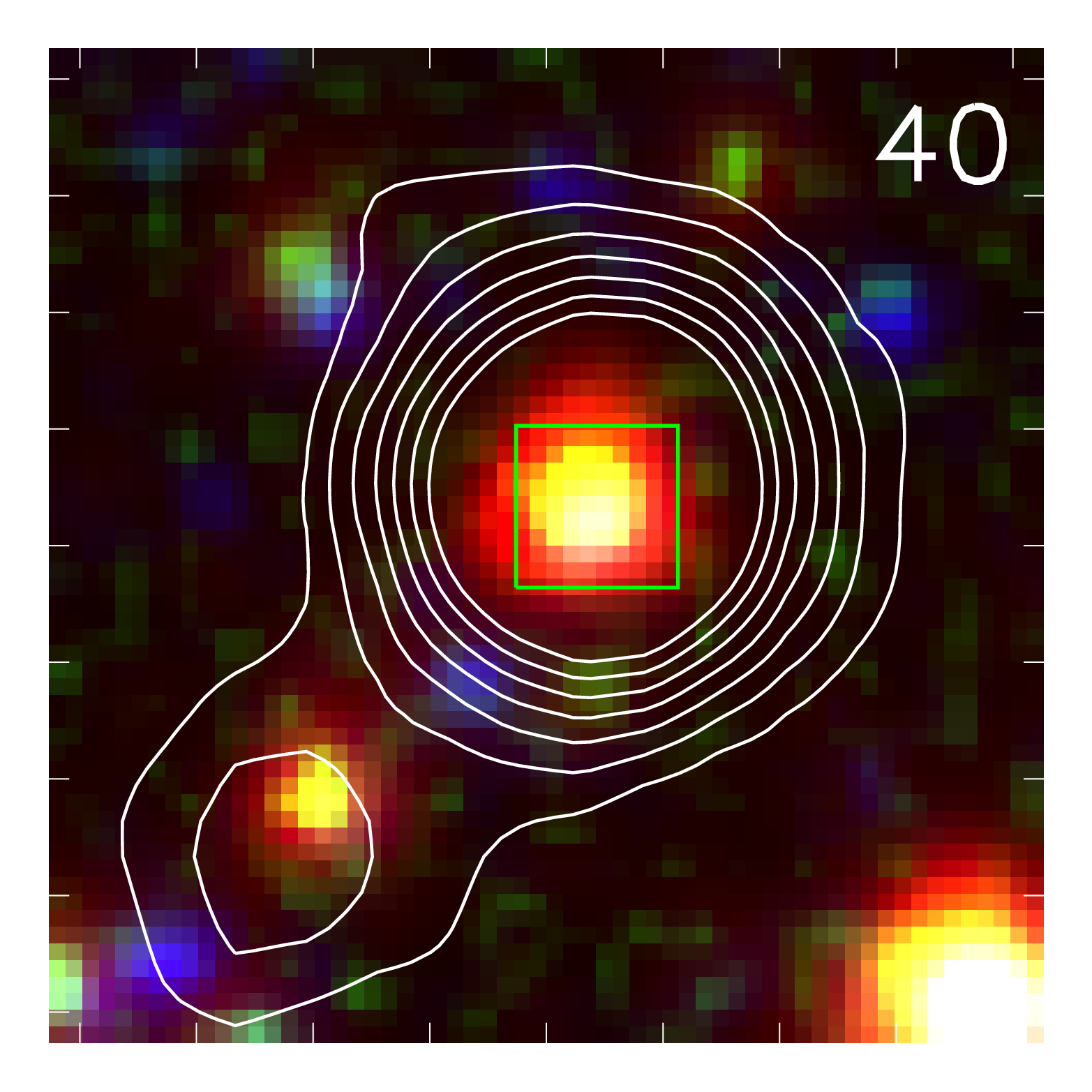,width=0.8in,angle=0}
\psfig{file=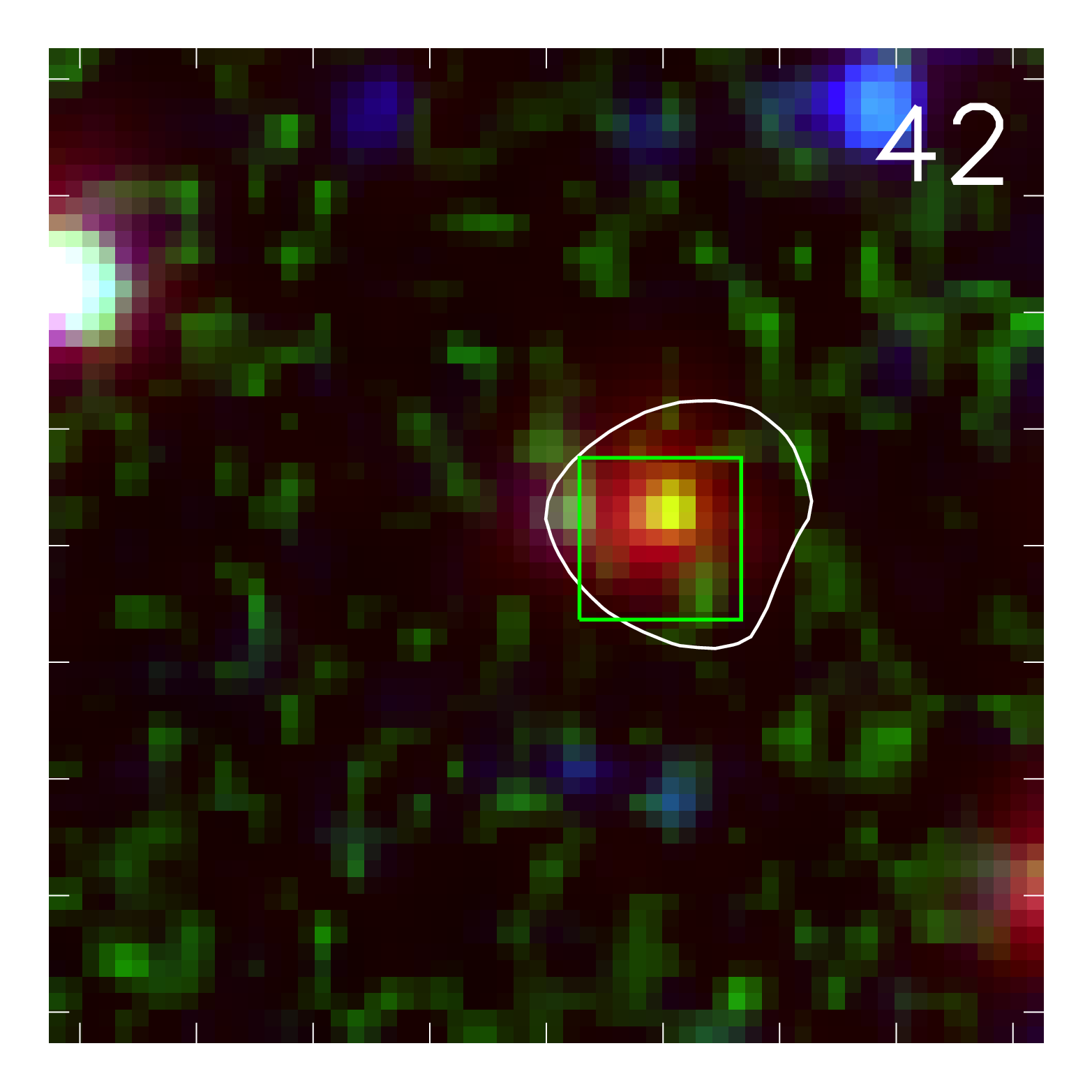,width=0.8in,angle=0}
\psfig{file=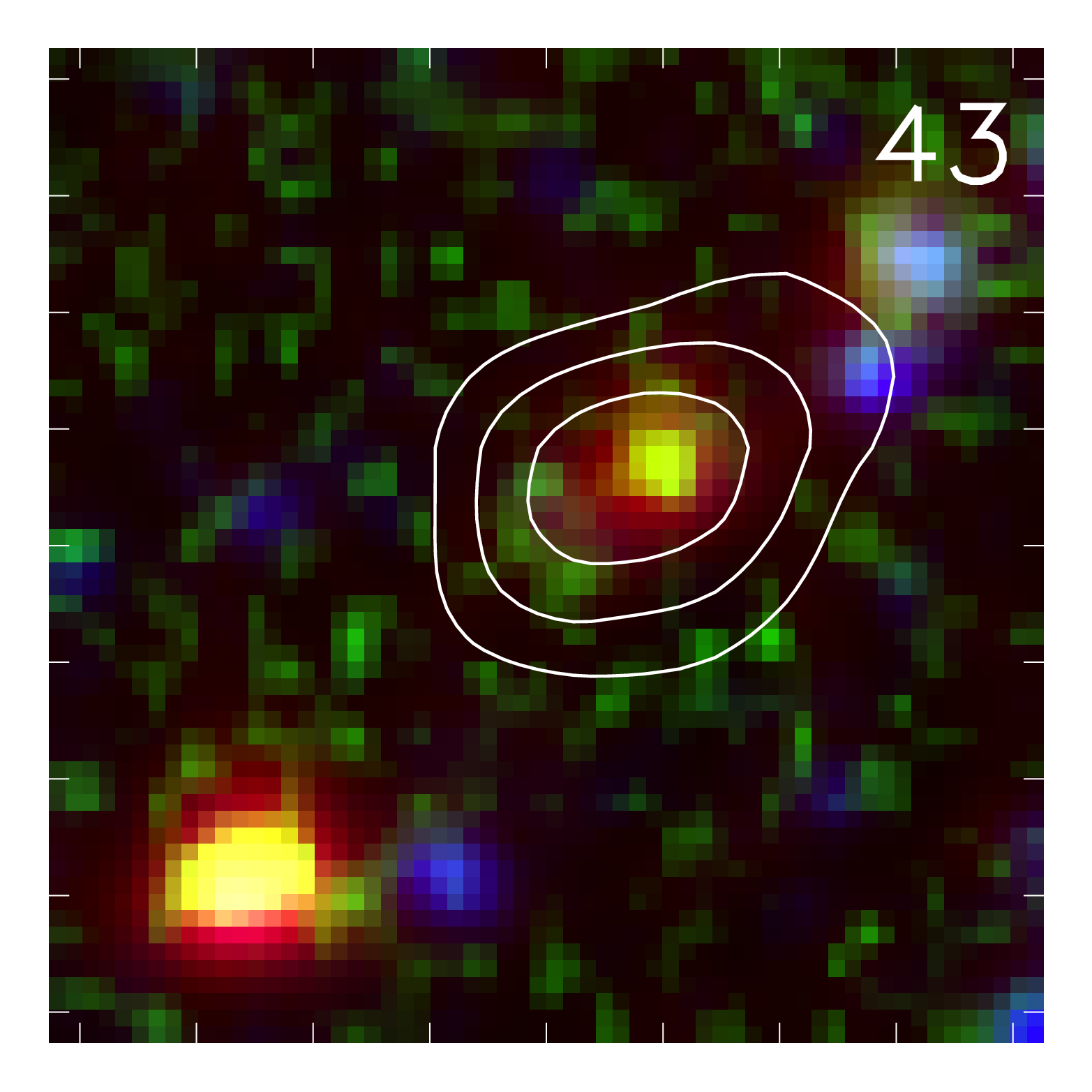,width=0.8in,angle=0}
}

\centerline{
\psfig{file=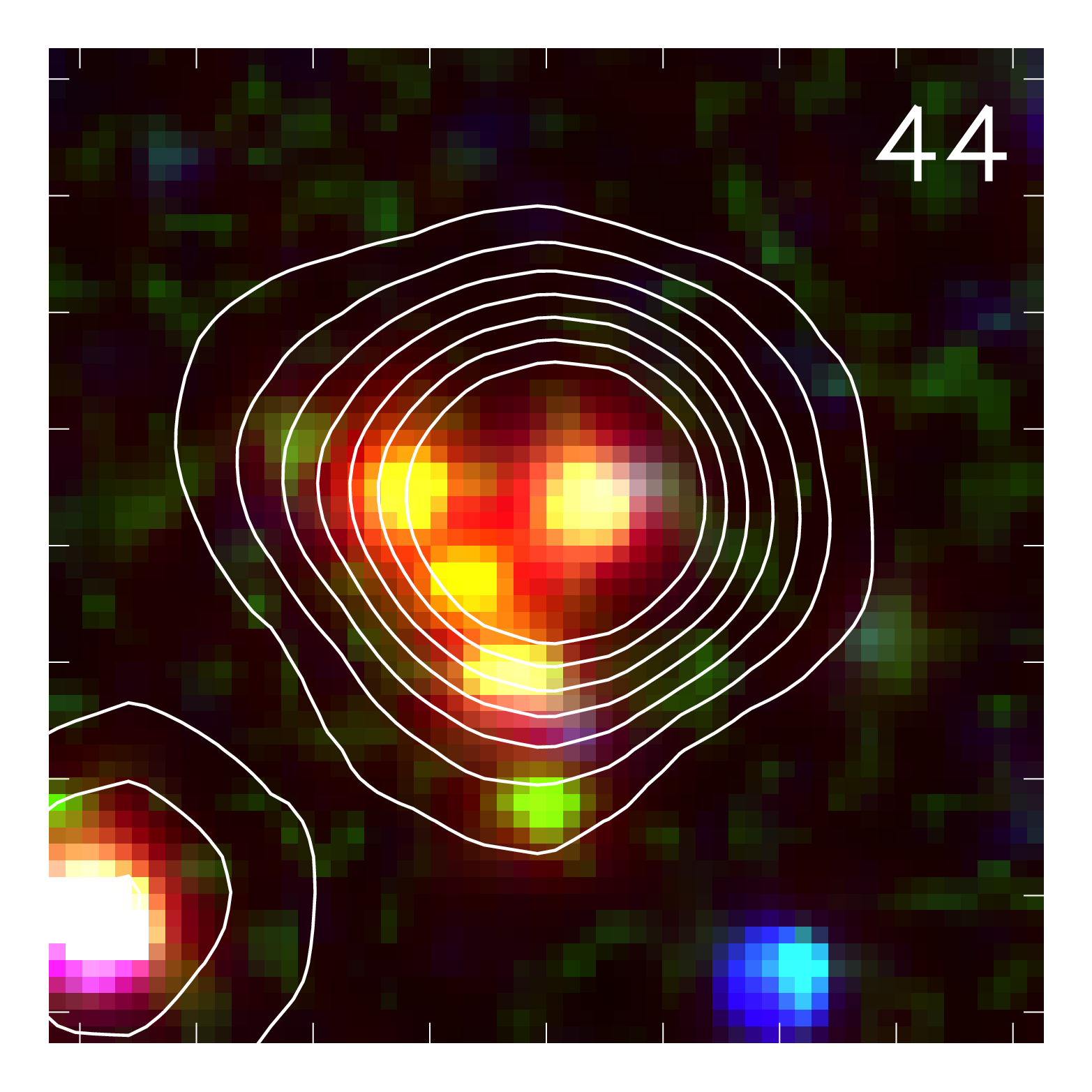,width=0.8in,angle=0}
\psfig{file=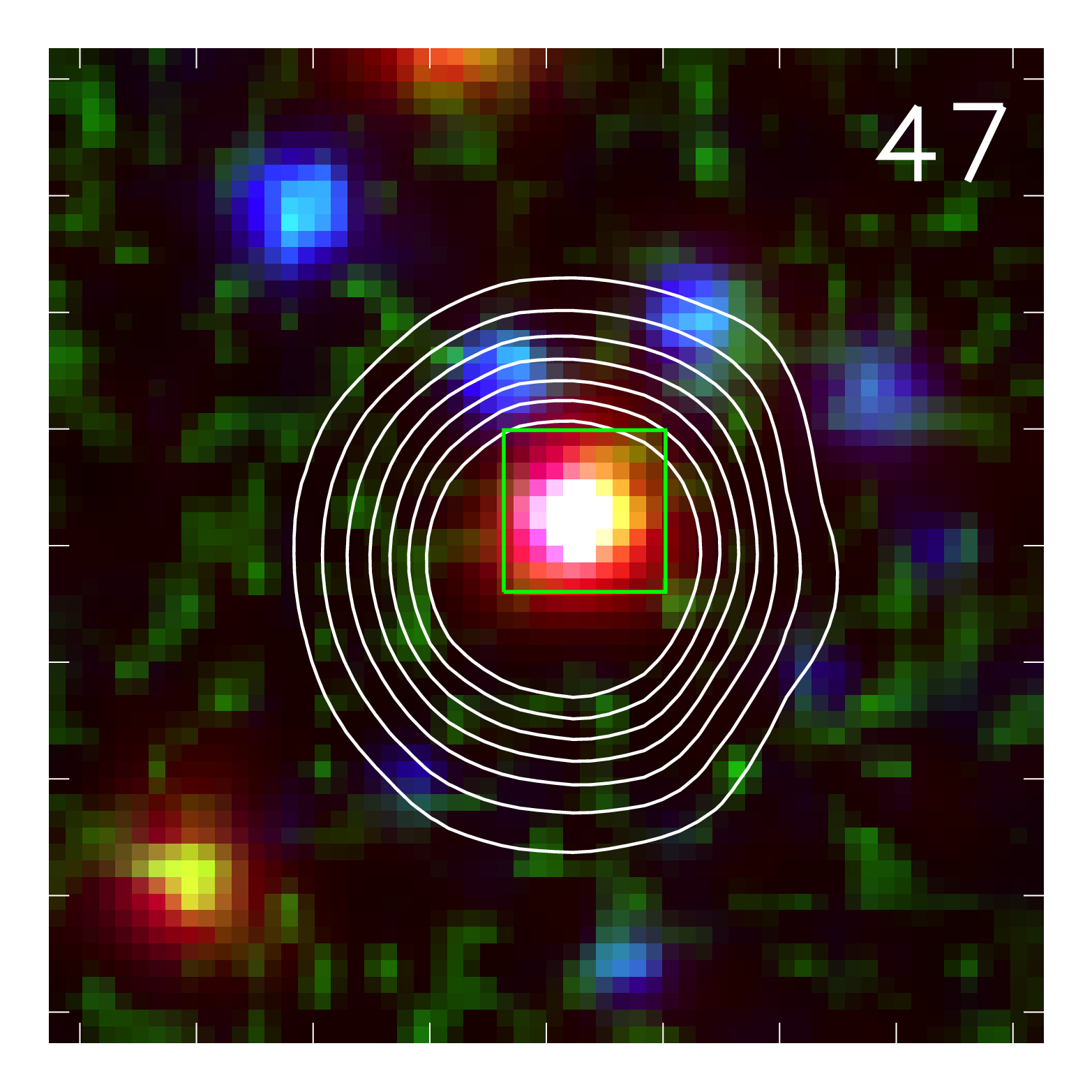,width=0.8in,angle=0}
\psfig{file=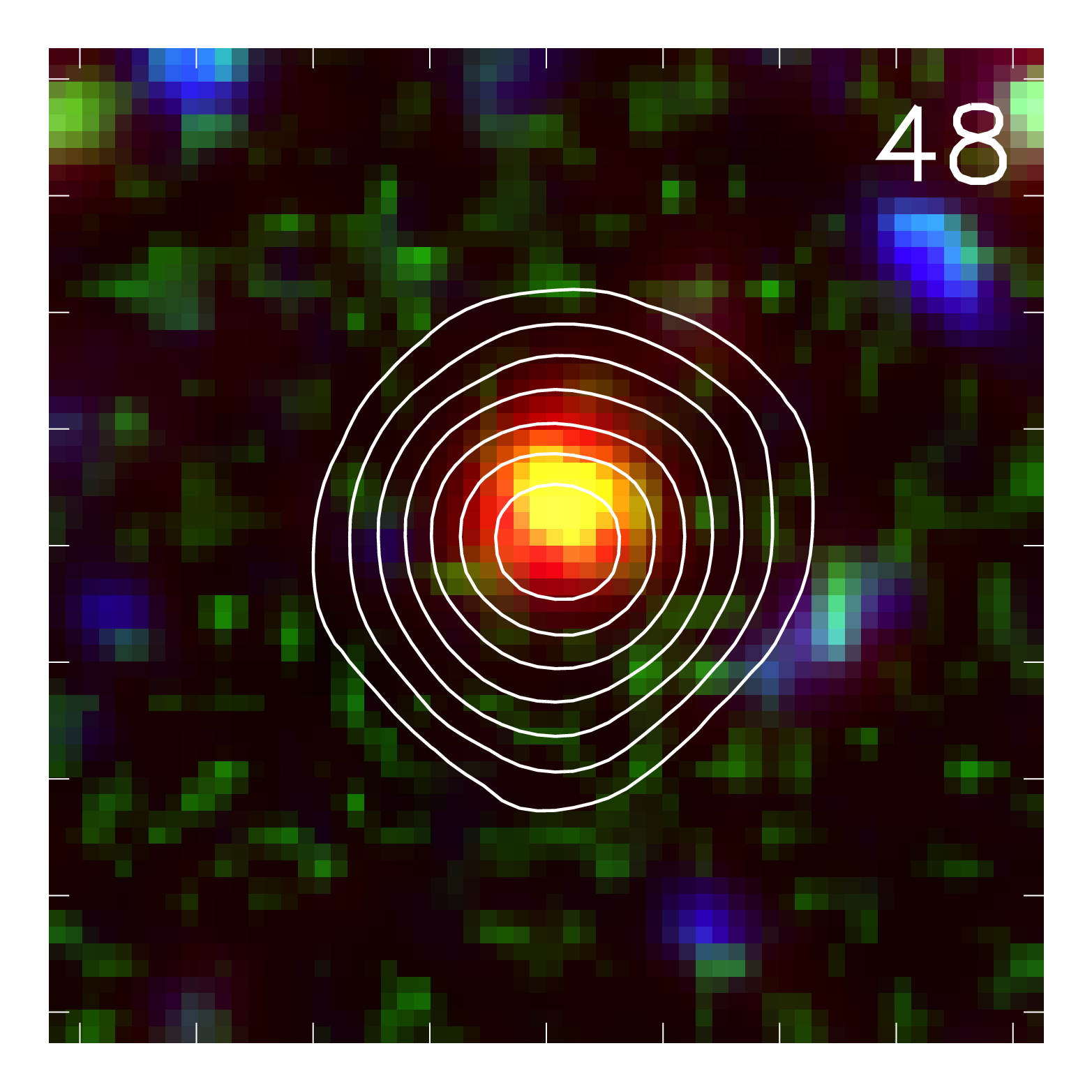,width=0.8in,angle=0}
\psfig{file=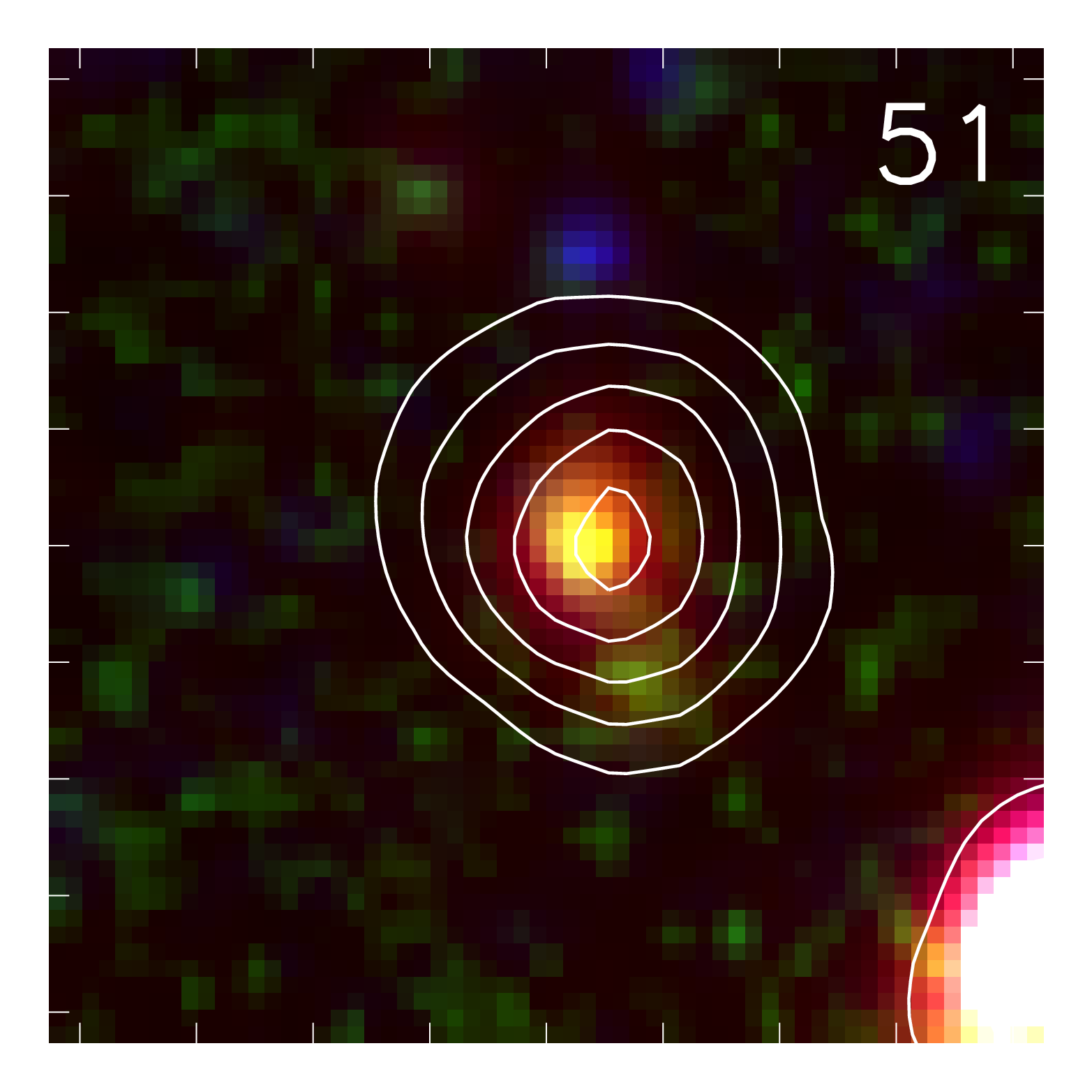,width=0.8in,angle=0}
\psfig{file=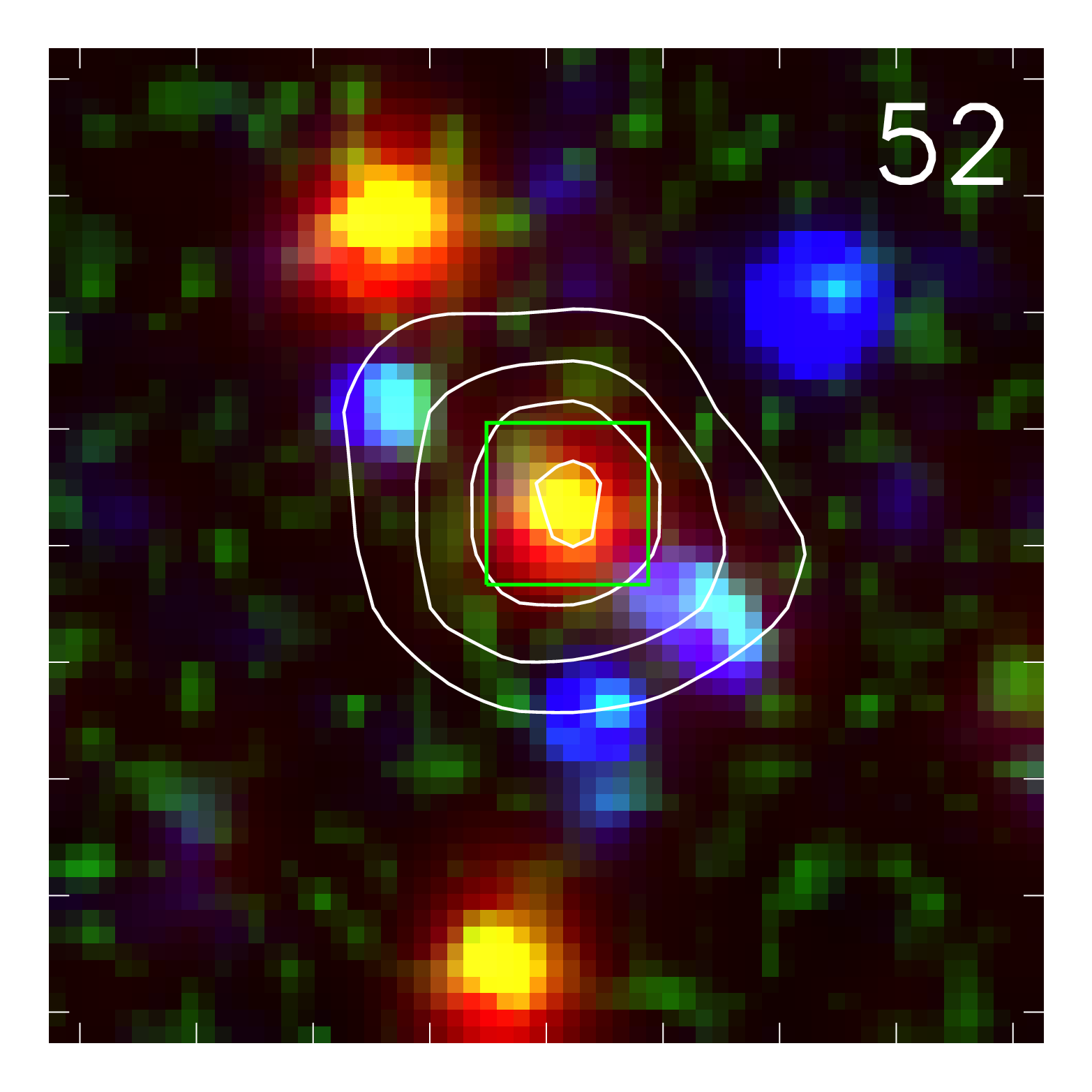,width=0.8in,angle=0}
\psfig{file=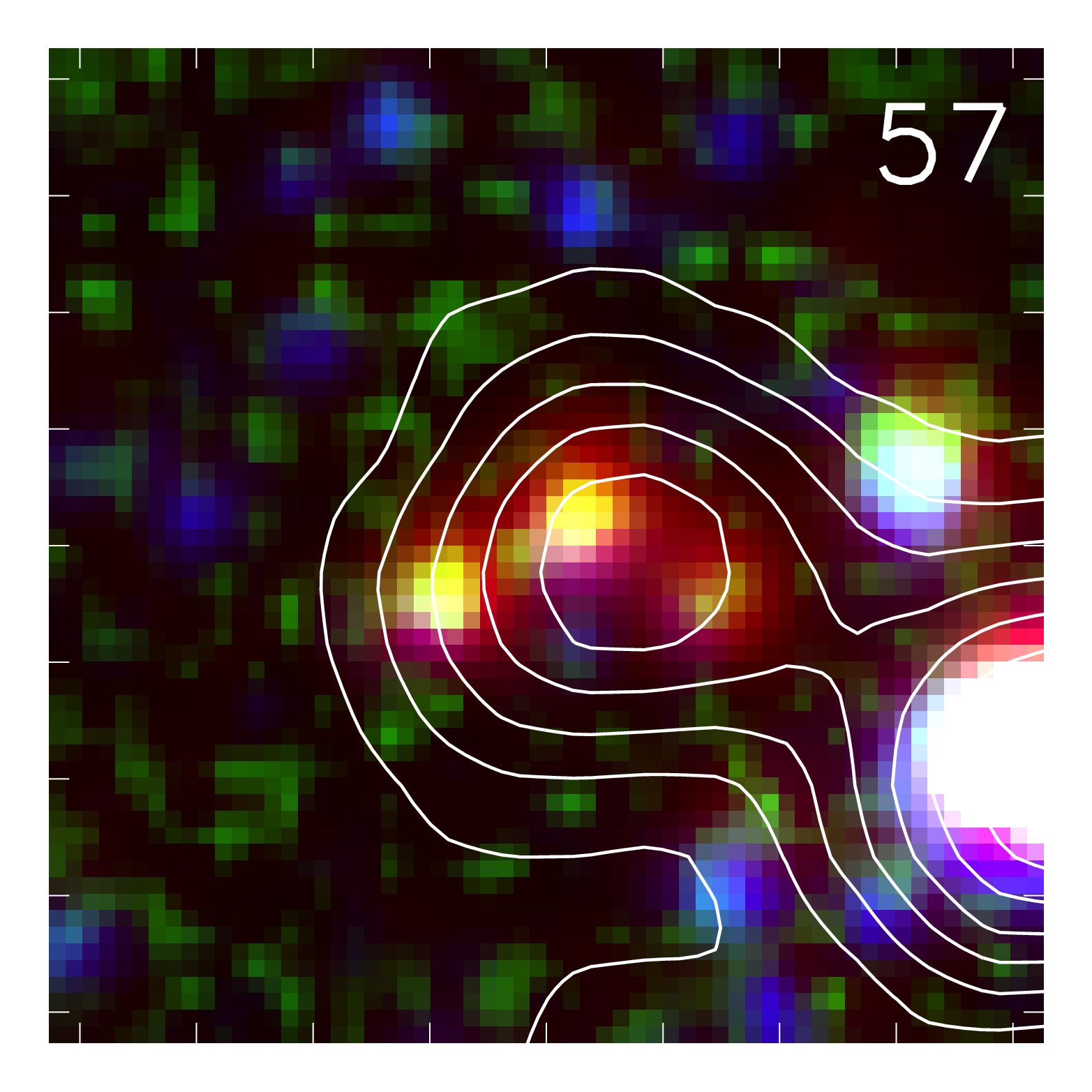,width=0.8in,angle=0}
\psfig{file=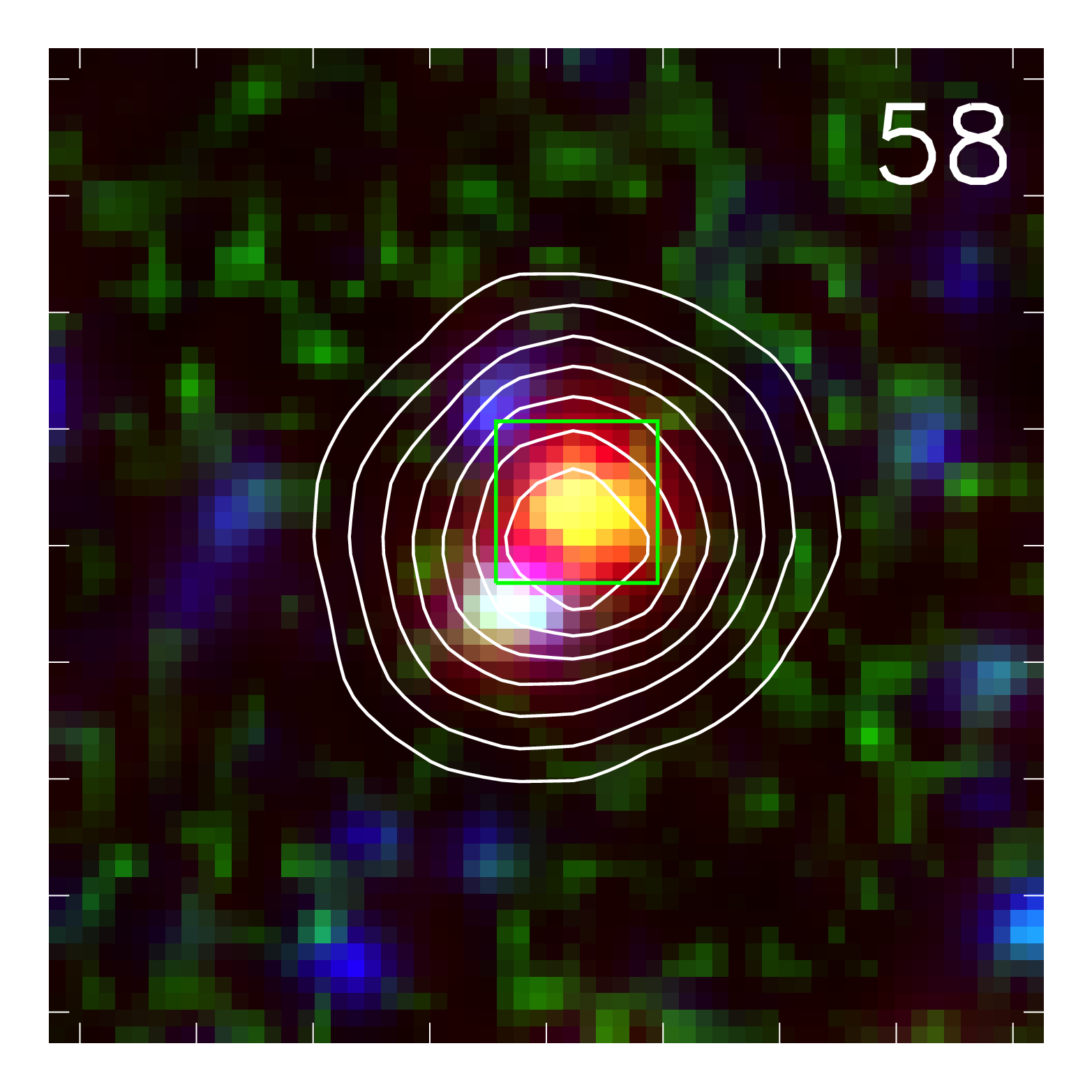,width=0.8in,angle=0}
}
\caption{\small True-color $VK4.5\mu{\rm m}$ images of the 31 SPIRE/SCUBA-2--detected cluster members with 24-$\mu$m contours overlayed (starting at 3\,$\sigma$ of the sky noise and incremented by 1\,$\sigma$).  The far-infrared/submm sources are typically fairly luminous and exhibit red colors, with roughly half of them showing close companions on scales of $\ls $\,30--50\,kpc.   Each panel is 18.6$''$ square (160\,kpc at $z=$\,1.6)  and the tick marks are every 2$''$ with North top and East to the left.  The green squares mark those sources with radio counterparts, the SCUBA-2 detected sources are 29, 37, 40, 57 and 58 and we show the ALMA 870-$\mu$m continuum map from Simpson et al.\ (in prep) as blue contours on the panel for source 37 (starting at 3\,$\sigma$ and incremented by 2\,$\sigma$).}

\end{figure*}

Building upon early mid-infrared studies with {\it ISO} (e.g.\ Coia et al.\ 2005), the MIPS instrument on-board {\it Spitzer} allowed the first deep, truly panoramic mid-infrared searches for obscured starburst activity in distant clusters (e.g.\ Geach et al.\ 2006; Fadda et al.\ 2008; Haines et al.\ 2009; Biviano et al.\ 2011).  These studies found an increasing frequency of mid-infrared  starbursts on the outskirts of clusters out to $z\sim $\,0.5. This is believed to represent the continued in-fall of star-forming galaxies into the clusters and their transformation into passive spheroids.   Surveys of higher-redshift clusters trace the continuing increase in the   star formation in clusters to at least $z\sim $\,1 (Bai et al.\ 2007; Saintonge et al.\ 2008; Finn et al.\ 2010; Koyama et al.\ 2010; Webb et al.\ 2013) and potentially beyond (e.g.\ Koyama et al.\ 2013; Kubo et al.\ 2013). 

Beyond $z\sim 1$ the mid-infrared provides an increasingly unreliable estimate of total star formation in the most obscured systems (e.g.\ Calzetti et al.\ 2006; Hainline et al.\ 2009) and so we instead need to employ far-infrared and submillimeter (submm) tracers.  Small-area submm  surveys of the cores of massive clusters at $z\sim 0.2$--0.5  have confirmed that, apart from activity  associated with a few central cluster galaxies (Edge et al.\ 1999), these regions are indeed devoid of dusty, ultraluminous galaxies (ULIRGs).  However, submm surveys of  the core regions of $z>$\,0.5--1 clusters, have uncovered a population of ultra/luminous infrared galaxies in these environments (Best 2002; Webb et al.\ 2005; Noble et al.\ 2012).  These surveys used SCUBA to map $\sim $\,2$'$-diameter ($\sim $\,1\,Mpc) fields in the cores of around a dozen clusters at $z\sim $\,1 and found a substantial number of ultraluminous starburst galaxies. The resulting 850-$\mu$m counts in these regions exceed the  field  by  $\sim $\,3--4\,$\times$ (Best 2002; Noble et al.\ 2012), suggesting that obscured starburst activity is increasingly common in cluster cores at $z\sim $\,1.  Indeed the {\it Herschel} satellite has expanded the number of  sensitive far-infrared surveys for ULIRGs  in  distant clusters, strengthening the evidence for a continued rise out to at least $z\sim $\,1, with a few rare examples at $z\sim $\,1.4--1.6 (Popesso et al.\ 2010; Santos et al.\ 2013; Pintos-Castro et al.\ 2013).

With the SPIRE instrument (Griffin et al.\ 2010) on {\it Herschel} and the new SCUBA-2 camera (Holland et al.\ 2013) on the JCMT we can  significantly improve our understanding of ultraluminous activity in the most distant clusters. These instruments allow sensitive far-infrared/submm surveys over wide areas necessary to determine if there is a continued increase in the starburst population outside cluster cores (as seen in the mid-infrared studies at lower redshift).  We have therefore exploited new deep submm and far-infrared imaging from SCUBA-2 and SPIRE of the field around the cluster Cl\,0218.3$-$0510 at $z=$\,1.62 to investigate the active galaxy populations in this region.    

Cl\,0218.3$-$0510 was first identified through both an overdensity of red galaxies in the SWIRE survey's {\it Spitzer} IRAC imaging of the UKIDSS UDS field and as a potentially distant X-ray source in {\it XMM}-Newton imaging (Papovich et al.\ 2010; Finoguenov et al.\ 2010; Tanaka et al.\ 2010).  It has been  extensively studied (Pierre et al.\ 2012; Rudnick et al.\ 2012; Papovich et al.\ 2012; Tadaki et al.\ 2012), most notably  Tran et al.\ (2010) used 24-$\mu$m imaging from {\it Spitzer} MIPS to identify an apparent increase in the star-forming fraction in the highest density environments, the reverse of what is universally seen in clusters at lower redshifts.   
However, the interpretation in terms of star formation  of the mid-infrared emission from galaxies in this 
$z\sim$\,1.62 structure  is particularly difficult as the MIPS 24-$\mu$m filter samples not only the redshifted polycyclic aromatic hydrocarbon (PAH) feature at 7.7$\mu$m, which is expected to trace star formation, but also the 9.7-$\mu$m Silicate absorption feature which is  strong in obscured AGN or particularly dense starbursts.   The goal of this work is therefore is to build upon these earlier studies by undertaking a survey of the far-infrared and submm emission from luminous star-forming galaxies within the cluster and its surroundings and so investigate the evolutionary state of the galaxy populations in this structure.  

We note that two days before this paper was accepted, a similar analysis was presented by Santos et al.\ (2014), using similar SPIRE, MIPS, photometric and [O{\sc ii}] samples.  The main differences between the two analyses are that we include SCUBA-2 in our far-infrared analysis, radio and MIPS catalogs in our deblending and an independent photometric redshift analysis.  Santos et al.\ come to broadly similar conclusions to those presented here.

In \S2 we describe the new and archival observations used in our analysis.  In \S3 we present our analysis and results, while in \S4 we discuss these and give our main conclusions.  In our analysis we assume a cosmology with $\Omega_{\rm M}$\,$=$\,0.27, $\Omega_\Lambda$\,$=$\,0.73 and $H_0$\,$=$\,71\,km\,s$^{-1}$\,Mpc$^{-1}$, giving an angular scale of 8.6\,kpc\,arcsec$^{-1}$ at $z=$\,1.62 and an age of the Universe at this redshift of 4.0\,Gyr.  All quoted magnitudes are on the AB system and errors on median values are derived from bootstrap resampling.

\section{Observations, Reduction and Analysis}

Our analysis involves both SCUBA-2 850\,$\mu$m observations and {\it Herschel} SPIRE maps at 250--500\,$\mu$m to provide a census of the far-infrared/submm population in our survey area.  To locate the counterparts to these sources (where the  maps have relatively poor spatial resolution) we also use observations at higher spatial resolution in the radio and mid-infrared from the VLA and {\it Spitzer} respectively.  Finally, to determine if these counterparts are probable members of the $z=$\,1.6 structure in this field we use redshift information from spectroscopic and photometric redshifts, as well as narrow-band imaging of this field.  We now describe each of these elements of the analysis.

\subsection{SCUBA-2}

As part of the SCUBA-2 Cosmology Legacy Survey (S2CLS) observations  with SCUBA-2 were obtained in Band 2 and upper-Band 3 weather conditions ($\tau_{\rm 225\,GHz}=$\,0.05--0.10) between 2011 October and 2013 February. To-date these total 130\,hours of on-sky integration. The field center for the S2CLS UDS observations is 02\,17\,49.2, $-$05\,05\,54  (J2000) to ensure maximum overlap with the existing UKIDSS/UDS coverage of this field (Lawrence et al.\ 2007). The S2CLS observations employ a  3300$''$-diameter {\sc pong} pattern to uniformly cover the full field.

The 195  individual 40\,min 850-$\mu$m scans are reduced using the   map-maker from the {\sc smurf} package (Jenness et al.\ 2011; Chapin et al.\  2013).  This involves the raw data being first flat-fielded using ramps bracketing every science observation and then scaling the data to units of pW. The dynamic iterative map maker assumes that the signal recorded by a bolometer is a linear combination of: i) a common mode signal dominated by atmospheric water and ambient thermal emission; ii) the astronomical signal (attenuated by atmospheric extinction); and iii) a noise term, reflecting any additional signal not accounted for by i) or ii). The map maker attempts to solve for these  components, refining the model until convergence is met, an acceptable tolerance has been reached, or a fixed number of iterations has been exhausted. This yields  time series signals for each bolometer that ought to comprise just the astronomical signal, corrected for extinction, plus noise. 

Filtering of the time series is performed in the frequency domain, with band-pass filters equivalent to angular scales of $\theta=$\,2--120$''$. The reduction also includes the usual filtering steps of spike removal ($>$\,10-$\sigma$ deviations in a moving boxcar) and DC step corrections.  The signal from each bolometer's time series is then projected onto a map, using the scan pattern, with the contribution to a given pixel weighted according to its time-domain variance (which is also used to estimate the $\chi^2$ tolerance in the fit derived by the map maker). Throughout the iterative map making process, bad bolometers (those significantly deviating from the model) are flagged and do not contribute to the final map.  Maps from independent scans are co-added in an optimal stack using the variance of the data contributing to each pixel to weight spatially-aligned pixels.  The average exposure time per 4$''$ pixel in the central $\sim$\,0.9\,degree-diameter region is $\sim$\,1.5\,ks. Finally, since we are interested in (generally faint) extragalactic point sources, we apply a beam matched filter to improve point source detectability, resulting in a map  with a noise level of 1.9\,mJy rms.   

For calibration we adopt the sky opacity relation for SCUBA-2 which has been obtained by fitting extinction models to hundreds of standard calibrators observed since the commissioning of SCUBA-2 (Dempsey et al.\ 2013).  The flux calibration for SCUBA-2 data has been examined by analysing all flux calibration observations since Summer 2011 until the date of observation. The derived beam-matched flux conversion factor (FCF) has been found to be reasonably stable over this period, and the average FCFs agree (within error) with those derived from the subset of standard calibrators observed on the nights of the observations presented here. Therefore we have adopted the canonical calibration of ${\rm FCF}_{850}=$\,537$\pm$26\,Jy\,beam$^{-1}$\,pW$^{-1}$ here. A correction of $\sim$\,10 per cent is included in order to compensate for flux lost due to filtering in the blank-field map. This is estimated by inserting a bright Gaussian point source into the time stream of each observation to measure the response of the model source to filtering.

%
%
\setcounter{figure}{2} 
\begin{inlinefigure}\vspace{6pt}
\centerline{\psfig{file=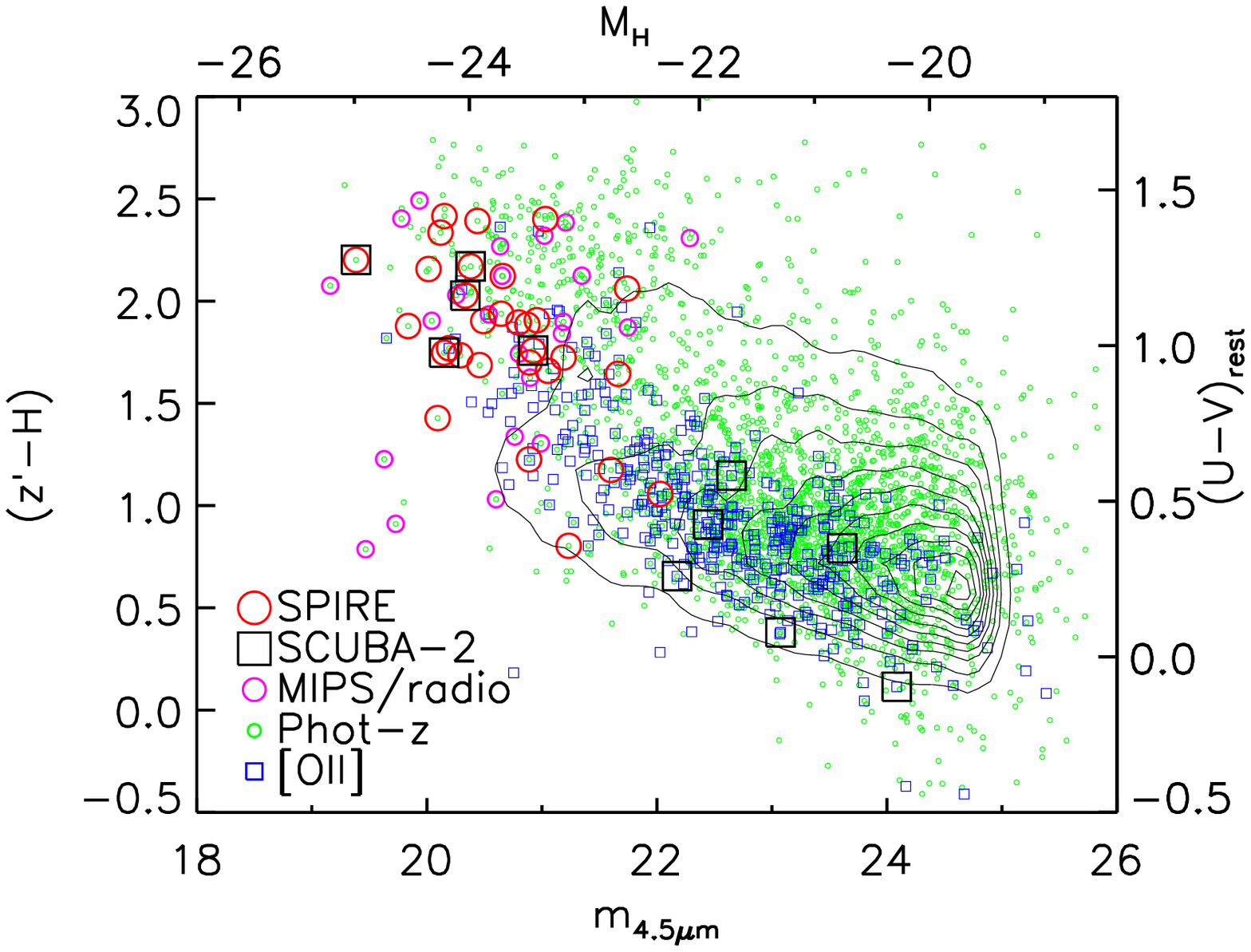,width=4.0in,angle=90}}
 \vspace{6pt}
\noindent{\small {\small \sc Fig.~3 --- }  The observed $(z'-H)$--$4.5\mu{\rm m}$ distribution for the confirmed and probable cluster members identified by their different selection techniques (see \S2.4).  These observables roughly map to $(U-V)$--$M_{\rm H}$ in the restframe at $z\sim 1.6$.   Most of the photometric members define a blue cloud, which is also seen in the [O{\sc ii}] emitters, with the remainder inhabiting a more luminous and redder clump at $(z'-H)\gs$\,1.5. The far-infrared/submm-detected  and the MIPS/radio sources are distributed very differently, with most of them having redder colors and brighter restframe $H$-band luminosities than the less-active populations. We also mark the six statistically-associated [O{\sc ii}] counterparts to five of the SCUBA-2 sources, which have much bluer colors than the bulk of the SPIRE/SCUBA-2 counterparts and which we exclude from our analysis (see \S2.5).  Finally, we compare the distribution of galaxies to that predicted by the Millennium simulation (see \S3.1) using the galaxy evolution model of Font et al.\ (2008), where the number density of sources in the theoretical model is shown as contours starting at 5 per cent of the peak density and incremented by 10 per cent.  While the model successfully reproduces the colors of the bulk of the star-forming galaxies, it clearly underpredicts the number of the reddest and brightest galaxies, both passive and active.

} 
\end{inlinefigure}

To identify the submm sources, we search the beam-convolved  signal-to-noise ratio map  for pixels above a threshold $>\Sigma_{\rm thresh}$. If a pixel is found, we record the peak-pixel sky coordinate, flux density and noise and mask-out a circular region equivalent to $\sim$\,1.5\,$\times$ the size of the 15$''$ beam at 850\,$\mu$m.  We then reduce $\Sigma_{\rm thresh}$ by a small amount and  repeat the search. The catalog limit, below which we no longer trust the reality of ``detections'' is  the signal-to-noise level at which the contamination rate due to false detections (expected from pure Gaussian noise) exceeds 5 per cent, corresponding to a significance of $\sigma\sim$\,3.75 (we confirm this using the simulations described next). We detect 97 discrete point sources above this significance within the survey area used in this analysis (Figure~1).

The completeness of our catalog is estimated by injecting artificial point sources into a blank map with the same noise properties as our real map. To create this map we apply a jackknife approach and randomly invert half of the individual time series  before co-addition (e.g.\ Weiss et al.\ 2009).  The recovery rate of 10$^5$ sources  (inserted in groups of ten with fluxes drawn from a uniform distribution) then gives the completeness function. These simulated maps also allow us to estimate the noise-dependent flux boosting that occurs for sources with true fluxes close to the noise limit of the map, and so we can determine  a statistical correction to de-boost the fluxes measured of  sources in the real map; typically this correction is $\mathcal{B}<$\,10\%. Finally, the source detection algorithm is applied to each of the jackknife maps with no fake sources injected in order to evaluate the false positive rate, which we find to be 5 per cent, in agreement with the false detection rate expected for a map of this size assuming fluctuations from pure Gaussian noise.

We also stack both the SCUBA-2 maps and the SPIRE maps described below,  to derive statistical measurements of the typical far-infrared/submm luminosities of the less active cluster populations.  This stacking involves extracting thumbnail maps around each source to be stacked and then median-combining these, before measuring the average source flux for the sample.

\subsection{Radio and MIPS 24$\mu$m}

In our analysis we are seeking to identify far-infrared/submm sources which are potential members of the $z=$\,1.6 structure around Cl\,0218.3$-$0510.  The spatial resolution of the far-infrared/submm data is typically poor (15--30$''$ FWHM) and several of these maps are also confused.  Hence to  locate the sources of  emission at these wavelengths we need to use observations with higher spatial resolution, but in wavebands where there is likely to be a correspondence with the far-infrared/submm emission.   The most commonly employed proxies for this purpose are the radio and mid-infrared (e.g.\ Ivison et al.\ 1998; 2004).  We therefore use catalogs of sources detected at 1.4\,GHz with the VLA and at 24\,$\mu$m using {\it Spitzer}.

UDS20 is a VLA 1.4-GHz survey of the UDS field and the reduction, analysis and the full catalog of sources from the survey are presented in Arumugam et al.\ (2013).  The observations comprise a mosaic of 14 pointings covering a total area of $\sim$\,1.3\,degrees$^2$ centered on the UDS and build upon the earlier A-array observations of this field in Simpson et al.\ (2006).  UDS20 employs 160-hrs of integration with the VLA in A, B and C--D configurations at 1.4\,GHz, yielding an almost constant rms noise of $\sim $\,10\,$\mu$Jy across the full field ($<$\,8\,$\mu$Jy at the field center) and a beam size of 1.8$''$ FWHM.  There are $\sim$\,5,100 sources detected at $>$4-$\sigma$ significance within the UDS. 

The UDS was also surveyed in the mid-infrared for the {\it Spitzer}  Legacy Program SpUDS (PI: J.\ Dunlop).  The SpUDS imaging covers the entire UDS survey area using both the IRAC (3.6, 4.5, 5.8 and 8\,$\mu$m) and MIPS (24$\mu$m) cameras. Reduced maps  and source catalogs released by SpUDS  can be found online\footnote{\tt http://irsa.ipac.caltech.edu/data/SPITZER/SpUDS/}.   The MIPS image has an effective 1-$\sigma$ depth of $\sim$\,8\,$\mu$Jy and we use the  SpUDS catalog based on this image in our search for counterparts to the far-infrared/submm sources in this field.  The IRAC imaging from SpUDS is also an important addition in the derivation of photometric redshifts for sources in the UDS field (see \S 2.4).

\subsection{SPIRE/Herschel}

SPIRE 250- 350- and 500-$\mu$m observations of the UDS were taken as part of the {\it Herschel} Multi-tiered Extra-galactic Survey (HerMES) guaranteed time program (Oliver et al.\ 2012).  The total exposure time for the UDS was 37.8\,ks broken into 5.4-ks blocks.  For each observation, we retrieved the Level 2 data product from the {\it Herschel} ESA archive and aligned and coadded the maps. The final combined maps reach 1-$\sigma$ noise levels of 2.3, 2.0 and 2.7\,mJy at 250, 350 and 500\,$\mu$m respectively (see Oliver et al.\ 2012).  To match the SPIRE maps to the astrometric reference frame of the radio map, we stacked the SPIRE maps at the VLA radio positions, centroiding the stacked emission and deriving shifts of $\Delta < $\,1.5$''$ which we apply to each map.

These deep SPIRE maps have coarse beams and are heavily confused, hence to measure reliable far-infrared fluxes we need to deblend them. We employ catalogs of {\it Spitzer} 24-$\mu$m and 1.4-GHz VLA sources (detected at $>$\,5\,$\sigma$) and remove any sources within 1.5$''$ of each other as duplicates) to construct a master ``prior'' catalog to deblend the SPIRE maps.  We use a Monte Carlo algorithm to deblend the maps and give a short description of this here (it is described in full in Swinbank et al.\ (2014), which also describes the extensive tests and simulations employed to confirm its reliability).\footnote{The stacked and deblended images and catalogs of deblended sources are all available on http://www.astro.dur.ac.uk/$\sim$ams/HSODeblend/}  We split the area up into smaller sub-areas for deblending and fit beams at the positions of sources from  our prior catalog, optimising their amplitudes to best match the map.  To ensure we do not ``over-deblend'' the longer wavelength maps (where the beam is larger), when deblending the 350-$\mu$m map, we only include sources from the prior catalog which are detected at $>$\,2\,$\sigma$ at 250\,$\mu$m as priors for the 350\,$\mu$m deblending (and similarly, for the 500$\mu$m map we only include sources detected at $> $\,2\,$\sigma$ at 350\,$\mu$m).  Appropriate errors and limits for non-detections are derived from simulations.  Swinbank et al.\ (2014) describe the simulations and comparisons with published catalogs which are used to confirm the reliability of their analysis. For UDS these indicate 3-$\sigma$ detection limits of 9.2, 10.6 and 12.2\,mJy at 250, 350 and 500\,$\mu$m respectively (comparable to those derived from the XID deblending procedure employed by HerMES, see Roseboom et al.\ 2010).

%
%
\setcounter{figure}{3}
\begin{figure*}[tbh]
\centerline{
\psfig{file=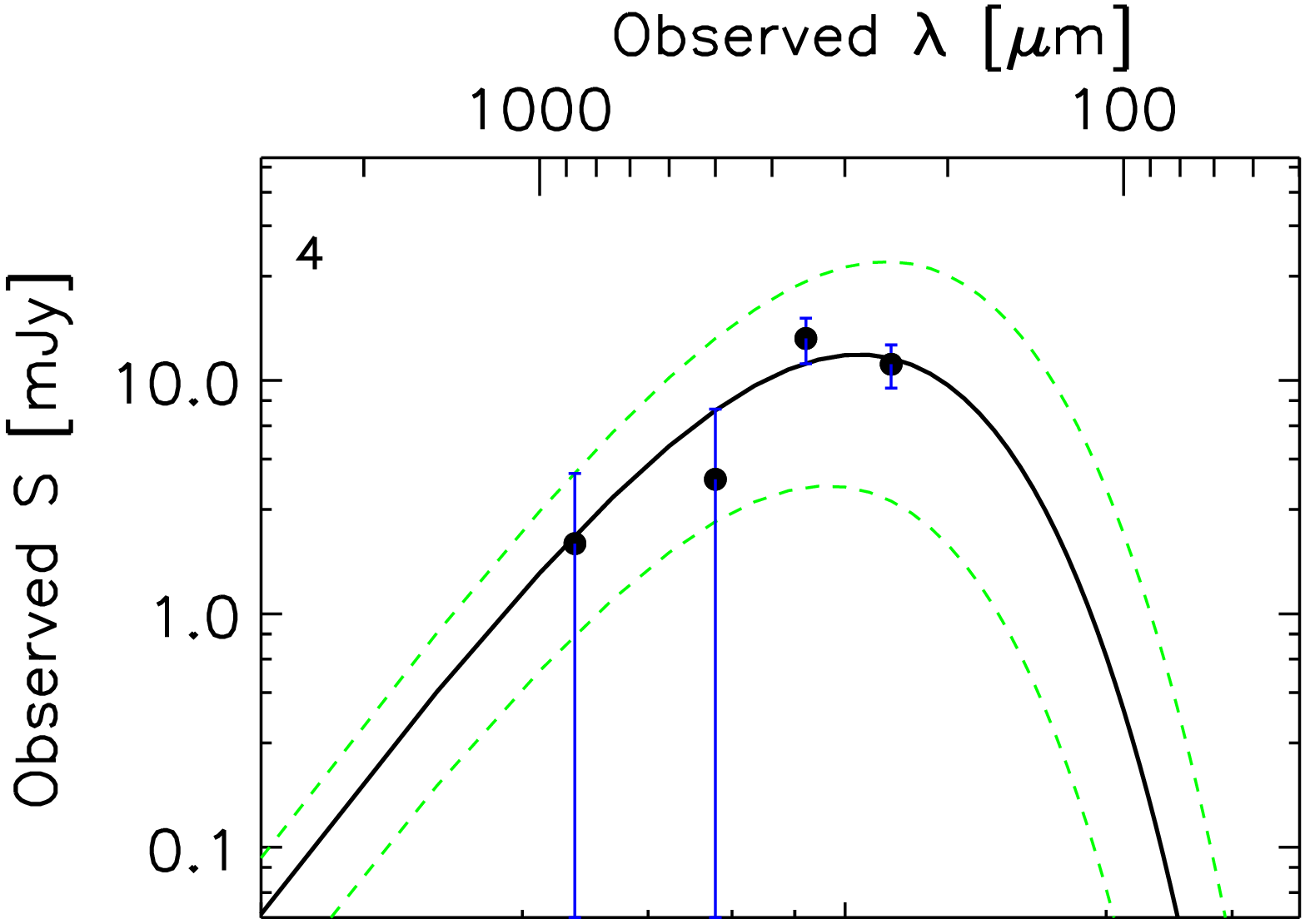,width=1.5in,angle=90}
\psfig{file=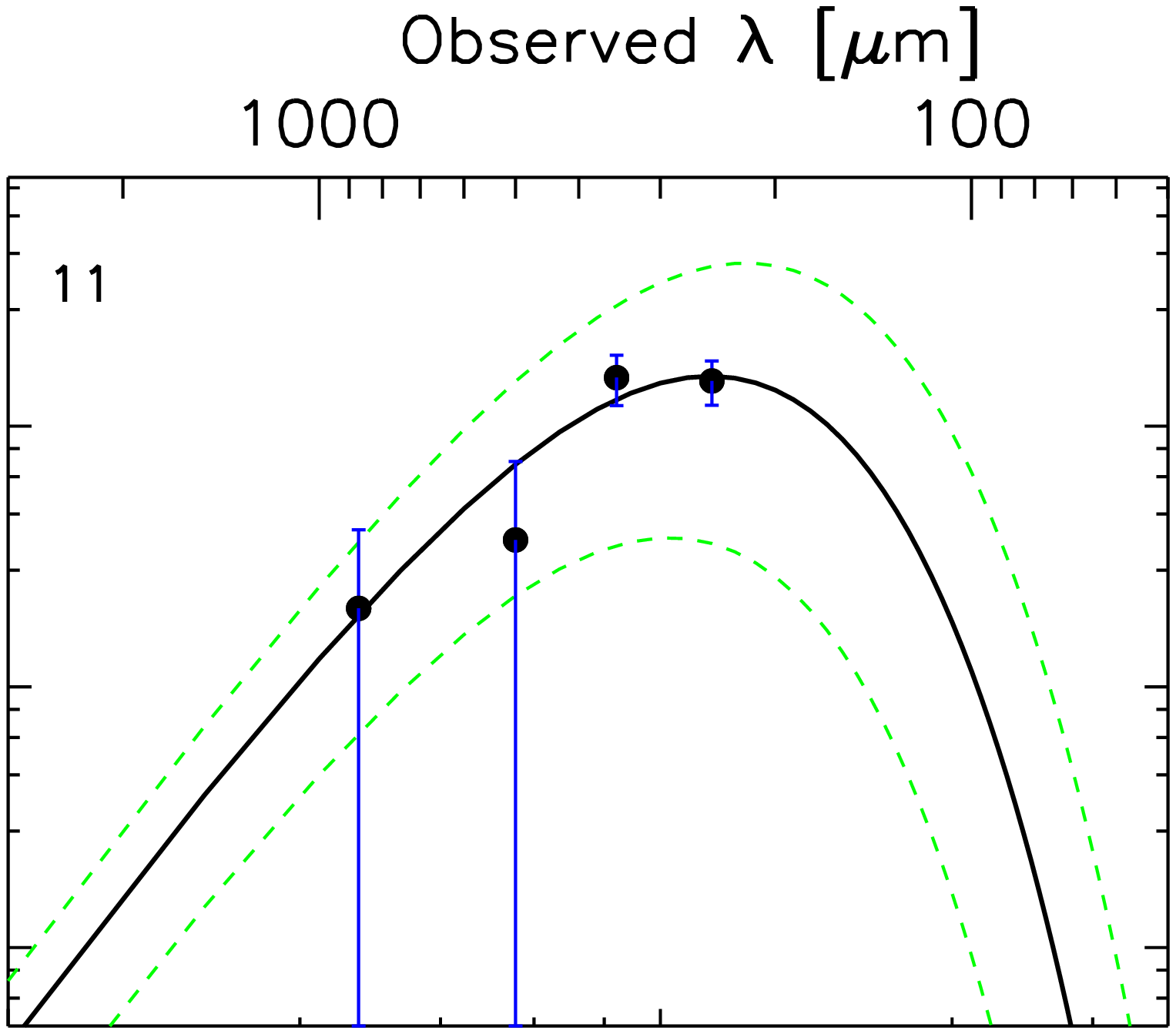,width=1.5in,angle=90}
\psfig{file=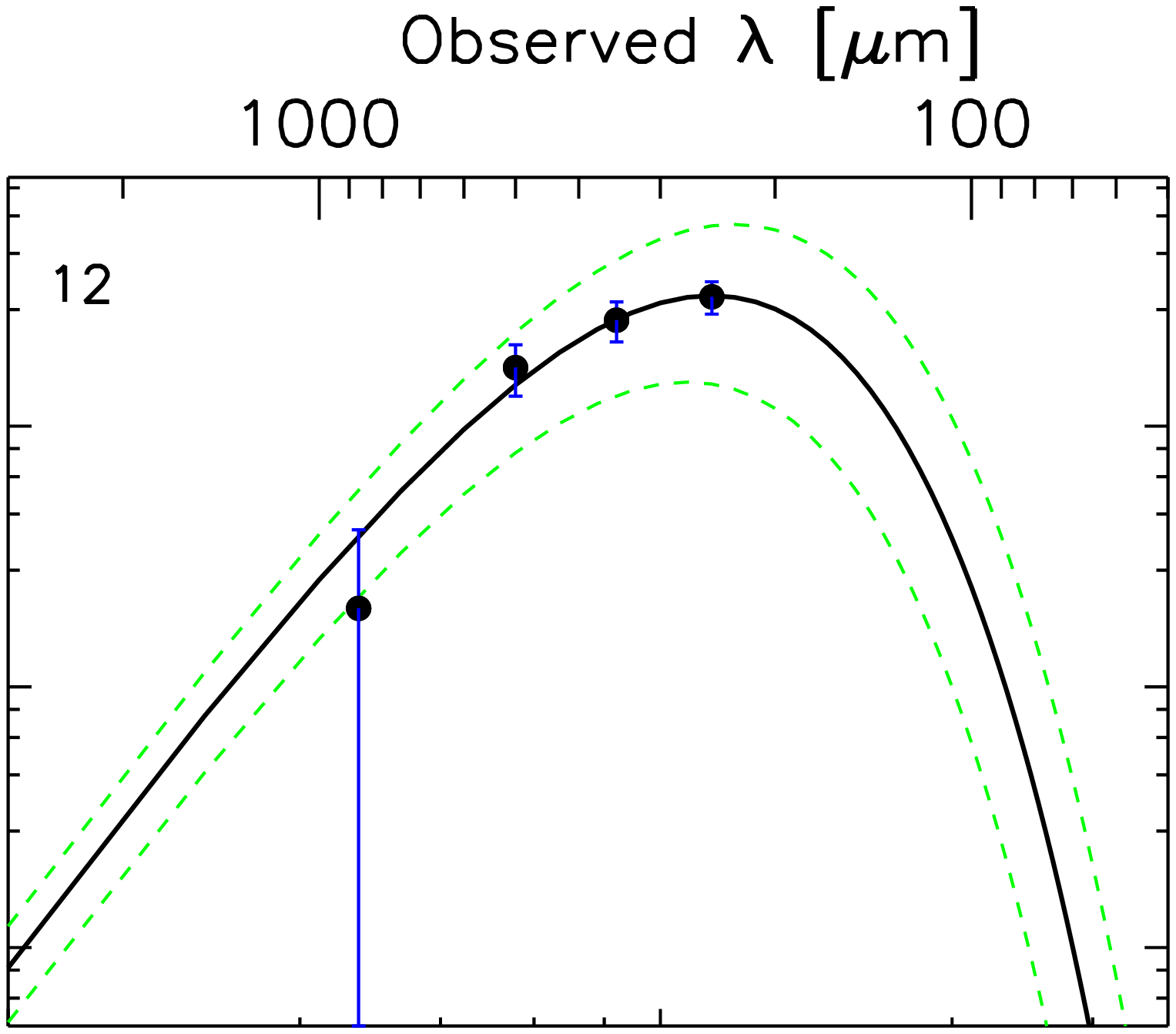,width=1.5in,angle=90}
\psfig{file=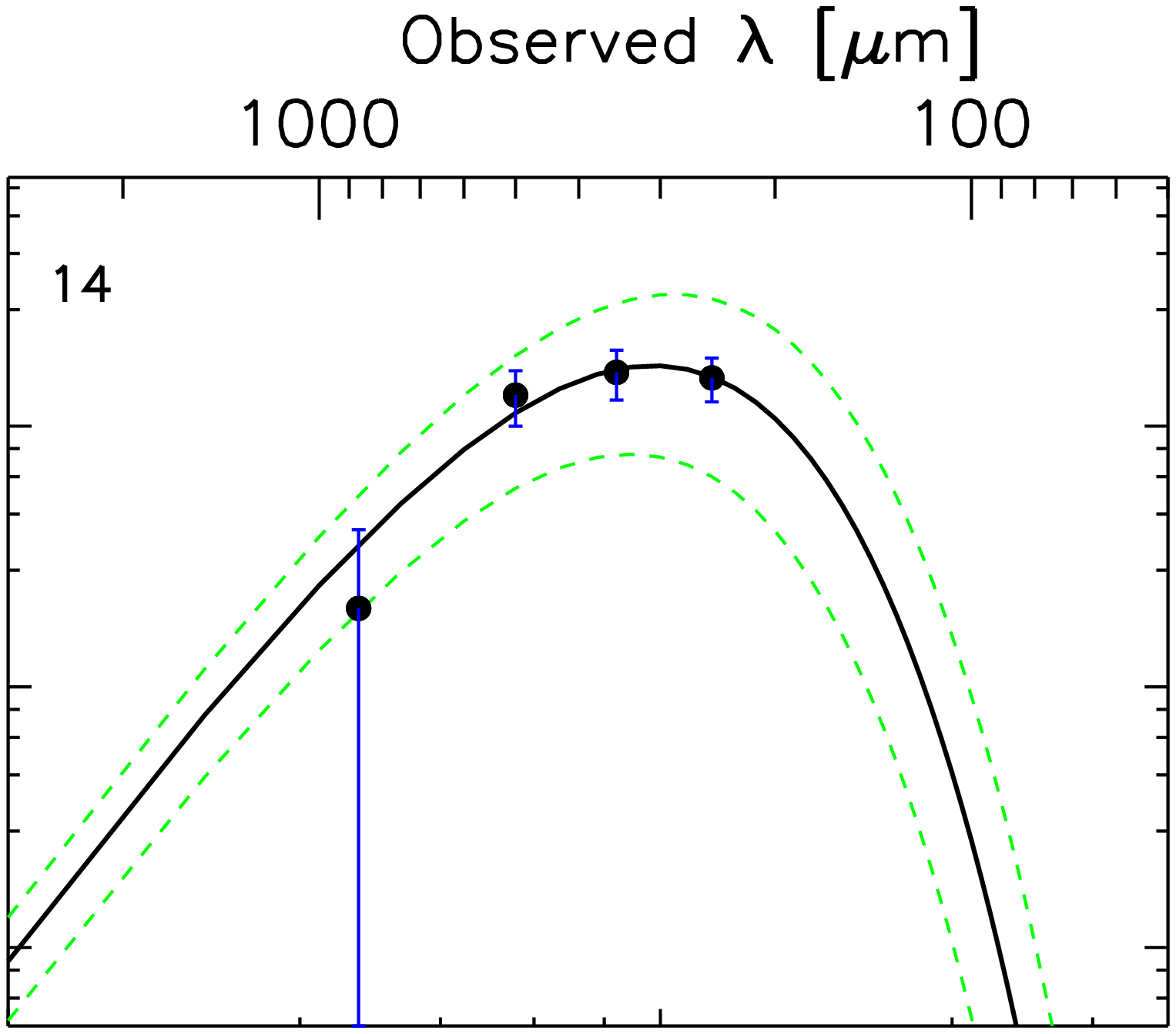,width=1.5in,angle=90}
}\centerline{
\psfig{file=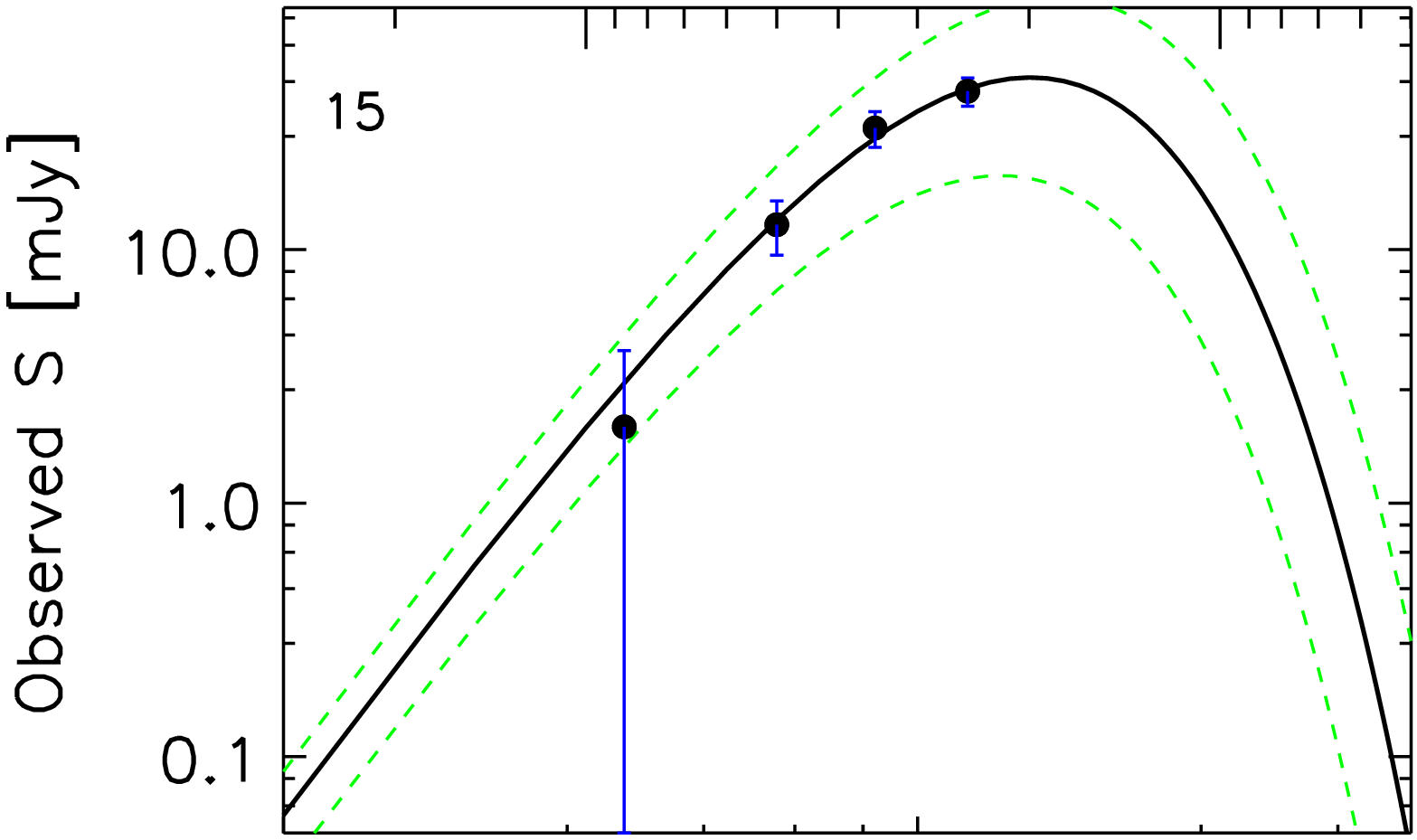,width=1.5in,angle=90}
\psfig{file=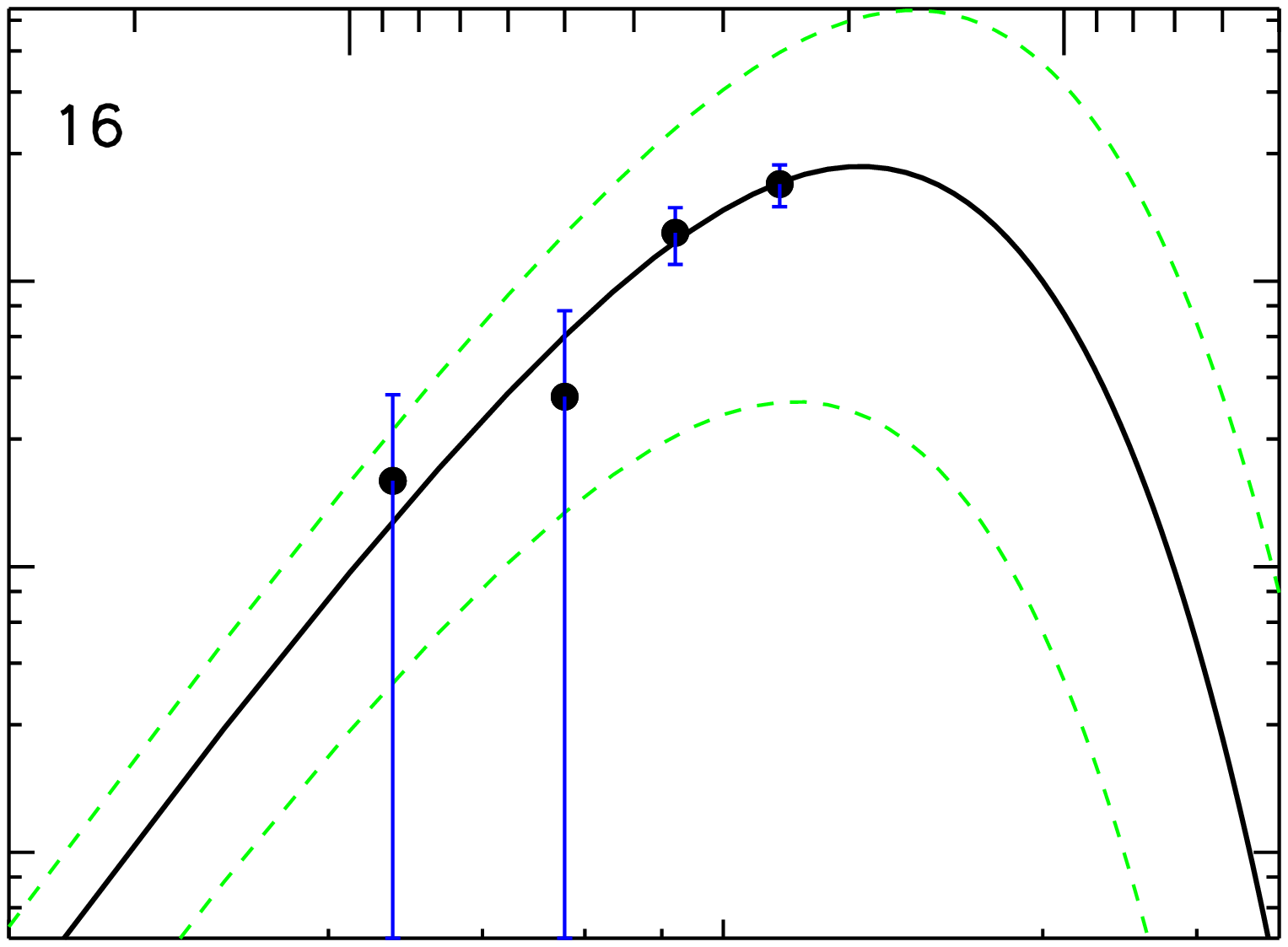,width=1.5in,angle=90}
\psfig{file=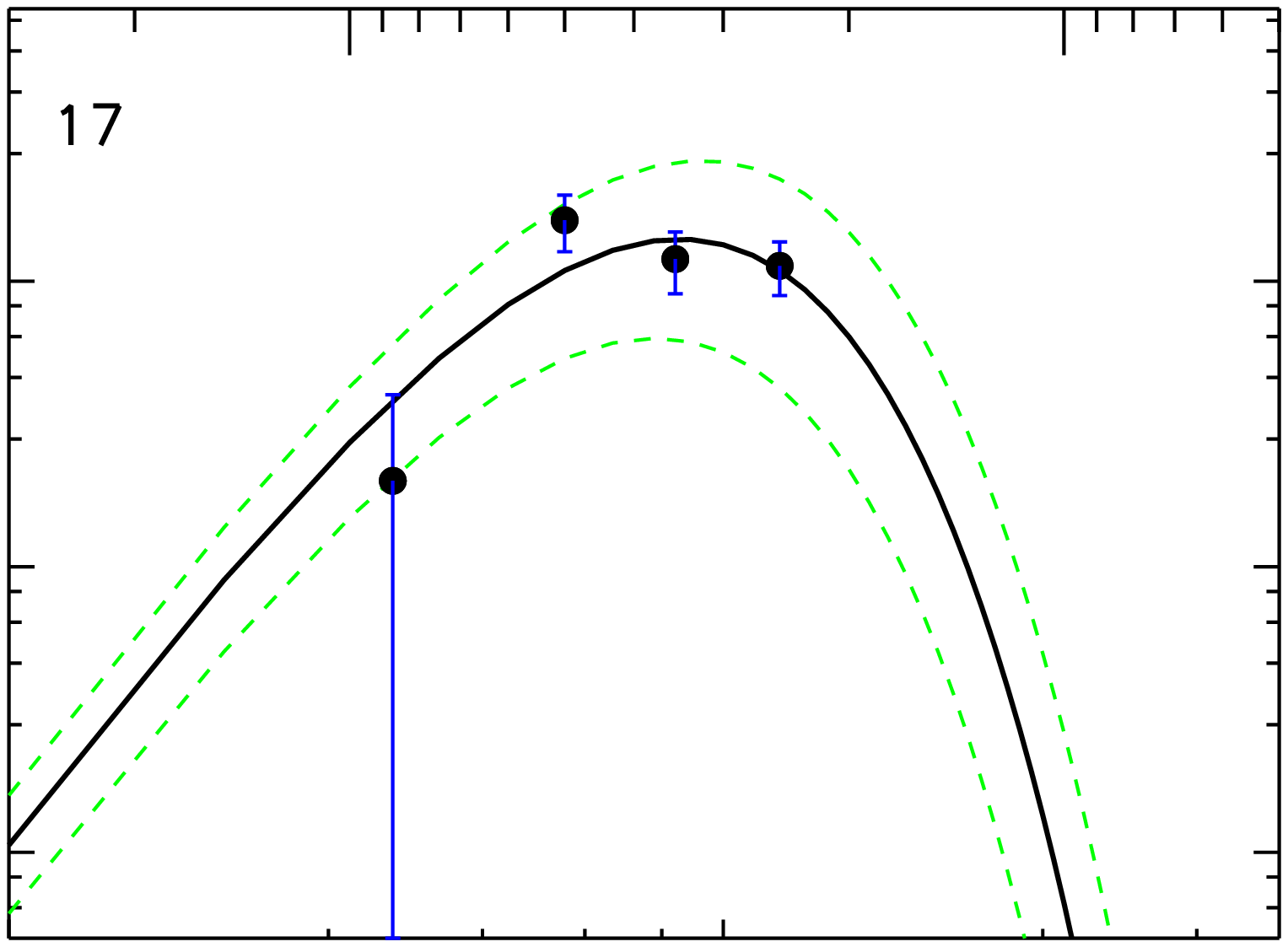,width=1.5in,angle=90}
\psfig{file=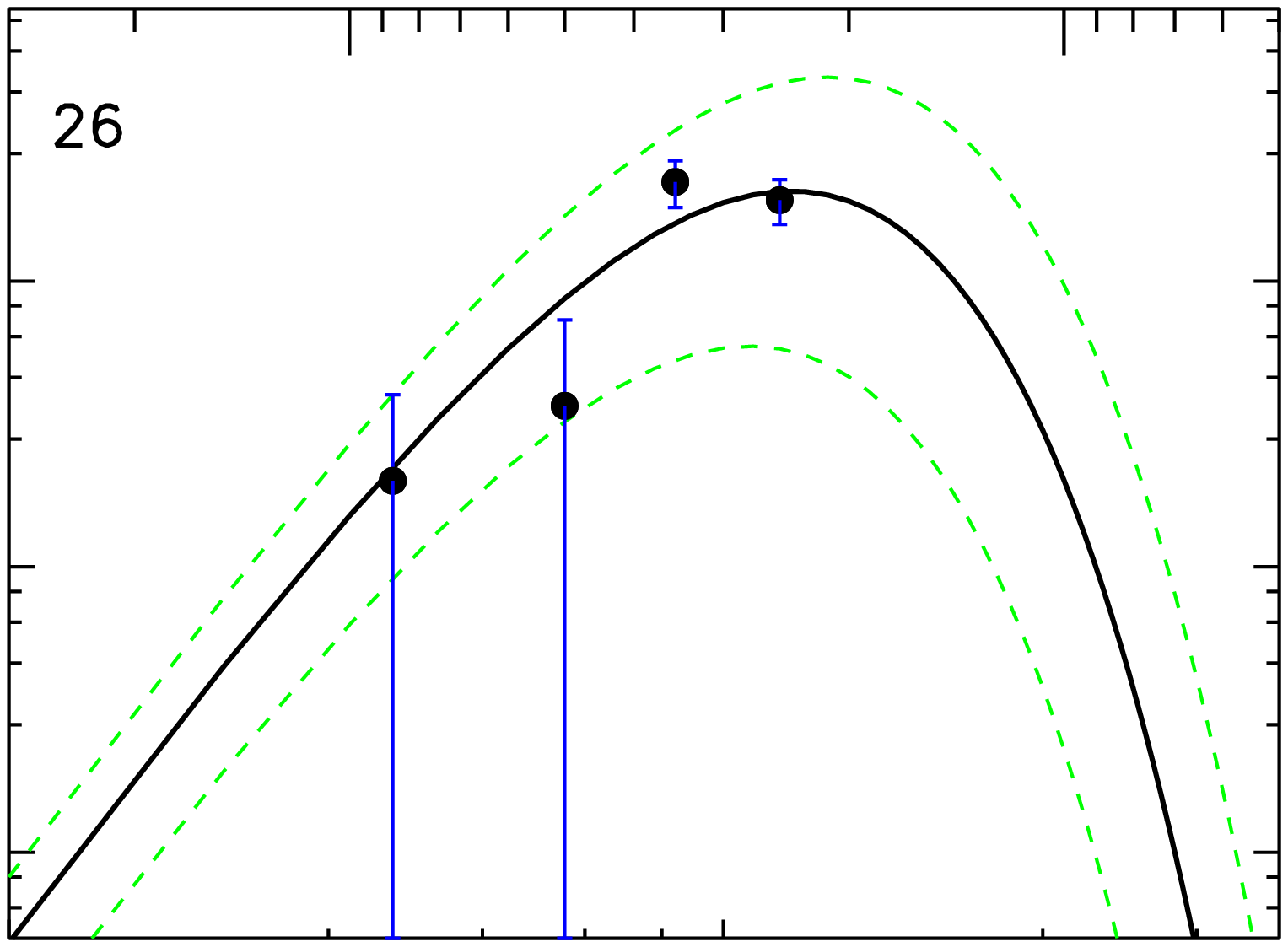,width=1.5in,angle=90}
}\centerline{
\psfig{file=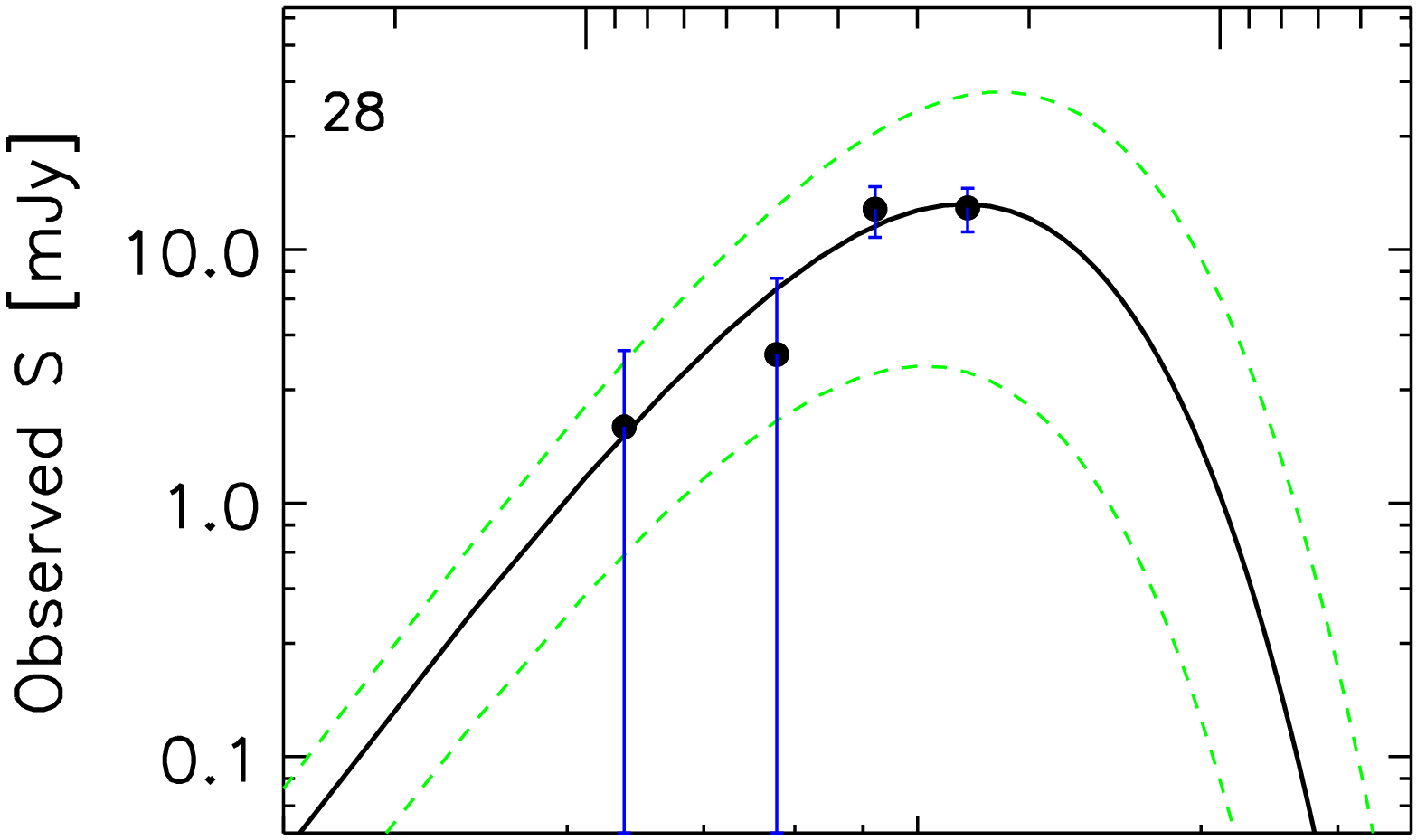,width=1.5in,angle=90}
\psfig{file=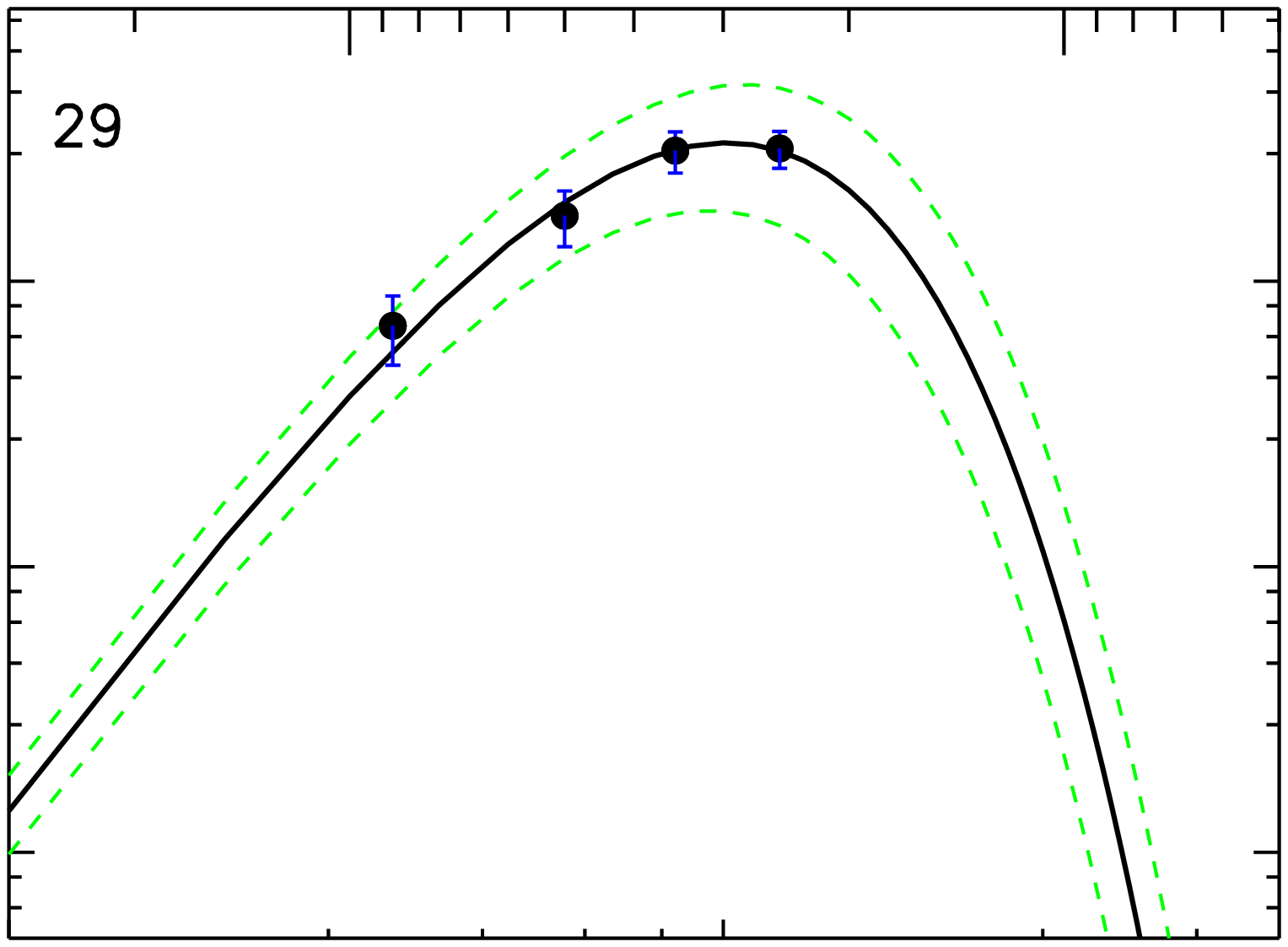,width=1.5in,angle=90}
\psfig{file=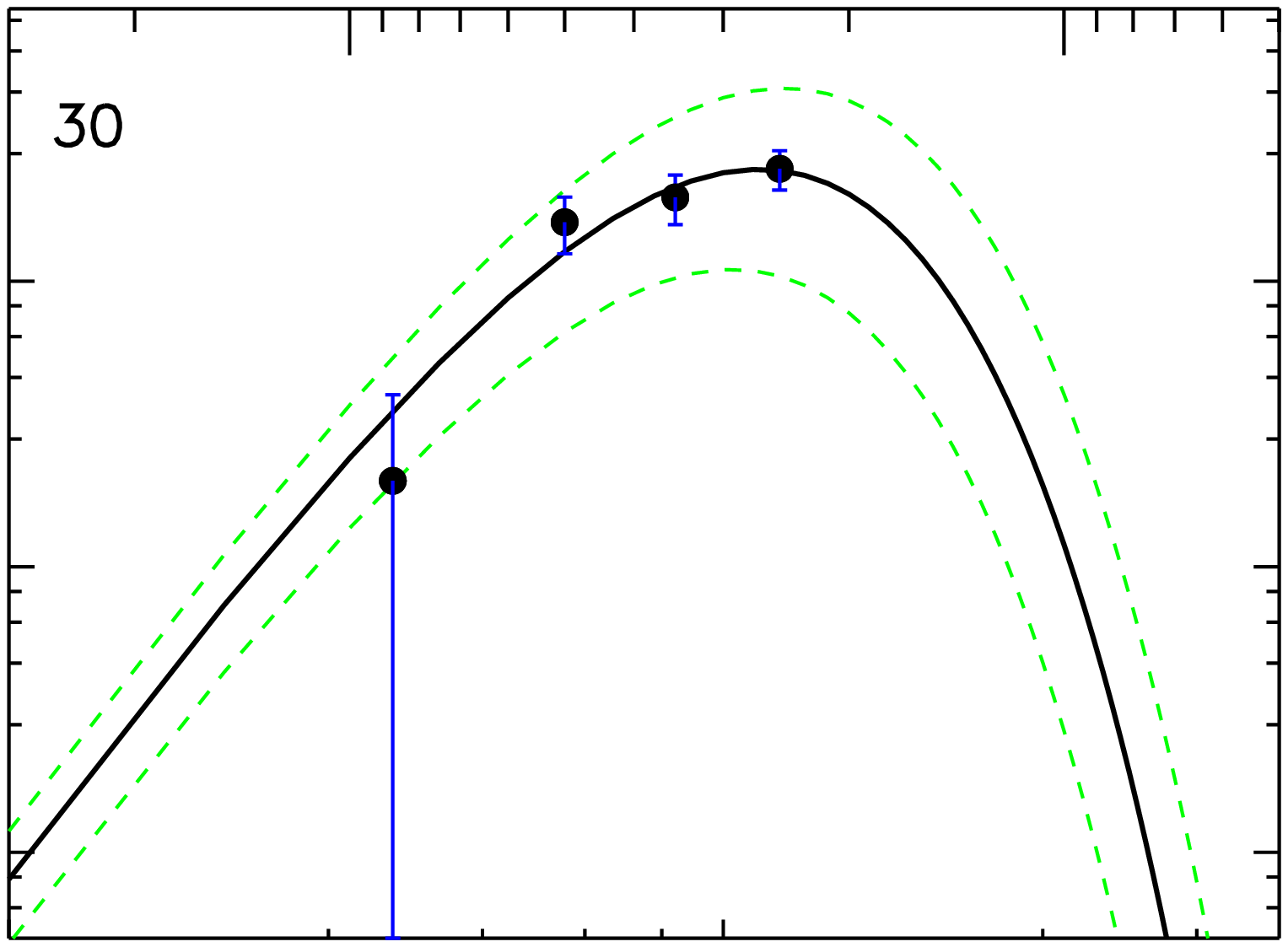,width=1.5in,angle=90}
\psfig{file=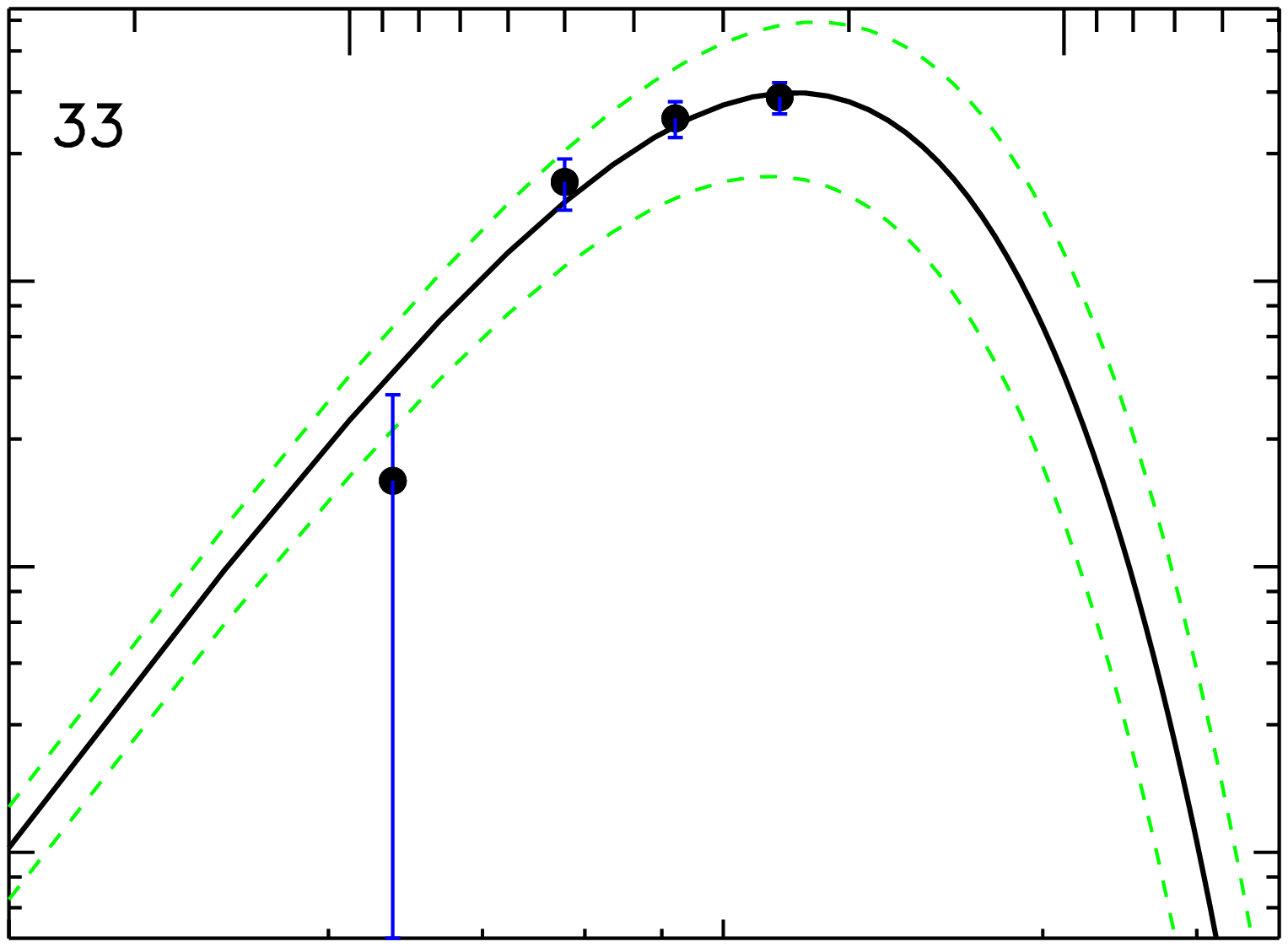,width=1.5in,angle=90}
}\centerline{
\psfig{file=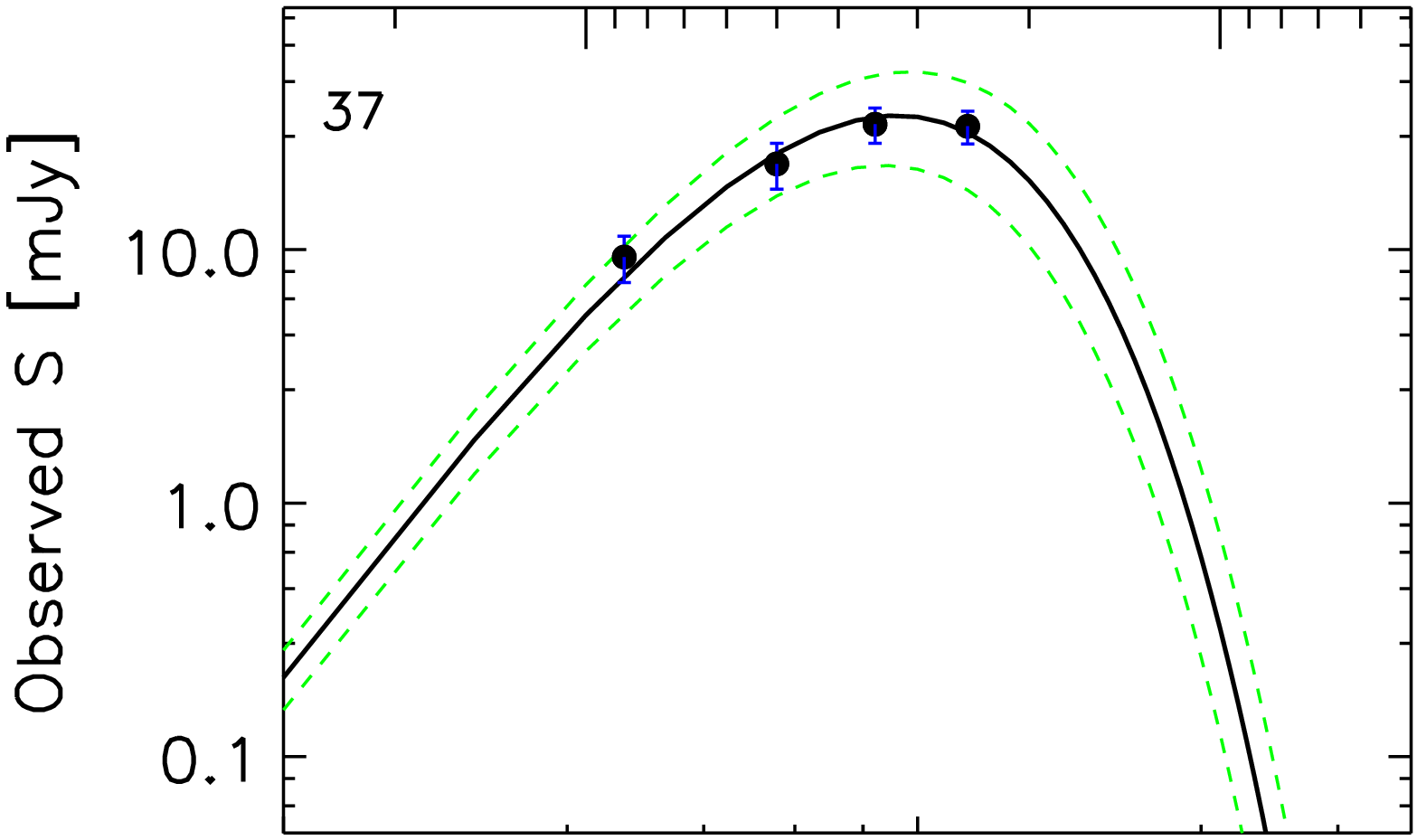,width=1.5in,angle=90}
\psfig{file=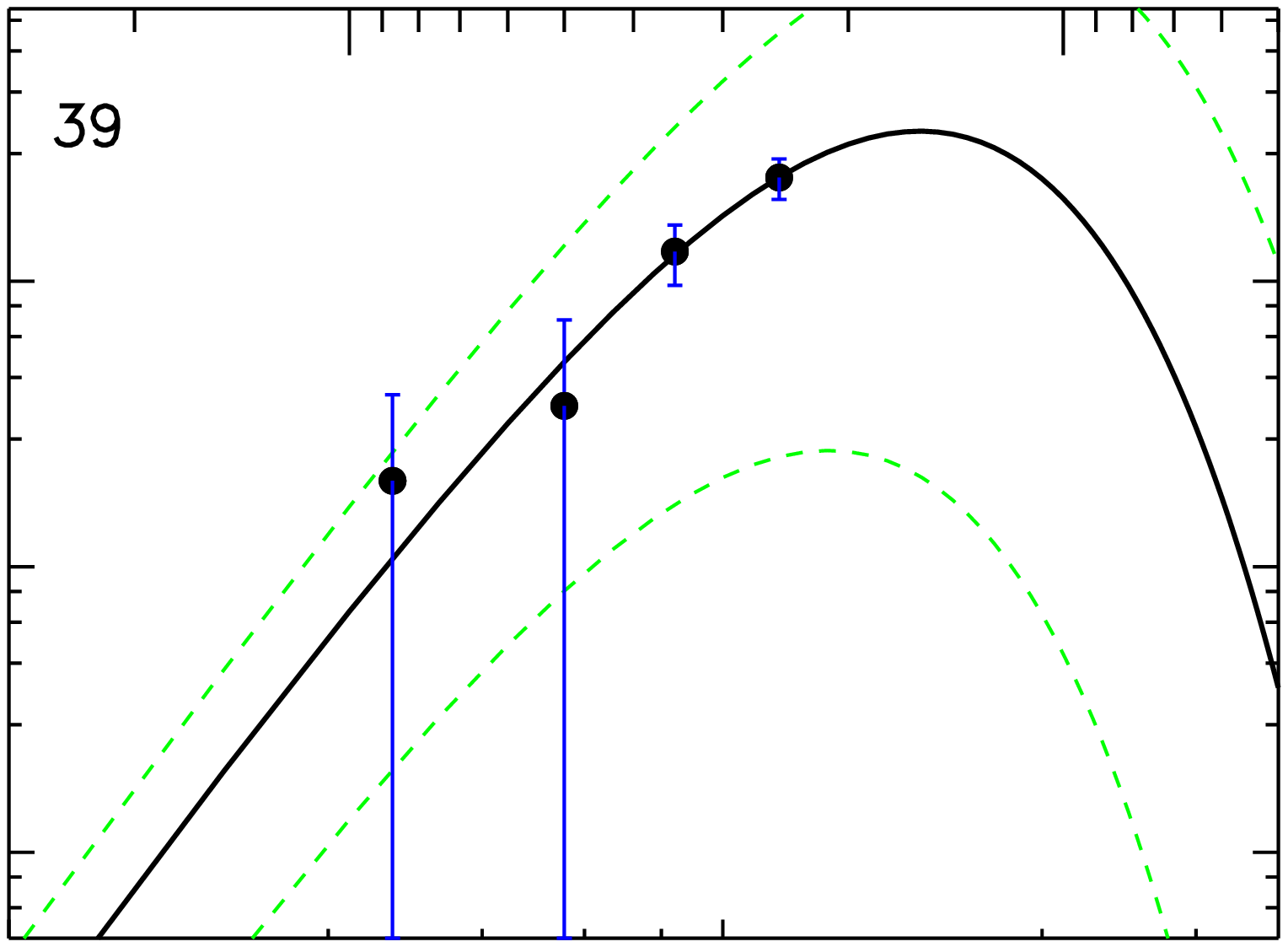,width=1.5in,angle=90}
\psfig{file=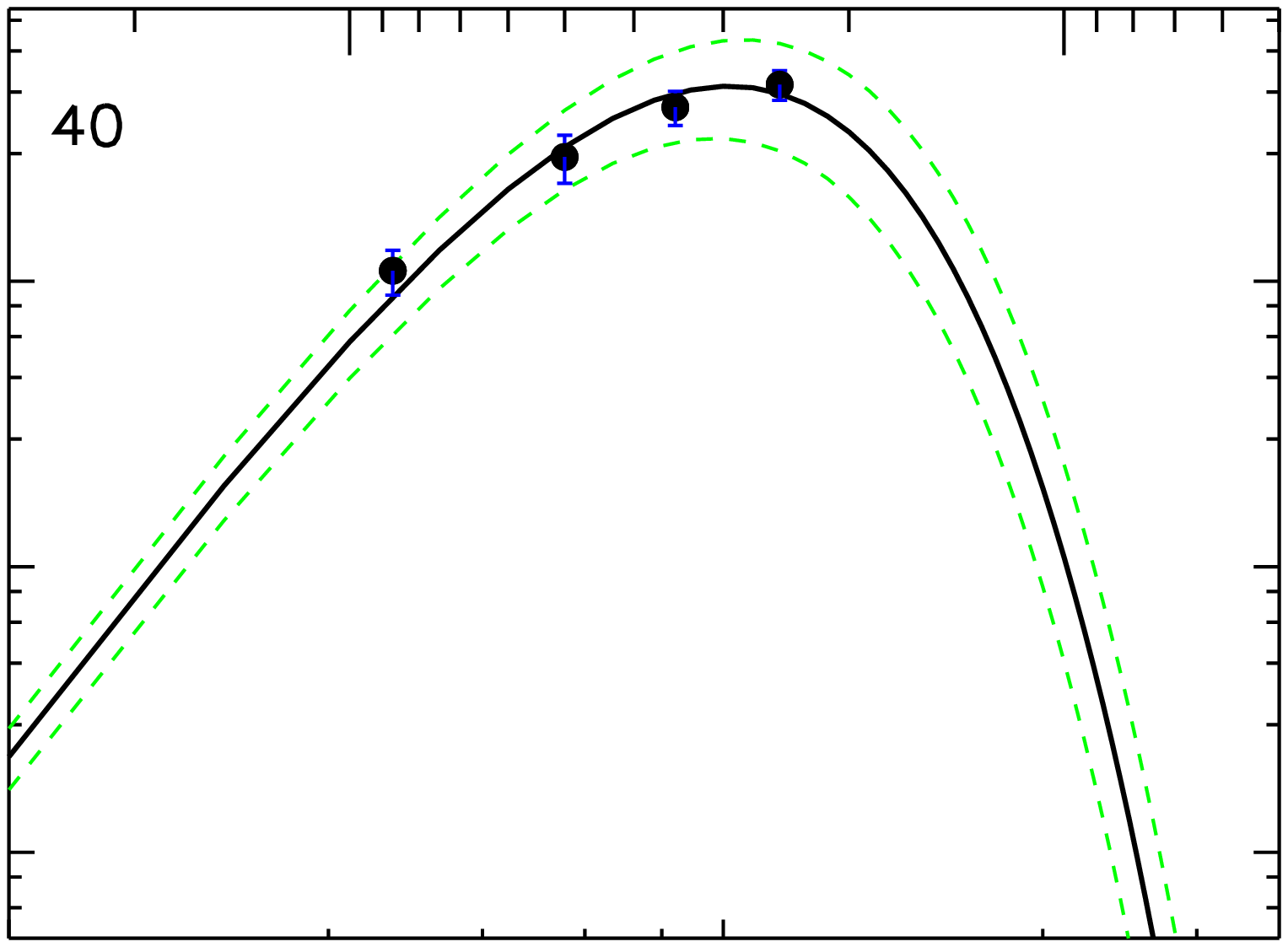,width=1.5in,angle=90}
\psfig{file=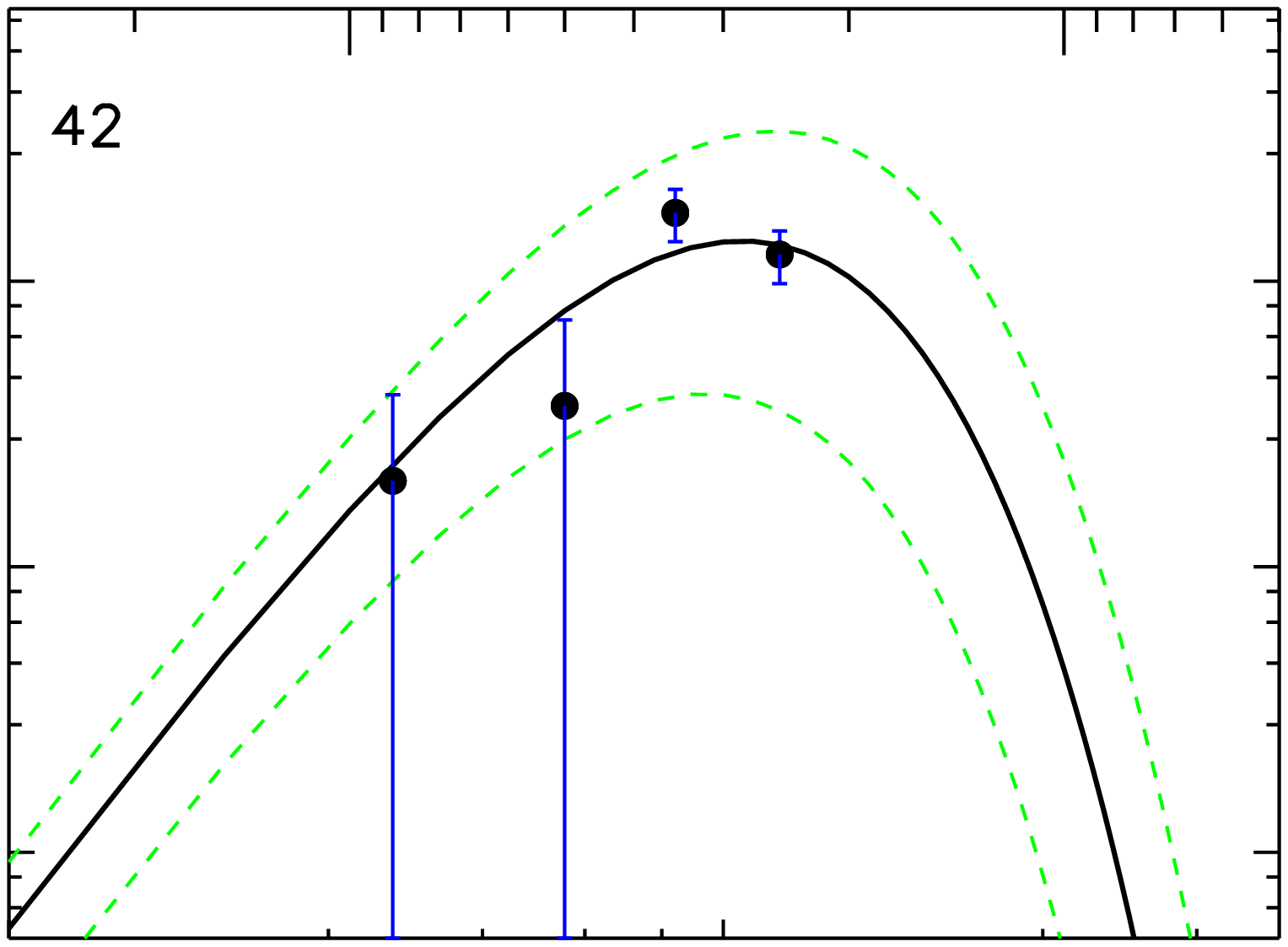,width=1.5in,angle=90}
}\centerline{
\psfig{file=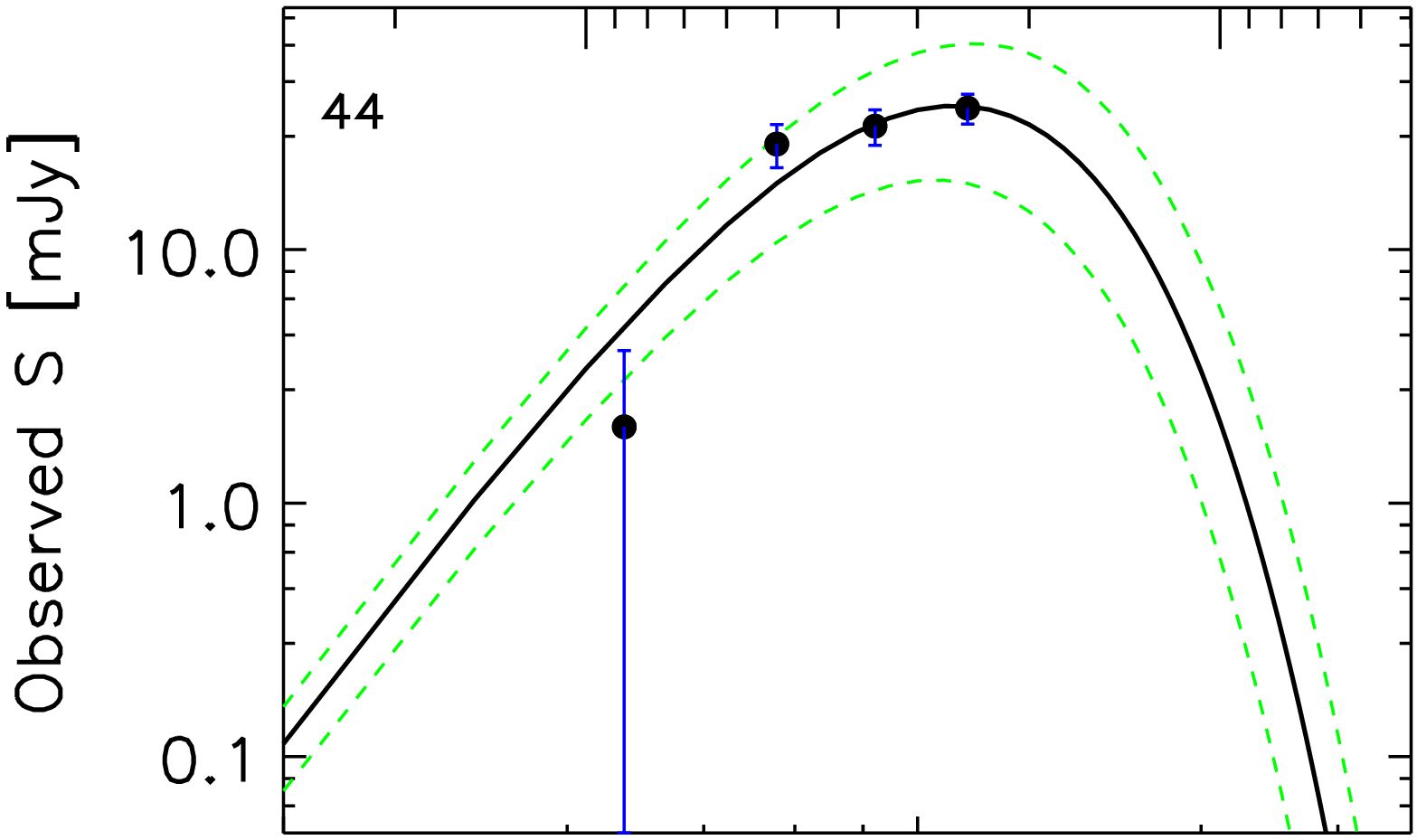,width=1.5in,angle=90}
\psfig{file=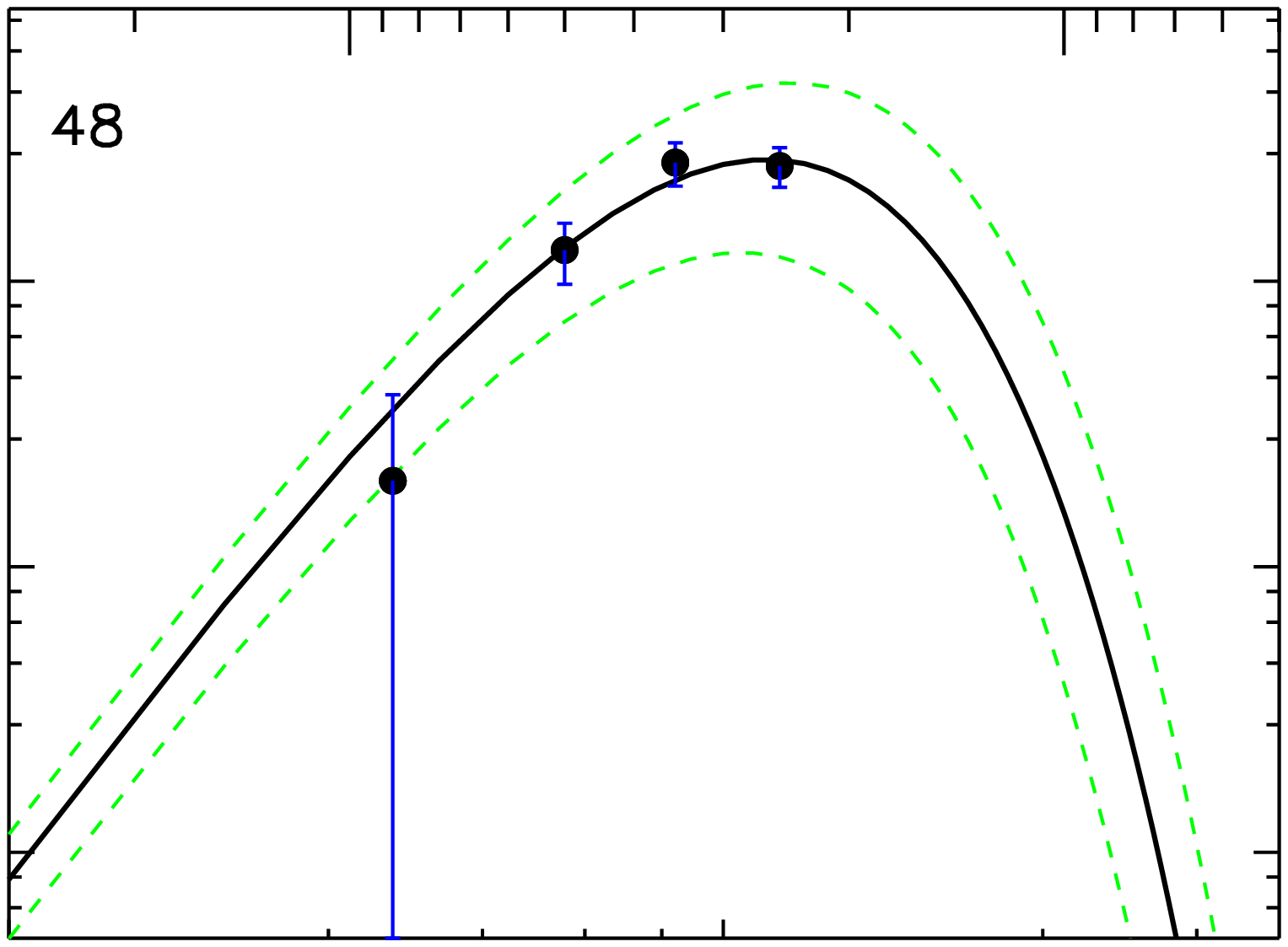,width=1.5in,angle=90}
\psfig{file=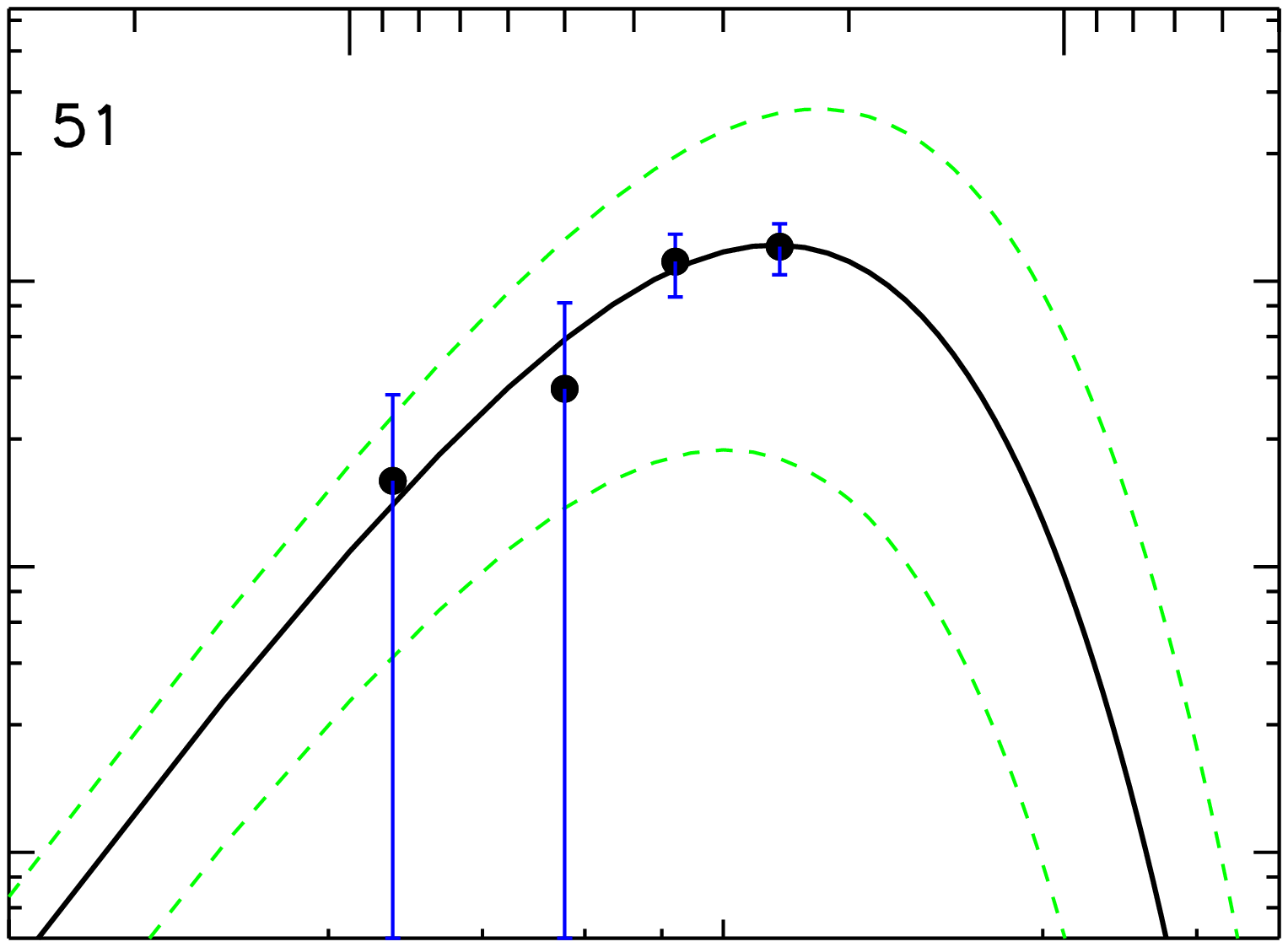,width=1.5in,angle=90}
\psfig{file=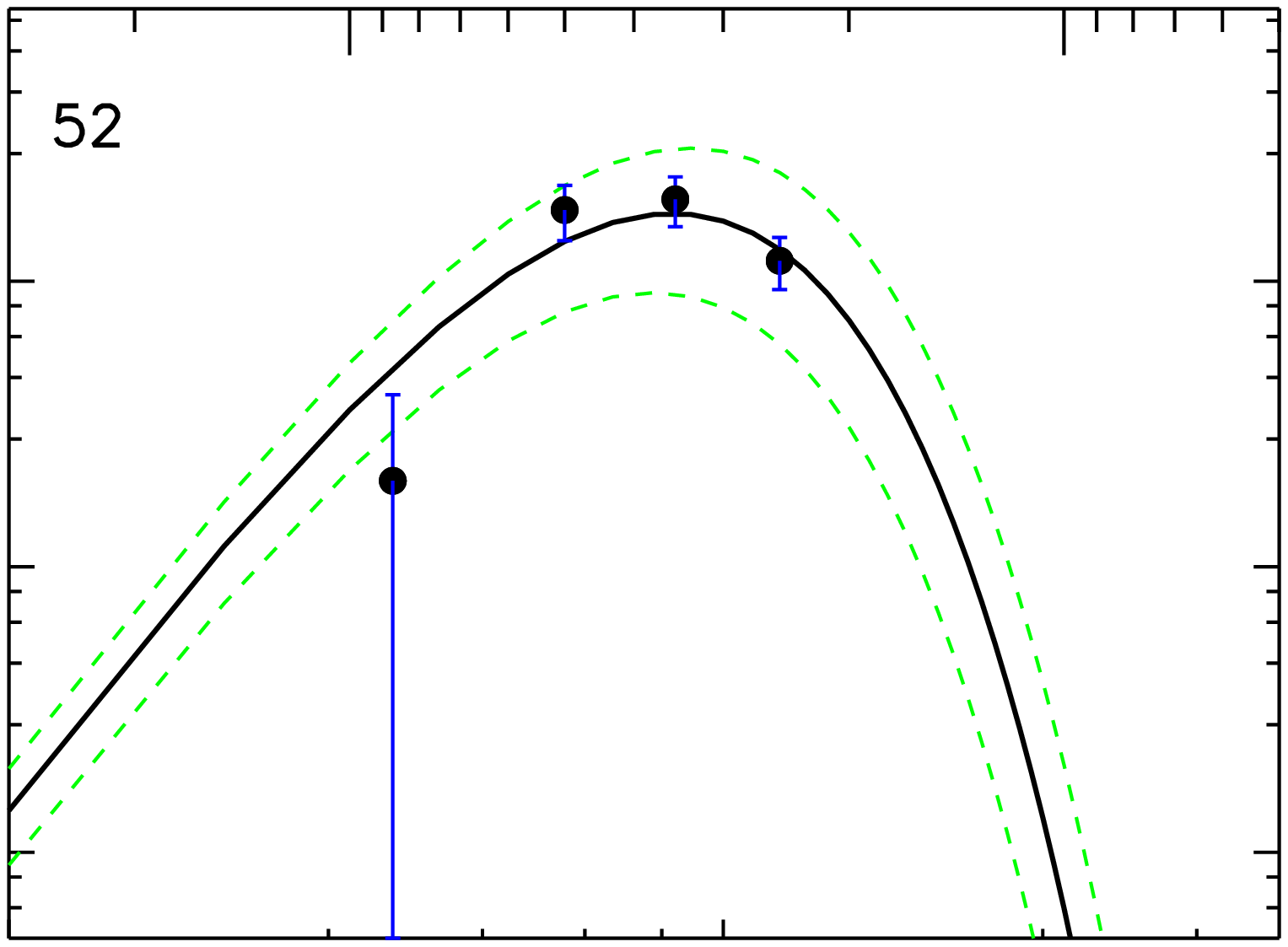,width=1.5in,angle=90}
}\centerline{
\psfig{file=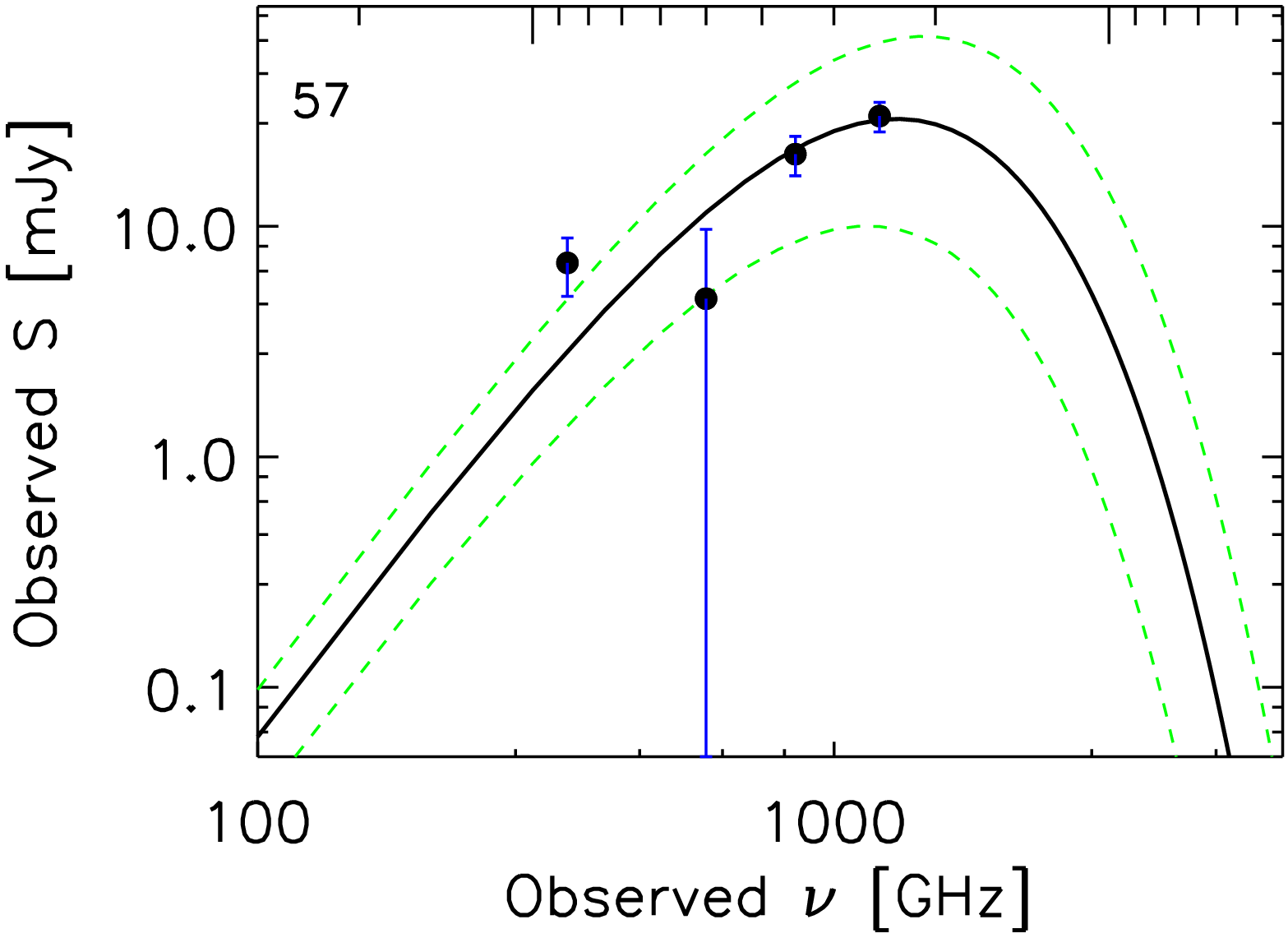,width=1.5in,angle=90}
\psfig{file=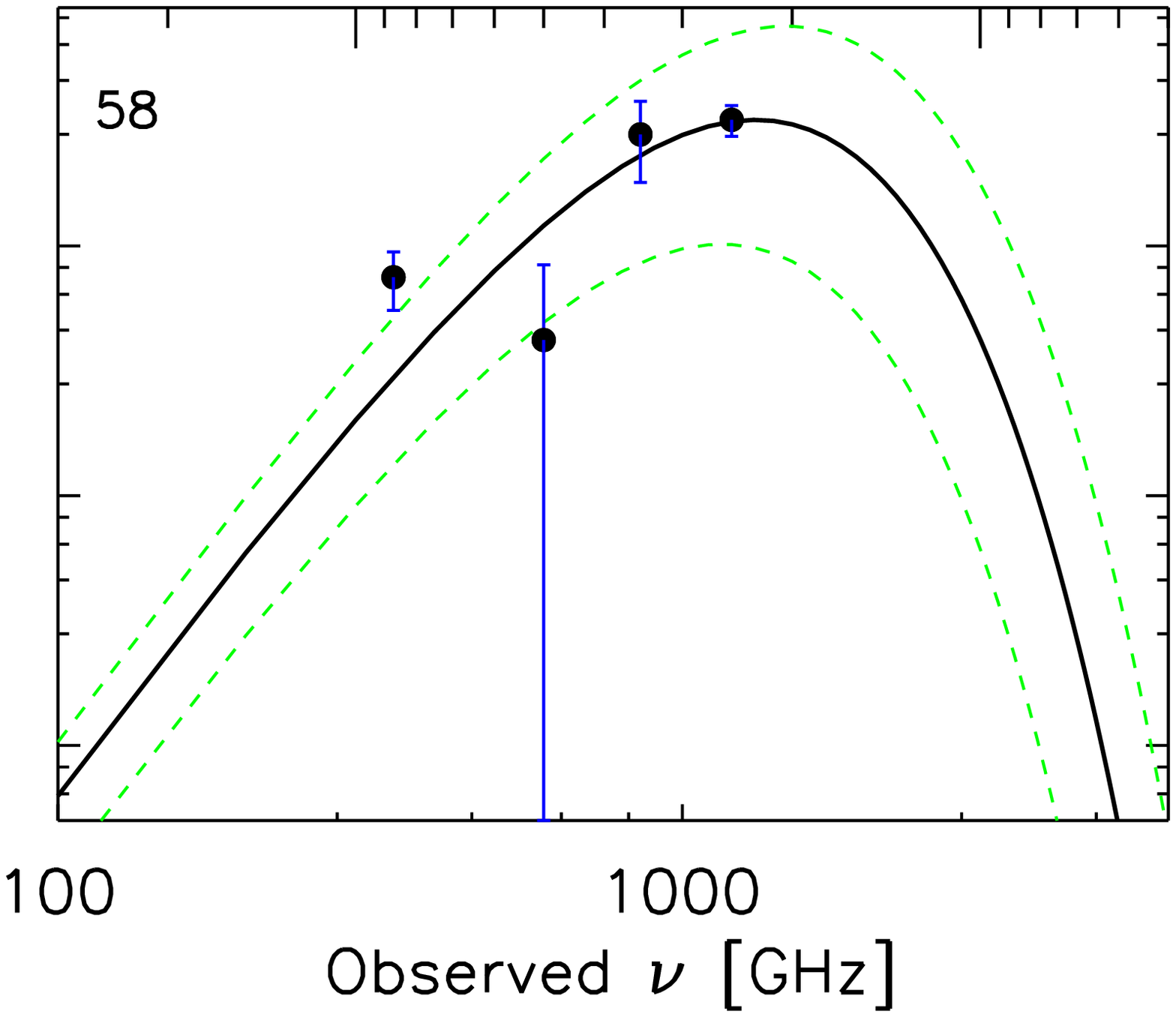,width=1.5in,angle=90}
\psfig{file=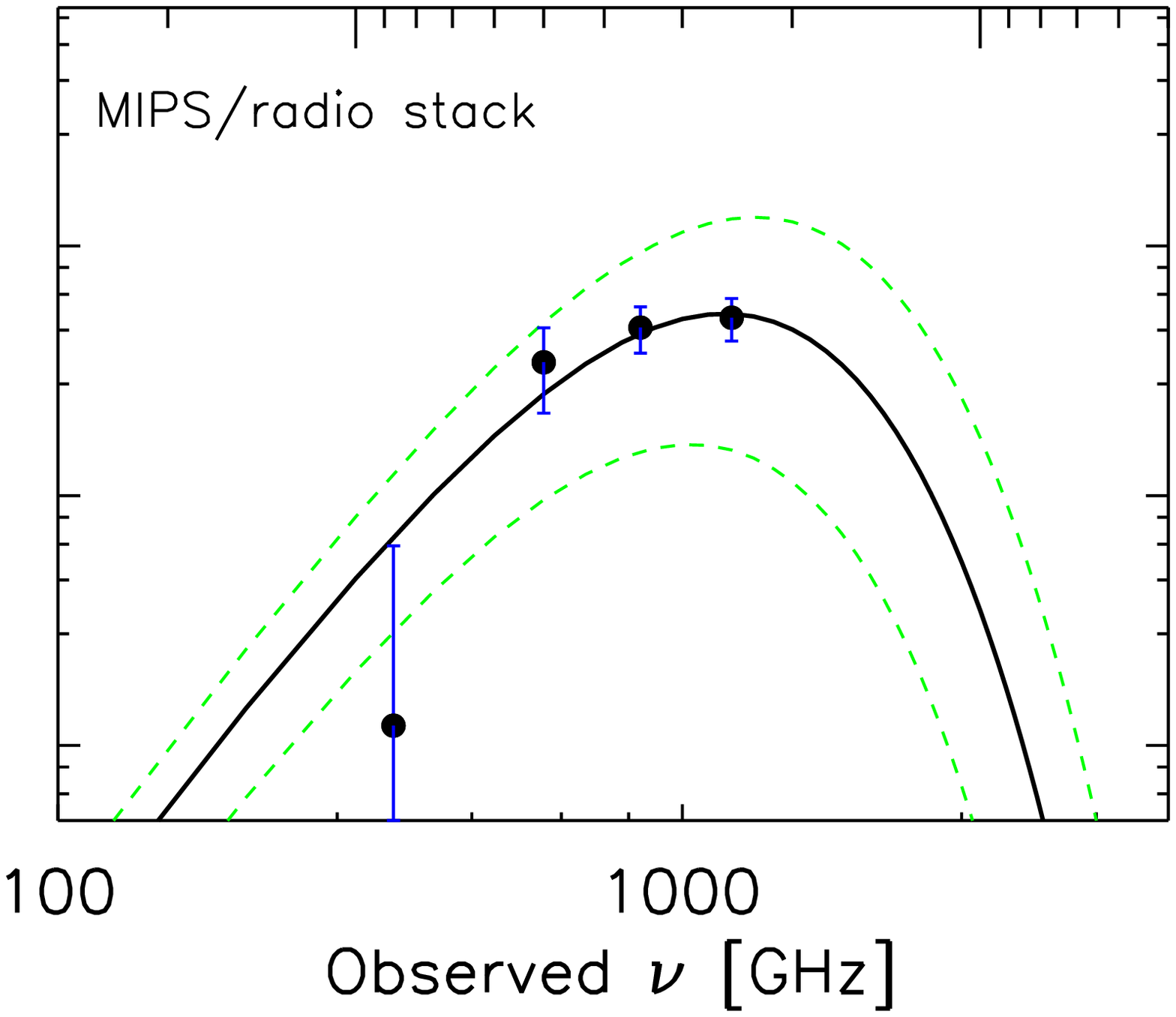,width=1.5in,angle=90}
\psfig{file=f4x.ps,width=1.5in,angle=90}
}
\smallskip

\caption{\small The far-infrared/submm SEDs for the 22 cluster members with two or more detections in the SPIRE and SCUBA-2 wavebands (see also Table~1).  We fit a modified black body model, with $\beta=$\,1.5, to the SPIRE 250, 350 and 500\,$\mu$m and SCUBA-2 850\,$\mu$m data points (non-detections are plotted at a flux corresponding to 1\,$\sigma$). The median luminosity and temperature of this sample is $L_{\rm bol}=(1.7\pm 0.3)\times 10^{12} $\,L$_\odot$ and $T_{\rm d}=33.0\pm 1.2$\,K. The best-fit models are shown by the solid curves and the 1-$\sigma$ limits are shown by dotted curves.  In addition we show in the lower right the equivalent fits to the stacked far-infrared/submm emission from those MIPS/radio cluster members which are not individually detected in more than one band and from the narrow-band [O{\sc ii}] sample. As expected these samples show lower bolometric luminosities than the individually detected sources with:  MIPS/radio, $L_{\rm bol}=0.4^{+0.6}_{-0.3}\times 10^{12} $\,L$_\odot$ and $T_{\rm d}=33\pm 6$\,K; [O{\sc ii}], $L_{\rm IR}=0.04^{+0.07}_{-0.03}\times 10^{12} $\,L$_\odot$ and $T_{\rm d}=29\pm 5$\,K.
}
\end{figure*}

\subsection{Redshifts}

To support our analysis we have gathered redshift information for confirmed and probable members of the $z=$\,1.6 structure from a number of sources.

\subsubsection{\it Confirmed members: [O{\sc ii}] Narrow-band imaging}

Tadaki et al.\ (2012)  studied the star-forming galaxy population in the $z=$\,1.6 structure through their redshifted [O{\sc ii}]\,3727 emission, as detected in  narrow-band imaging (NB973) with SuprimeCam/Subaru taken by Ota et al.\ (2010).   The NB973 filter ($\lambda_c = $\,9755\AA, $\Delta \lambda$ = 202\AA\ FWHM, using the $z_R$ filter for continuum correction) covers [O{\sc ii}] line emission at $z=$\,1.590--1.644.  This corresponds to $\sim$\,6000\,km\,s$^{-1}$ in the restframe, which is sufficient to encompass the spectroscopically-identified members (see next section) without including galaxies at significantly different redshifts. The stacked NB973 image has a total integration time of 5.5\,hrs in 1.0$''$ seeing, reaching a 5-$\sigma$ limiting magnitude of 25.4 in a 2.0$''$ diameter aperture and covers a total area of 830 arcmin$^2$, corresponding to a survey volume of $1.4 \times 10^5$\,Mpc$^3$.  More details of the data reduction and  analysis can be found in Tadaki et al.\ (2012) and Ota et al.\ (2010).

We adopt the survey region from Tadaki et al.\ (2012), which is roughly centered on the peak of the $z=$\,1.6 structure,  to define the area used in our study (Figure~1).  This allows us to compare the obscured star-forming population in the structure with their survey of the less-obscured population across  a range in local galaxy density.   We employ the Tadaki et al.\ catalog,  but with a simple optical/near-infrared color-selection applied to remove contamination, instead of the photometric redshifts used in Tadaki et al.\ (2012), to reduce possible biases against obscured sources.  The resulting catalog has  a total of 441 likely [O{\sc ii}]-emitters in the $z=$\,1.6 structure which we include in our analysis.  

\subsubsection{\it Confirmed members: UDSz and archival spectroscopy}

The UDS field has significant archival spectroscopy, including surveys specifically targeting the members of the $z=$\,1.6 cluster (Papovich et al.\ 2010;  Tanaka et al.\ 2010), AGN in the field (Pierre et al.\ 2012) and the general galaxy population (Simpson et al.\ 2012 and references therein).  One significant element of the latter is the recently-completed UDSz ESO Large Programme (PI: O.\ Almaini). UDSz obtained spectra for $\sim$\,3,500 $K$-band selected sources across the full UDS field using the VLT/VIMOS and FORS2 spectrographs.  A description of the target selection, reduction and analysis of UDSz is given in Bradshaw et al.\ (2013), with further details provided in Almaini et al.\ (in prep).

For our analysis we rely on a compilation of redshifts from UDSz, supplemented by a small number of unpublished redshifts (S.\ Chapman, priv.\ comm.) and the archival spectroscopy from Papovich et al.\ (2010), Tanaka et al.\ (2010) and Pierre et al.\ (2012).

\subsubsection{\it Probable members: UDS photometric redshifts}

Finally our analysis employs photometric redshifts for galaxies in the UDS to isolate possible cluster members.  While much less precise than the redshift information provided from the samples described above, the photometric redshifts obviously benefit from their uniform coverage of the field as well as the relatively deeper limiting magnitude to which redshift information can be obtained.

The photometric redshifts used here are described in Hartley et al.\ (2013) [see also Lani et al.\ (2013) and Simpson et al.\ (2012)]. The photometric basis of these redshifts is the deep near-infrared imaging of the 0.8-degree$^2$ from the Ultra-Deep Survey (UDS) element of the UKIRT Infrared Deep Sky Survey (UKIDSS, Lawrence et al.\ 2007) using WFCAM on  UKIRT. The DR8 release was used, with imaging reaching median depths of $J=24.9$, $H=24.2$, $K=24.6$ ($5\sigma$, 2-$''$ apertures)\footnote{\tt http://surveys.roe.ac.uk/wsa/}.  These near-infrared images are supplemented by the IRAC mid-infrared observations from the SpUDS {\it Spitzer} Legacy Programme (see \S2.2), as well as optical imaging covering the whole field with Suprime-Cam on Subaru.  The latter data were obtained by Furusawa et al.\ (2008) for the Subaru/{\it XMM-Newton Deep Survey} and comprise $B$, $V$, $R_c$, $i'$ and $z'$ data reaching 5-$\sigma$ limits (in 2$''$ diameter aperture)  of $B = 27.6$, $V = 27.2$, $R = 27.0$, $i' = 27.0$ and $z' = 26.0$.  The full UDS was also imaged in the $u$-band with MegaCam on CFHT (Almaini et al.\ in prep.) to a 5-$\sigma$ depth of $u'=26.75$.  Together these observations provide $uBVR_ci'z'JHK$, plus 3.6- and 4.5-$\mu$m photometry for a $K\leq 24.5$ limited sample 

The photometric redshift estimates are based on {\sc eazy} (Brammer, van Dokkum \& Coppi 2008). This code was applied in a two-step process where an initial calibration of the  redshifts was derived for a sample of $\sim $\,2,100 galaxies with reliable spectroscopic redshifts (archival and from UDSz, see \S2.4.2) and excluding AGN, to determine zero-point offsets between the photometric system of UDS  and that used for the templates employed in {\sc eazy}, deriving offsets of $\leq$\,0.05\,mag in all filters, except for $u'$.     The  templates library used in the fitting consist of the standard {\sc eazy} templates, plus a slightly-reddened version ($A_V=$\,0.1) of the bluest star-forming template, employing the Pei (1992) parametrization of the Small Magellanic Cloud (SMC) extinction law.  Based on this calibration they obtained a normalised median absolute deviation estimate of the dispersion of $\sigma_{\rm NMAD}(z_{\rm spec}-z_{\rm phot}/(1+z_{\rm spec}))=0.031$ for the spectroscopic training set.   More details of the analysis and testing of the photometric redshift catalog can be found in Hartley et al.\ (2013).

\subsection{Matching}

%
%
\setcounter{figure}{4} 
\begin{figure*}[tbh]
\centerline{\psfig{file=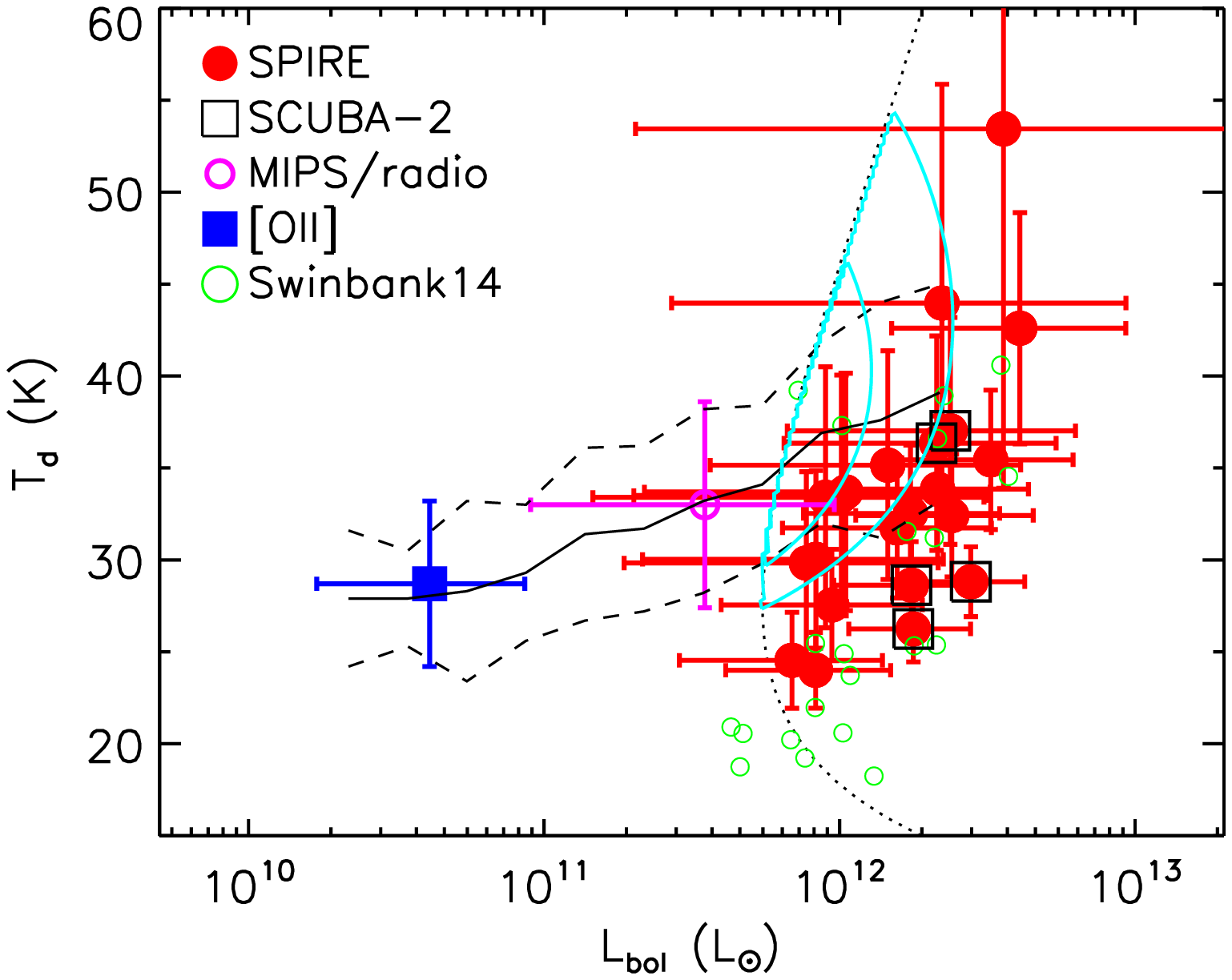,width=3.5in,angle=90}~\psfig{file=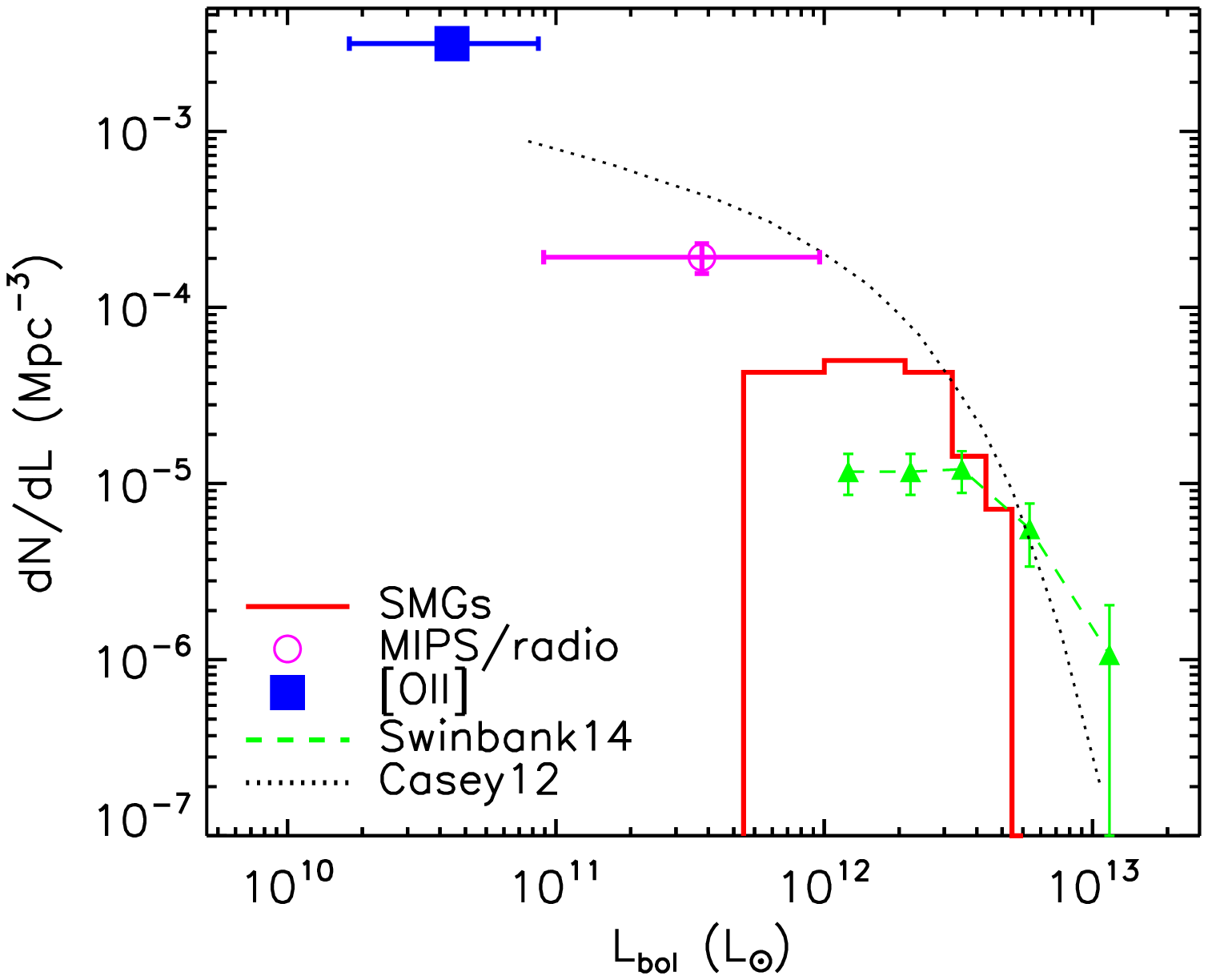,width=3.5in,angle=90}}
\caption{\small {\it (left)}  The dust temperature--luminosity ($T_{\rm d}$--$L_{\rm bol}$) relation for the active galaxies in the structure around Cl\,0218.3$-$0510.  We plot the individual SPIRE and SCUBA-2 detected ULIRGs and the results from the fits to the stacked fluxes for the narrow-band [O{\sc ii}] emitters and the SPIRE/SCUBA-2 undetected MIPS/radio sources.   We also show the luminosity limit at $z=$\,1.6 as a function of dust temperature for sources at the limit of the SPIRE 250-$\mu$m map (9.2\,mJy 3\,$\sigma$, dotted line), the trend found by Symeonidis et al.\ (2013) for  $z\ls$\,1 {\it Herschel} galaxies and its dispersion (solid and dashed lines) and contours showing the density distribution predicted by the model of the local $T_{\rm d}$--$L_{\rm bol}$ relation in Valiante et al.\ (2009) after applying our 250-$\mu$m flux limit.  The apparent trend at $z\ls$\,1 appears to agrees well with the less-active populations at $z=$\,1.6, but both it and the Valiante distribution appear to predict higher typical $T_{\rm d}$ than observed for the most active, ultraluminous, systems.  We caution that the Symeonidis et al.\ sample is flux-limited (in contrast our survey is effectively volume limited) and this complicates the derivation of this trend. {\it (right) } The luminosity function for SPIRE/SCUBA-2 detected photometrically-selected cluster galaxies along with the points denoting the volume density and average $L_{\rm bol}$ from stacking analysis of the [O{\sc ii}] narrow-band emitter population in the cluster and those MIPS/radio source members which are not individually detected in two or more SPIRE/SCUBA-2 bands.  We compare this distribution to the field luminosity functions for a $z=$\,1.2--1.6 250$\mu$m-selected sample from Casey et al.\ (2012) and from an ALMA-identified  870-$\mu$m-selected sample at $z\sim$\,2  from Swinbank et al.\ (2014), finding rough agreement.   In both plots we have scaled down the luminosities of the comparison samples by 1.25$\pm$0.01 to reflect the difference in luminosity measurements based on template fitting (as used in the comparisons) and the simple modified black body fits employed here.} 
\end{figure*}

As stated above, for reasons of uniformity, we restrict our analysis to the  region of UDS covered by the Tadaki et al.\ (2012) narrow-band [O{\sc ii}]  survey:  a 0.45\,$\times$\,0.55 degree patch demarcated by a rectangle with vertices  [34.250 ,$-$5.450] and [34.800, $-$5.000], see Figure~1.  This is roughly centered on the peak of the galaxy density in the $z=$\,1.6 structure (Figure~1) and also covers lower-density regions in the outskirts of this  structure.   Within this region from our input catalogs, there are 97 850-$\mu$m sources, 402 1.4-GHz sources, 891 MIPS 24-$\mu$m sources, and 441 [O{\sc ii}] emitters. 

To determine the membership of the structure we define a spectroscopic range of $z=$\,1.61--1.66, using the limits of the overdensity in the field in a redshift histogram of the available spectroscopy.  With this definition of membership there are 14 spectroscopic sources from UDSz in the structure (of which four have matches in the [O{\sc ii}] sample).   The archival spectroscopy of this field yields another 17 (after removing two which are duplicated in UDSz), ten of these have matches  in the [O{\sc ii}] sample.  Hence in total we have 31 spectroscopic members of the structure in our survey area, of which 14 are [O{\sc ii}] emitters.

We are forced to adopt a broader range in photometric redshift than that used for the spectroscopic sample (and hence also broader than the dispersion in redshift expected for virialised members of a cluster) due to the uncertainties in the photometric redshifts.  This choice is a trade-off between minimising the contamination by interlopers and maximising the number of true cluster members selected.   For the photometric redshift boundaries we aim for a  conservative cut (to reduce false-matches) and so select $z_{\rm ph}=$\,1.59--1.71.  We find that 15/31 (48 per cent) of the UDSz/archival members have photometric redshifts in this range and based on the average source densities across $z=$\,1.3--2.0 we estimate potential contamination in our sample by interlopers of 40--60 per cent (Figure~1).  Hence the interloper fraction roughly balances the incompleteness due to the photometric redshift errors.    This selection criteria results in 2793 photometric members of the structure from the UDS photometric catalog of Hartley et al.\ (2013)  within our survey region.  Of these, 207 are in the [O{\sc ii}] sample, 10 are in the archival spectroscopic sample and 12 are in the UDSz spectroscopic sample.  Thus combining the spectroscopic, [O{\sc ii}] and photometric samples there are 3031 members or probable members of the $z=$\,1.6 structure within our survey area.

As stated earlier, to circumvent the poor resolution of the submm catalogs of this region from SCUBA-2 and SPIRE, as is standard in this field we make use of more precisely located proxies to identify likely sources for this emission:  radio and MIPS sources (e.g.\ Biggs et al.\ 2011).   These catalogs are exploited in different ways for SCUBA-2 and SPIRE.  Due to its depth and the 850-$\mu$m source counts, the SCUBA-2 map is not confused and so we simply use the source catalog from these data and then associate detected sources with MIPS or radio counterparts using a probabilistic test, following Biggs et al.\ (2011).   We adopt a definition of robust counterparts of a matching probability of $P\le $\,0.05 and search for counterparts within a maximum 8$''$ search radius (Ivison et al.\ 2007). Based on recent ALMA observations of submm sources from similar resolution data, we expect this probability cut to provide a relatively pure ($>$\,80 per cent correct) but incomplete sample of identifications for the submm sources (Hodge et al.\ 2013).  In contrast, the deeper SPIRE maps of the UDS field are confused and so we derive fluxes or limits on the emission using the deblended catalog described in \S2.3.  This process then results in far-infrared/submm fluxes or limits for the emission from each source in the combined radio and MIPS catalog.

To identify which sources are members of the $z=$\,1.6 structure, we then match the MIPS and radio samples to the catalogs of confirmed or probable members ([O{\sc ii}], spectroscopic and photometric redshifts), with search radii which reflect the uncertainty in the positions of sources (due to resolution and signal-to-noise issues) and the relative astrometric alignment of the different samples.   We therefore match to the closest counterpart in the membership catalog within a 1$''$ radius  (which should yield $<$\,1 per cent false-positive matches).     We find 58 members or probable members which are either radio detected or have bright MIPS counterparts: 28 MIPS, 19 radio and 11 MIPS-and-radio.  Of these, one MIPS source has an archival spectroscopic redshift (it is an AGN), six sources are narrow-band [O{\sc ii}] emitters (one radio, five MIPS, of which two MIPS are also members based on their photometric redshifts), and the remaining 51 sources are members based on their photometric redshifts  (33 MIPS detected, 29 radio, 11 both MIPS and radio).   

Of the 58 MIPS- or radio-selected members, 31 have detectable 250-$\mu$m emission in the SPIRE map and 22 are detected in two or more bands so can have spectral energy distributions (SEDs) fitted, details of these sources are given in Table~1. Five of these  also  match SCUBA-2 submm sources.\footnote{A further SCUBA-2 source matches a non-FIR-detected MIPS counterpart. Indeed matching the SCUBA-2 sample directly to the [O{\sc ii}] narrow-band emitter catalog we find another five SCUBA-2 sources with apparent statistical associations, $P\leq $\,0.01, with [O{\sc ii}] emitters (one matching a close pair of [O{\sc ii}] emitters). However, the photometric properties of these emitters, blue colors and low luminosities, are characteristic of the bulk of the [O{\sc ii}] population, rather than the redder and more dusty submm sources (as shown by the MIPS/radio-identified examples, Figure~3).  For this reason we believe that these [O{\sc ii}]--submm associations are likely to indicate that the [O{\sc ii}] sources are companions to the true submm emitter and so we cannot use these to investigate the properties of that population.  However, we find no evidence for an excess of red/luminous sources in the immediate vicinity of these five apparent [O{\sc ii}]-submm pairs, nor do they have obvious SPIRE counterparts, and so while they could represent cold, low-luminosity cluster members we have chosen to discard them from our analysis.}  By chance one of these five SCUBA-2 sources (\#37 in Table~1) was included in a high-priority Cycle~1 ALMA program to image the  thirty bright submm sources from the S2CLS  map of the UDS.  These Band 7 (870\,$\mu$m) continuum observations reach 1-$\sigma$ depths of $\sim$\,0.2\,mJy and confirm that  the MIPS/radio source with a photometric redshift of $z_{\rm phot}=$\,1.67 we had identified is indeed a bright submm galaxy (see Figure~1).   This  provides a useful test of the reliability of the identification process used in our analysis.   Full details of the ALMA program are given in Simpson et al.\ (in prep).

Finally, we have also stacked the emission in the SPIRE and SCUBA-2 maps from the individually undetected MIPS/radio cluster members and the [O{\sc ii}] narrow-band sample. The MIPS/radio sample are stacked on the original SPIRE maps, while for the [O{\sc ii}] sample we have stacked these sources on the residual maps constructed by removing all of the deblended sources (using a 5-$\sigma$ MIPS/radio-based prior catalog).  These stacked fluxes yield detections in all three SPIRE bands for both samples which are also reported in Table~1.

\section{Analysis and Results}

We plot the spatial and redshift distributions of the various populations inhabiting the $z=$\,1.6 cluster in Figure~1.  The spatial distribution of cluster members shows two dense clumps, with weaker structures spread across the whole of our survey region.   The scale of this structure is commensurate with that expected for the progenitor of a massive cluster of galaxies at the present day (e.g.\ Governato et al.\ 1998; Chiang et al.\ 2013).  The photometric redshift distributions display modest peaks in both the photometric and far-infrared/submm-detected populations at the cluster redshift.  To test the significance of the peak in the far-infrared-detected sample within the redshift window we use to define membership, $z=$\,1.59--1.71, we perform 10$^5$ Monte Carlo simulations perturbing all of the photometric redshifts by their estimated errors and determining how frequently  the number of sources in this redshift range exceeds that observed.  We find that the observed peak occurs less than $\leq $\,1 per cent of the time by random chance.  In addition, we note that given  the potential contamination of the photometric member sample, it is reassuring that a generally similar structure is also seen in the spatial distribution of [O{\sc ii}]-emitters.  However, in this regard we also note that the more active members of the structure: the far-infrared/submm galaxies, as well as the MIPS and radio sources, are somewhat less concentrated and we discuss this in more detail in \S3.3.

\subsection{SEDs and multi-wavelength properties}

Figures 2 and 3 illustrate the colors and luminosities of the different cluster populations.   Figure 2 shows a true-color  $VK4.5\mu{\rm m}$ representation of 160-kpc regions around the 31 far-infrared/submm-detected cluster members.  We see that the far-infrared/submm sources are typically bright with relatively red colors. Figure~2 also shows that around half of the far-infrared/submm sources have potential companions on projected scales of $\ls $\,30--50\,kpc, with the 24-$\mu$m emission from the system frequently offset in the direction of these companions.   However, we see little correlation between $L_{\rm bol}$ and the presence or separation of these companions.

%
%
\setcounter{figure}{5} 
\begin{inlinefigure}\vspace{6pt}
\centerline{\psfig{file=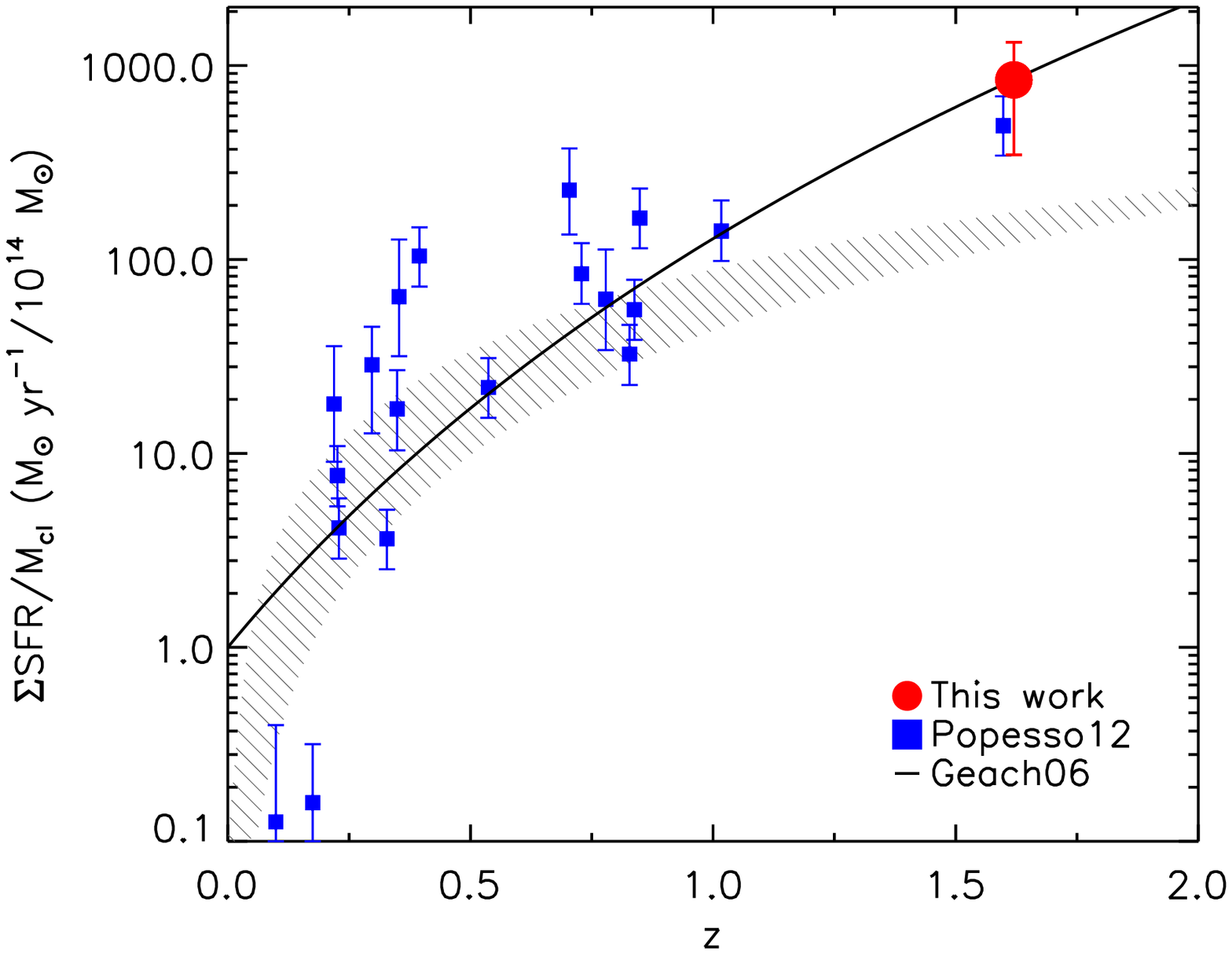,width=4.0in,angle=90}}
\vspace*{-3mm}
 \noindent{\small {\small \sc Fig.~6 --- } The evolution of the mass-normalised SFR (for galaxies more luminous that $L_{\rm FIR}=$\,10$^{11}$\,L$_\odot$) in clusters and groups as a function of redshift.
We plot Cl\,0218.3$-$0510 and the data  from Popesso et al.\ (2012).  This shows that this $z=$\,1.62 cluster extends the trend for higher mass-normalised SFRs out to the highest redshifts.   The  cross-hatched region shows the fitted trend from Popesso et al.\ (2012) for clusters of galaxies and the solid line shows the $(1+z)^7$ evolution proposed by Geach et al.\ (2006), based on the field evolution of LIRGs found by Cowie et al.\ (2004).    We see that the strong evolution implied by the latter model is a better fit to the high-redshift systems than the trend proposed by Popesso et al.  The lower error on the Cl\,0218.3$-$0510 data point indicates the reduction in integrated SFR which occurs if we remove the brightest source from the sample (to reflect the potential contamination of the sample by unrelated sources), while the upper error shows the result of including lower luminosity star-forming galaxies in the structure from the narrow-band [O{\sc ii}] survey of Tadaki et al.\ (2012).
} 
\end{inlinefigure}

Figure~3 shows the distribution of probable cluster members on the $(z'-H)$--4.5\,$\mu$m plane (this combination of filters corresponds to restframe $(U-V)$--$M_H$ at $z\sim $\,1.6).  This
confirms the tendency for the far-infrared/submm sources to be brighter and redder than the typical cluster member:  at any given color the far-infrared/submm sources are some of the brightest members of the population.  Moreover, the far-infrared/submm sources are preferentially associated with the brightest members of the cluster population, with the SCUBA-2-detected sources being amongst the most luminous cluster galaxies at restframe $H$-band.   This figure also highlights three MIPS-detected sources which are luminous at restframe $H$-band, but relatively blue, these are likely to be AGN-dominated systems.

To see how our observations compare to theoretical expectations, we also plot in Figure~3 the predicted colors and luminosities of galaxies taken from the Millennium database (Springel et al.\ 2005), using the prescription for galaxy evolution from Font et al.\ (2008).  This model is an adaptation of the Bower et al.\ (2006) galaxy formation recipe, including the influence of  stripping on the extended gas halos of galaxies in high-density regions (groups and clusters).   To match our survey we select all halos at $z=$\,1.6 with masses within 10 per cent of the observed mass of Cl\,0218.3$-$0510 (see \S3.2),  and then identify those galaxies with $M_H\leq -19$ which are members of these halos.  We contour the number density of galaxies as a function of their $(U-V)$ color and restframe $H$-band absolute magnitude, from 5 per cent of the peak density in increments of 10 per cent.   We see broad agreement between the observed and predicted distributions for the bluer and fainter star-forming galaxies.  However, it is clear that the model predictions do not reproduce the redder and brighter galaxies, both passive and dusty star-forming.  We expand on this theoretical comparison below.

In Figure~4 we show the far-infrared/submm SEDs for the 22 galaxies which are individually detected in two-or-more bands by SPIRE or SCUBA-2.  We fit a modified black body spectrum with $\beta=$\,1.5 to each source to derive the restframe dust temperature, $T_{\rm d}$, and integrate the fit to derive the bolometric luminosity, $L_{\rm bol}$, both assuming the source is at $z=$\,1.6.  We derive a median luminosity of this population of $L_{\rm bol}=(1.7\pm 0.3)\times 10^{12}$\,L$_\odot$ and a dust temperature of $T_{\rm d}=$\,33.0$\pm$1.2\,K, with the five SCUBA-2 detected SPIRE sources being slightly more luminous, but cooler: $L_{\rm bol}=(2.1\pm 0.3)\times 10^{12} $\,L$_\odot$ and $T_{\rm d}=$\,29$\pm$4\,K. Details of the fits to the 22 individual sources are given in Table~1.  We note that there is no measurable correlation between $L_{\rm bol}$ and 24-$\mu$m flux, meaning that the MIPS fluxes of these $z=$\,1.6 ULIRGs cannot be used to reliably infer their far-infrared luminosities or star-formation rates (c.f.\ Tran et al.\ 2010).  We also note that these modified black body fits are expected to give lower luminosities than fitting template SEDs to the data and so  we have scaled down the template luminosities from the literature by a factor 1.25$\pm$0.01 (Swinbank et al.\ 2014) when comparing to our black body fits in Figure~5.  Finally, we warn that interferometric studies of single-dish submm sources have shown that a significant fraction of these comprises blends of multiple, fainter sources  (e.g.\ Wang et al.\ 2011; Barger et al.\ 2012; Hodge et al.\ 2013).  If this same effect is present in our sample it will influence those measurements where the multiple submm galaxies (SMGs) are not all detected in the mid-infrared/radio wavebands used in our prior catalog.  This will affect the derived luminosities more than the temperatures, as the $T_{\rm d}$--$L_{\rm bol}$ relation means that any lower-redshift, lower-luminosity and hence cooler components will have dust SEDs which peak in the observed frame at similar wavelengths to the more distant, more luminous and thus warmer components. 

We also show in Figure~4 the equivalent fits to the stacked far-infrared/submm emission from those MIPS/radio cluster members which are not individually detected in two or more bands and from the narrow-band [O{\sc ii}] emitter sample.   As expected these samples show lower bolometric luminosities than the individually detected sources.  The MIPS/radio sources have $L_{\rm bol}=0.4^{+0.6}_{-0.3}\times 10^{12} $\,L$_\odot$ and $T_{\rm d}=$\,33$\pm $6\,K, and the  [O{\sc ii}] emitters: $L_{\rm bol}=0.04^{+0.07}_{-0.03}\times 10^{12} $\,L$_\odot$ and $T_{\rm d}=$\,29$\pm$5\,K.   

In Figure~5 we take the distribution of dust temperatures and bolometric luminosities for the various cluster populations and compare these to the trends found in both the local Universe and at higher redshifts.  We plot on the individually detected SPIRE/SCUBA-2 sources and the stacked results from the undetected MIPS/radio cluster members and the [O{\sc ii}] population.   To compare to these we overplot the temperature--luminosity trend derived for $z\sim $\,0 {\it Herschel} galaxies by Symeonidis et al.\ (2013), who also adopted $\beta=$\,1.5,  the  selection limit  as a function of dust temperature expected for a source with a constant 250-$\mu$m SPIRE flux at $z=$\,1.6, which roughly defines the selection boundary for our sample and the  distribution of $T_{\rm d}$--$L_{\rm bol}$ predicted for the local {\it IRAS} population after application of our 250-$\mu$m flux limit (Valiante et al.\ 2009; see also Chapin et al.\ 2009). We also plot on the distribution of temperatures/luminosities derived for a sample of high-redshift, ALMA-detected submm galaxies from Swinbank et al.\ (2014).   We see from these that the SPIRE- and SPIRE/SCUBA-2-detected cluster members have a similar range in luminosity and temperature to the field population at high redshift, with both populations having slightly cooler temperatures, $\Delta T_{\rm d}\sim $\,5.3$\pm$1.6\,K than implied by an extrapolation to high luminosities of the low-redshift {\it Herschel} or {\it IRAS} temperature--luminosity trends.   In contrast the stacked SED fits for the less-active cluster populations are in good agreement with the low-redshift trend, suggesting a closer similarity in the dust properties between lower-luminosity sources at high and low redshift, than is seen for the ultraluminous population.

\subsection{Mass-normalised integrated SFR}

Figure~5 also shows the bolometric luminosity function for the various cluster populations from our direct detections and stacks.  For this calculation we have adopted a survey volume of 1.4$\times $10$^5$\,Mpc$^3$ from  Tadaki et al.\ (2012).  This corresponds to a 50\,Mpc cube and hence is likely to be an upper limit to the actual volume spanned by the photometrically-defined cluster members (e.g.\ Chiang et al.\ 2013).  As a result we expect our number densities will be under-estimated,  although this statement is complicated by the competing effects of incompleteness and contamination in our sample (see \S2.5).  Nevertheless, as we use the same volume for all three cluster populations, this uncertainty should not strongly affect the {\it relative} number densities derived for the three sub-samples.

As expected, Figure~5 shows that the progressively less-active cluster members have correspondingly higher number densities.  For comparison we also plot two high-redshift samples:  a 250$\mu$m-selected field sample from Casey et al.\ (2012) lying at similar redshifts to our cluster sample, $z=$\,1.2--1.6,  and the ALMA-identified  $z\sim $\,2 870-$\mu$m-selected sample from Swinbank et al.\ (2014).   At the bright end, the luminosities and space densities of the individually far-infrared/submm-detected cluster sources are comparable to those derived for similar luminosity field populations at $z\sim $\,1.5, although the faint-end is relatively flat (for this reason the contrast of the cluster population against the field is greatest for the more luminous SPIRE/SCUBA-2 sources, Figure~1), while the median luminosities and number densities derived for the less active cluster populations broadly follow the form of the  $z\sim $\,1.5 field population from Casey et al.\ (2012).

%
%
\setcounter{figure}{6} 
\begin{inlinefigure}\vspace{6pt}
\centerline{\psfig{file=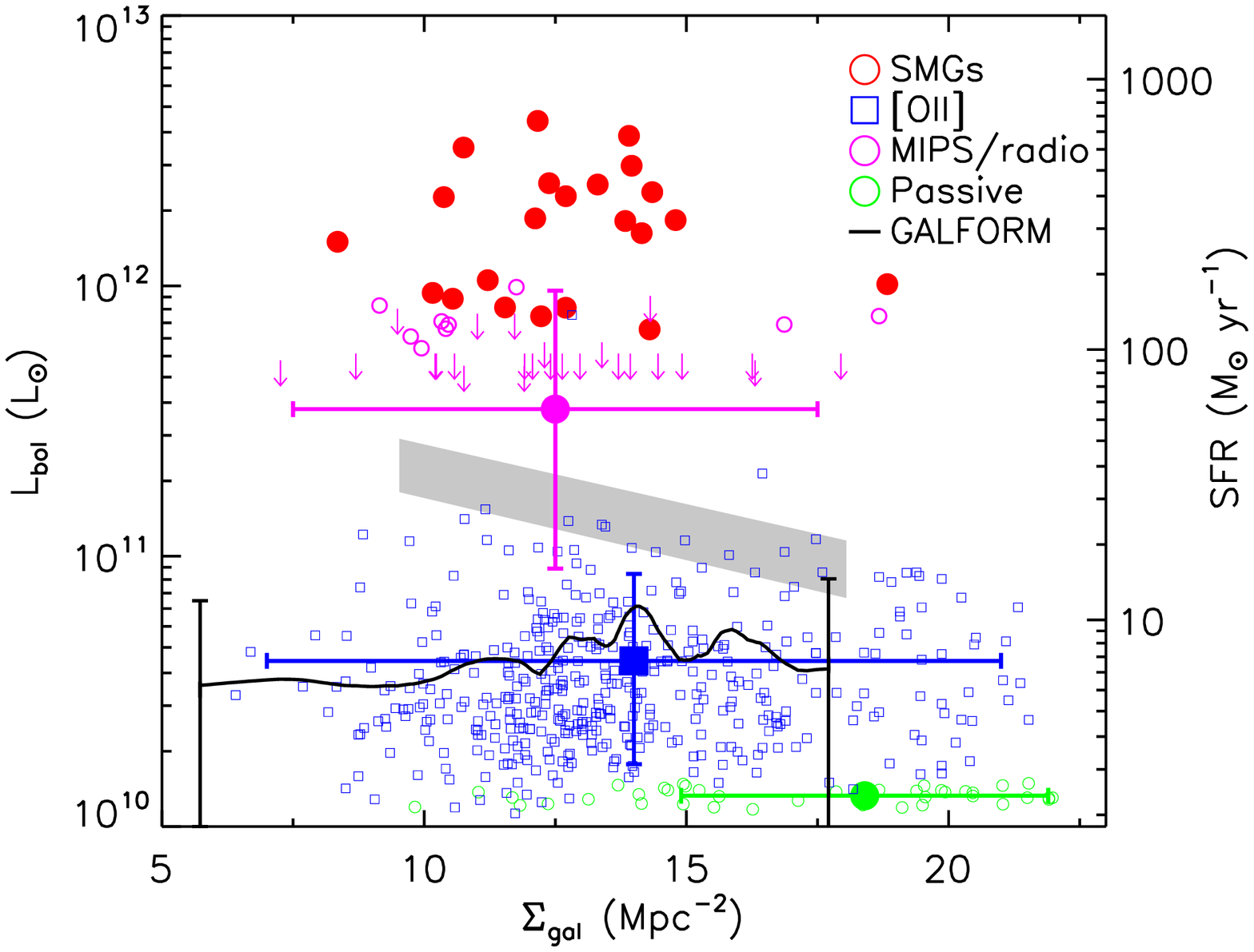,width=4.0in,angle=90}}
\vspace*{-3mm}
 \noindent{\small {\small \sc Fig.~7 --- } The variation in bolometric luminosity of sources with environment in the structure associated with Cl\,0218.3$-$0510, parameterised by the  galaxy density.  We show the individual SPIRE/SCUBA-2 selected members, the remaining far-infrared/submm-undetected MIPS/radio sources (and the median stack of the latter as a large filled circle) and the narrow-band [O{\sc ii}] population (with their individual $L_{\rm bol}$ determined from their [O{\sc ii}] luminosities and a scaling factor derived from the stacked detection of this sample from Figure~4 and plotted here as the large filled square).  The error bars on the stacked points represent the density range of each sample and the 1-$\sigma$ uncertainties on their $L_{\rm bol}$.  The local galaxy density is determined from the smoothed distribution of photometrically-selected probable cluster members shown in Figure~1.  We also plot as a gray wedge a linear fit and the 1-$\sigma$ uncertainties to the running mean trend of  $L_{\rm bol}$ for the star-forming populations with galaxy surface density, which shows a factor of $\sim $\,3 decline in characteristic $L_{\rm bol}$ over a factor of two range in local galaxy density.  We show the median density for the passive cluster galaxies plotted in Figure~8 (using the color selection of Papovich et al.\ 2012), where these galaxies have been given an arbitrary $L_{\rm bol}$ value.  It can be seen that the most active galaxies reside in a distinct environment from these passive systems.  Finally, we overplot the expected variation in mean star-formation rate in a matched population of galaxies in clusters at $z=$\,1.6 from the Millennium database using the Font et al.\ (2008) prescription for galaxy formation (we show representative error-bars on the trend).  We see that the model shows no decline in mean star-formation rate with environment in halos at this epoch, in contrast to the observations.    } 
\end{inlinefigure}

To investigate the overall level of star-formation activity in Cl\,0218.3$-$0510 compared to other clusters, we show in Figure~6 the variation in the mass-normalised, star-formation rate (SFR) in clusters as a function of redshift, compared to similarly derived far-infrared-detected samples from Popesso et al.\ (2012).  To determine the mass-normalised SFR for Cl\,0218.3$-$0510, we use the same criteria as employed by Popesso et al.\ (2012) and integrate the total bolometric emission in our sample from sources with luminosities above 10$^{11}$\,L$_\odot$ and lying within 2\,Mpc of the cluster center (see Figure~1), converting this to SFR following Kennicutt (1998).  We derive an integrated SFR, from the far-infrared/submm detected sources and MIPS/radio sources, of 650$\pm$370\,M$_\odot$\,yr$^{-1}$, where the uncertainty is dominated by the potential contamination from interlopers.  We caution that this emission arises from just five sources within the 2-Mpc region and so there is significant uncertainty in the total: 80 per cent is contributed by the two SPIRE-detected sources, 15 per cent from a single MIPS/radio source (undetected in SPIRE) and 5 per cent from two  [O{\sc ii}] emitters with inferred luminosities above 10$^{11}$\,L$_\odot$.   For the normalising mass we adopt the cluster mass estimate from Pierre et al.\ (2012) of 7.7$\times $10$^{13} $\,M$_\odot$.  Combining these we estimate a mass-normalised SFR within the central 2-Mpc of 800$\pm$500\,yr$^{-1}$ in units of 10$^{-14}$\,M$_{\odot}^{-1}$ and show this on Figure~6.    As can be seen from Figure~6, Cl\,0218.3$-$0510 lies at a higher redshift than any of the structures studied by Popesso et al.\ (2012) and also displays a higher mass-normalised SFR than any of their clusters (although close to the highest-redshift example from Popesso et al.).  The cluster also lies somewhat above the predicted evolutionary trend from Popesso et al.\ (2012), with the observed trend being closer to $(1+z)^7$ (Cowie et al.\ 2004; Geach et al.\ 2006) and supports claims of an order-of-magnitude enhancement in the level of star formation activity in overdense regions at $z\sim $\,1.5, compared to $z\sim $\,0.5--1.

%
%
\setcounter{figure}{7} 
\begin{figure*}[tbh]
\centerline{\psfig{file=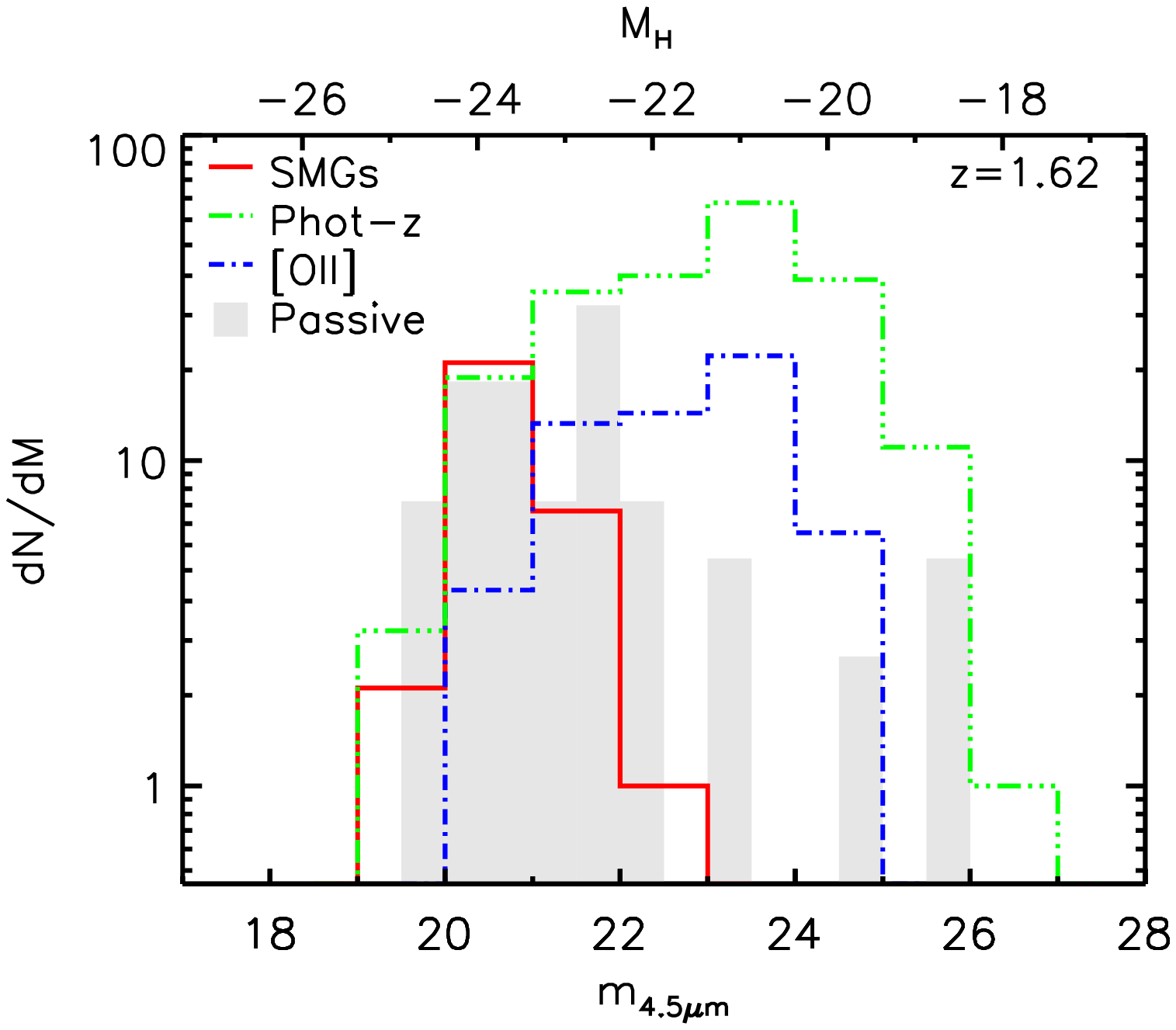,width=3.5in,angle=90}~\hspace*{-1cm}~\psfig{file=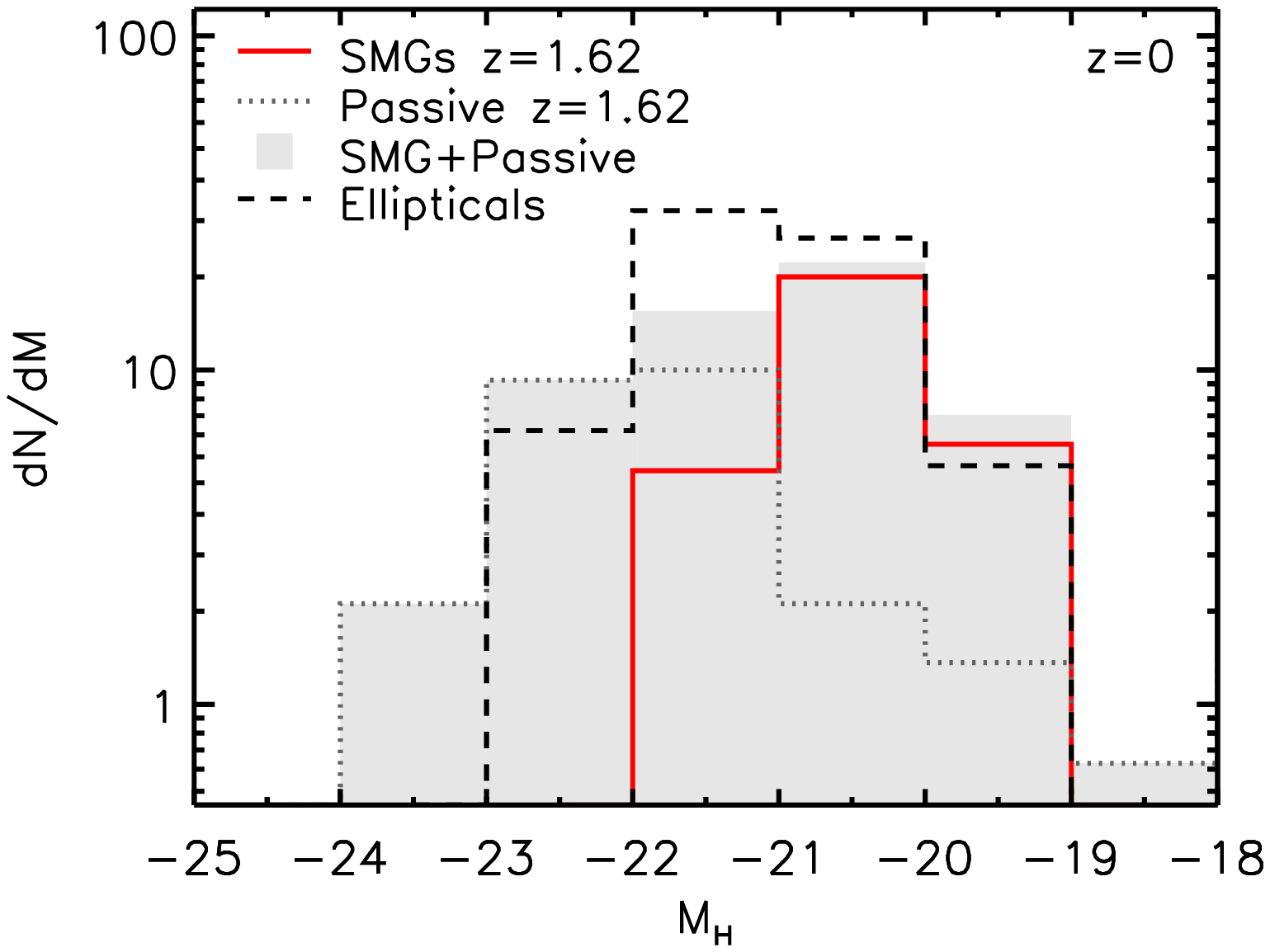,width=3.5in,angle=90}}
\caption{\small {\it (left) } The observed IRAC 4.5-$\mu$m magnitude distribution for the various cluster populations within the central 4-Mpc diameter region of the cluster:  SPIRE/SCUBA-2 detected, narrow-band [O{\sc ii}] and photometric-redshift selected.   We also plot in gray the distribution of passive cluster galaxies using the color selection of Papovich et al.\ (2012).  These are comparable in apparent restframe $H$-band luminosity to the far-infrared/submm-detected cluster ULIRGs, but they are expected to fade less to the present-day and so will correspond to intrinsically more luminous galaxies at $z\sim $\,0.  Both the passive and cluster
ULIRGs are brighter than the bulk of the [O{\sc ii}] and photometrically-selected samples. {\it (right) }  We show as the gray solid histogram the predicted $z\sim$\,0 combined distribution of the descendants of the SPIRE/SCUBA-2 and passive cluster members from the left-hand panel, compared to the observed  absolute $H$-band magnitude distribution of local elliptical galaxies from Poggianti et al.\ (2013).  The fading of the two populations is predicted to be $\Delta M_H\sim $\,3.3 for the SPIRE/SCUBA-2 sources assuming they are seen half way through a 100\,Myr burst at $z=$\,1.62 and $\Delta M_H\sim $\,1.4 for the passive populations adopting a formation epoch of $z\sim$\,2.5 consistent with the bulk of the field SMG population (Simpson et al.\ 2014).  This means that the present-day luminosities of the passive population at $z=$\,1.6 matches those of the brighter half of the local ellipticals, while the $z=$\,1.6 ULIRGs fade to become the fainter half of the elliptical population at $z\sim $\,0.   The reader should be aware that the relative normalisation of the distributions is arbitrary.   } 
\end{figure*}

\subsection{Environment and galaxy formation} 

As we saw from Figure~1, the projected distribution of the photometric members around Cl\,0218.3$-$0510 shows significant structure on the sky, which is also seen in the distribution of narrow-band [O{\sc ii}] emitters from the survey of Tadaki et al.\ (2012).  The good agreement between the distribution of [O{\sc ii}] emitting members, which should be relatively free from contamination by interlopers, and the photometric redshift sample suggests that the latter is a relatively clean tracer of the integrated galaxy density within the structure.   Both samples indicate that the structure includes two dense peaks and a number of less dense clumps of galaxies.   However, the spatial distribution of the more active cluster populations appears more uniform:  with most of these sources residing in intermediate- or low-density environments.

To provide a more quantitative comparison of the environments of the different populations we plot in Figure~7 the distribution of bolometric luminosity as a function of  galaxy density for the different cluster members.  For the  narrow-band [O{\sc ii}] population we have scaled their $L_{\rm bol}$ to agree with the stacked detection shown in Figure~4.  We use the  surface density of the photometric members, smoothed with a Gaussian kernel with a {\sc fwhm} of 1.7\,Mpc, from Figure~1 to define the environment on the grounds that it is independent of the populations we wish to test (although we note that two of our subsamples do require photometric redshifts to assign their membership) and it provides a robust measure of the global environment of the sources.  The alternative approach of determining the density using the N$^{\rm th}$-nearest neighbor (e.g.\ Tran et al.\ 2010) will tend to provide a more local measure of the density and as a result is more sensitive to contamination from interlopers than the smoothed density field.

Figure~7 shows that the  [O{\sc ii}] population extends to higher galaxy densities than the more active, far-infrared/submm population, as well as the remaining individually-undetected MIPS/radio sources -- with few of the most active galaxies lying in the highest density regions.  Hence,  comparing the  mean environmental density of the [O{\sc ii}] emitters we estimate a local density of 13.4$\pm$0.5\,galaxies per Mpc$^{2}$, versus  11.8$\pm$0.2\,Mpc$^{-2}$ for the combined sample of far-infrared/submm luminous galaxies. A Kolmogorov-Smirnov test indicates a 1.1 per cent chance that the two samples are drawn from the same underlying population. This trend  suggests that the highest density regions in this $z=$\,1.6 structure are devoid of the most actively star-forming galaxies, which instead are found in intermediate density environments.    To quantify the extent of this environment effect, if we fit a linear relation to the running mean  of $L_{\rm bol}$ for the star-forming populations with galaxy surface density, we find  a factor of $\sim $\,3 decline in characteristic $L_{\rm bol}$ over a factor of two range in local galaxy density.   Indeed, even though we didn't include them in the running mean, we  see that this trend of activity with environment extends to the color-selected passive galaxies, which are  preferentially found in the highest-density regions (Figure~7).

We  again compare our observations with the  theoretical expectations from the Millennium simulation in Figure~7.  We use the same selection for the model predictions as we adopted earlier, using the Font et al.\ (2008) prescription for galaxy evolution and then selecting all  galaxies with $M_H\leq -19$ which lie in halos at $z=$\,1.6 with masses within 10 per cent of the observed mass of Cl\,0218.3$-$0510. We then use the two-dimensional density of these galaxies, smoothed with a 1.7-Mpc FWHM Gaussian, to estimate the local galaxy densities for those  sources with  star-formation rates of $\geq$\,1\,M$_\odot$\,yr$^{-1}$ (derived from their predicted H$\alpha$ luminosities in the database), roughly matching the selection limit of the [O{\sc ii}] narrow-band survey of this field (Figure~7).   We assume a $\sim$\,50 per cent completeness in the selection of cluster members and so reduce the inferred number densities by a factor of two. Finally, we calculate the running mean of the star-formation rate as a function of local galaxy density and plot this onto Figure~7.   The model provides a reasonable match to the measured galaxy densities and mean star-formation rate (although in common with most theoretical galaxy formation models it struggles to match the most luminous star-forming systems).  However, it shows no significant variation in mean star-formation rate with environment, in contrast to the decline seen in the observations in the higher-density regions.  

Tran et al.\ (2010) and Tadaki et al.\ (2012) have both studied the environmental trends in star-forming galaxies within  Cl\,0218.3$-$0510 and have drawn conclusions which appear at first sight to be at variance with our findings.  Tadaki et al.\ found no evidence for a variation with environment in the typical SFR of star-forming galaxies identified in their [O{\sc ii}] narrow-band survey.  As can be seen from Figure~7, we see the same result using our estimates of environment, in that the SFRs for the typically low-luminosity [O{\sc ii}] population do not vary with environment,  with the trend for lower average SFRs being driven by the small number of individually more luminous far-infrared sources which are absent in the highest-density regions.    We conclude therefore that Tadaki et al.\ and our results are consistent and point to an increasing  environmental sensitivity in galaxies with higher SFRs.  

Turning to the Tran et al.\ (2010) study,  this used MIPS 24-$\mu$m observations of 17 sources from a sample of $\sim$\,100 photometric and spectroscopically identified members in the central regions of the cluster to infer an increase in the fraction of strongly star-forming galaxies with increasing local density.  This is in the opposite sense to the trend we see and we attribute this to one of three potential causes.  Firstly, our catalog of far-infrared/submm sources (drawn from a 30\,$\times$ larger area than that analysed by Tran et al.) could either suffer significant contamination from non-members, which would appear predominantly in the lower-density regions, or the incompleteness in our selection could mean we are missing far-infrared/submm bright members in the higher-density regions.  Both of these effects would reduce/remove any trend of increasing activity in higher-density regions.  However, we discount the latter as there are few candidate far-infrared/submm sources at any redshift seen in projection within the higher-density regions.  This leaves dilution due to contamination as a potential source of the disagreement, although we note that this would imply a much lower level of far-infrared/submm activity in this structure than suggested by either the trend in Figure~6 or the visibility of the structure in the redshift distribution of far-infrared/submm sources in Figure~1.  Hence we next  consider the other two explanations, the first of which arises from the fact that Tran et al.'s study relies on 24-$\mu$m luminosity (restframe 9\,$\mu$m) to infer SFRs. As we have discussed earlier this is an uncertain assumption at $z\sim$\,1.6 owing to the mix of continuum, PAH emission and silicate absorption included in the MIPS passband and indeed we see no correlation at all between 24-$\mu$m flux and far-infrared luminosity for the sources in our sample.  Hence the trend observed by Tran et al.\ could have its origin in variations in either SFR, AGN contribution or hot/cold dust luminosity ratio with environment (e.g.\ Rawle et al.\ 2012).     In addition, Tran et al.\ used the 10$^{\rm th}$-nearest neighbor in their sample to determine the galaxy density in their analysis, this provides a more local estimate of environment (on $\sim$\,100-kpc scales in the densest regions) than our use of the density field of photometrically-selected members smoothed on 1.6-Mpc scales, it is also more sensitive to contamination by interlopers in the galaxy sample.  Nevertheless, if we adopt their approach we find that some of the far-infrared/submm sources appear to reside in denser environments, especially when using a smaller number of neighbors to assess the environment (e.g.\ 5$^{\rm th}$-nearest).  Thus the behaviour that Tran et al.\ see may be driven by small-scale density enhancements in the very local environment of the most active galaxies, rather than providing an indication of enhanced activity in higher-density environments when defined on a larger scale.  Indeed, such small-scale associations of galaxies could  be responsible for triggering the activity in the far-infrared/submm sources through interactions and mergers.    We conclude that the discrepancy between our results and those in Tran et al.\ may have its origin in either the small size of their study, their reliance on restframe 9-$\mu$m luminosity to infer SFR or their use of a local measure of environment, which is more sensitive to transient, small-scale clustering than our global measure.

\section{Discussion and Conclusions}

While we have seen that there is an enhanced level of star formation activity in the $z=$\,1.6 cluster Cl\,0218.3$-$0510, compared to clusters at lower redshifts, we can also test if that activity is occuring in the densest regions of the cluster (as claimed by Tran et al.\ 2010).  Theoretical models suggest phase-space is conserved during the hierarchical collapse of structures, implying that the densest parts of the structure at high redshift will correspond to the densest regions in its descendant in the local Universe.  These regions are populated by some of the oldest and most massive elliptical galaxies (e.g.\ Smith et al.\ 2012), and if the ultraluminous activity we see at high redshift is similarly associated with the densest parts of the structures, this would be an indication that we are seeing the formation phase of the oldest elliptical galaxies within these systems.  

As Figure~7 illustrates, in this $z=$\,1.6 structure we actually see that the most active, ultraluminous star-forming galaxies inhabit lower-density environments, while a population of apparently passive galaxies has already been built up in the highest-density regions.  But are these galaxies passive because they have exhausted their gas reservoirs or has this gas been removed by environmental processes operating in these high-density regions?    The comparison with the Font et al.\ model in Figure~7 is informative in this regard.  Their model supplements the standard hierarchical suppression of star formation in satellite galaxies in high-density regions (as they are no longer the central galaxy of a halo of cooling gas) with additional stripping of extended gas reservoirs in their halos (termed ``strangulation''), but this is clearly not sufficient to suppress the predicted activity in these regions.   This failure is evident in both the lack of any variation in mean star-formation rate with environment in Figure~7 and through the relative absence of luminous, red galaxies in the model predictions shown in Figure~3.   We therefore suggest that the lack of activity in the passive population we observe in the highest-density regions of Cl\,0218.3$-$0510 is because these galaxies have exhausted their gas reservoirs, rather than because the gas has been removed via an externally-driven environmental process.   Other processes are thus needed to suppress the star formation in these galaxies, for example feedback from AGN, as suggested for the passive red galaxy population in a $z=$\,1.46 cluster by Hayashi et al.\ (2011), which could aid in the removal of their gas reservoirs. Overall we propose that we are not seeing the initial phase of formation of cluster ellipticals in Cl\,0218.3$-$0510, which must instead have occured at an earlier epoch, as suggested by studies of $z\gs$\,2 proto-clusters (e.g.\ Geach et al.\ 2005; Chapman et al.\ 2009).  

While the theoretical galaxy evolution models may not reproduce the details of the galaxy populations in this high-redshift cluster, we can still use the Millennium simulation to indicate the likely present-day descendant of this $z=$\,1.62 structure and so link our observations to local populations.  We search the simulation for  $z=$\,0 halos which had masses of $M_{\rm cl}\sim 7.7\times 10^{13}$\,M$_\odot$ at $z=$\,1.6 and derive a  mean mass of these of $M_{\rm cl}=(5\pm3) \times 10^{14}$\,M$_\odot$, where the error is the 1-$\sigma$ scatter.  This indicates that the cluster mass will grow by a factor of $\sim$\,6$\times$ on average between $z=$\,1.6 and the present-day.  So, can we empirically relate the galaxy populations we see at $z=$\,1.6 to those of similarly, massive clusters today?

We begin by investigating the relationship between the far-infrared/submm sources and the other cluster populations,  we show in Figure~8 the 4.5-$\mu$m magnitude distributions for the different  populations within a 4-Mpc diameter region centered on Cl\,0218.3$-$0510 (Figure~1).   As expected from Figure~3, the far-infrared/submm detected sources are brighter in the restframe $H$-band (median $H$-band absolute magnitude of $M_H=-$23.6$\pm$0.6) than the narrow-band [O{\sc ii}] emitters or the photometric-redshift sample (median $H$-band absolute magnitudes of $M_H=-$21.6$\pm$1.1 and $M_H=-$21.2$\pm$1.2 respectively).  We also compare these distributions to the passive cluster population in the $z=$\,1.6 structure, selected using the $zJ4.5\mu$m color criteria for passive galaxies from Papovich et al.\ (2012), which have $M_H=-$22.9$\pm$0.3.  Interestingly, these passive galaxies have comparable restframe $H$-band luminosities to the far-infrared/submm population.   However, the latter galaxies -- which are currently in a very active phase -- are expected to fade more than those galaxies which are already passive at $z=$\,1.6 and so will correspond to intrinsically lower-luminosity galaxies at the present day, assuming that neither population undergoes a subsequent phase of significant star formation or merging.  

To illustrate the possible subsequent evolution of these populations, we also show in Figure~8 a qualitative prediction for the relative absolute $H$-band magnitude distributions for the descendants of the SPIRE/SCUBA-2 and passive cluster members at $z\sim $\,0.  The fading of the far-infrared/submm sources is expected to be $\Delta M_H\sim 3.3$ assuming that they are seen half way through a 100\,Myr burst at $z=$\,1.6 (this is the canonical duration of SMGs, e.g.\ Hickox et al.\ 2012).  In contrast the passive population at $z=$\,1.6 will fade by only $\Delta M_H\sim $\,1.4,  adopting a formation epoch of $z\sim $\,2.5 consistent with the bulk of the field SMG population (Simpson et al.\ 2014).  Figure~8 compares these two faded populations to the observed  absolute $H$-band magnitude distribution of a volume-limited survey of local elliptical galaxies with morphological classifications from Poggianti et al.\ (2013).   We see that the effects of the differential fading of the ULIRGs and passive populations since $z=$\,1.6 means that the passive galaxies are a good match for the luminosities of the brightest $\sim $\,50 per cent of elliptical galaxies at $z\sim $\,0, while the ULIRG population from Cl\,0218.3$-$0510 has faded further and these galaxies now match the  fainter half of the distribution of ellipticals seen locally.  Given the different spatial distributions of these two populations within Cl\,0218.3$-$0510, we would also expect this difference in formation age to remain, with the more luminous ellipticals in the cores of massive clusters today being older than the somewhat less luminous systems further out (e.g.\ Smith et al.\ 2012).

The main conclusions of this work are:

\noindent$\bullet$  We combined SCUBA-2 submm and {\it Herschel} far-infrared imaging of a $\sim $\,0.25\,deg.$^2$ area containing the $z=$\,1.62 cluster Cl\,0218.3$-$0510.  We use these data, in conjunction with {\it Spitzer} mid-infrared and VLA radio observations and photometric and spectroscopic redshifts, to identify far-infrared/submm bright galaxies which are probable members of the structure around this high-redshift cluster.  We show that ALMA Cycle 1 observations of one of the proposed identifications confirms the submm emission arises from a MIPS/radio source with a photometric redshift of $z_{\rm phot}=$\,1.67.

\noindent$\bullet$ We find that these far-infrared/submm bright members have ULIRG-like luminosities and comprise some of the brightest and reddest cluster galaxies in the restframe optical/near-infrared.   Their restframe near-infrared magnitudes are comparable to the brightest passive galaxies seen in the core regions of the $z=$\,1.6 cluster, with $M_H\sim -23$.

\noindent$\bullet$ We determine the dust temperature--luminosity relation for the various classes of active galaxies in the cluster, using modified black body fits to the individual the far-infrared/submm-detected sources and fits to the stacked photometry of the less-active MIPS/radio sources and [O{\sc ii}]-emitters.   We find that the latter two samples of lower-luminosity sources, $L_{\rm bol}\ls $\,10$^{10-11} $\,L$_{\odot}$, lie on the local $T_{\rm d}$--$L_{\rm bol}$ relation, but that the far-infrared/submm-detected sources, with $L_{\rm bol}\gs $\,10$^{12} $\,L$_{\odot}$, are some  $\Delta T_{\rm d}\sim $\,5.3$\pm$1.6\,K cooler on average than comparable luminosity galaxies at $z\sim $\,0.  

\noindent$\bullet$ By integrating the total star-formation rate in all galaxy populations with bolometric luminosities $\geq $\,10$^{11}$\,L$_\odot$, we derive a total mass-normalised star-formation rate of 800$\pm$500\,yr$^{-1}$  (in units of 10$^{-14}$\,M$_{\odot}^{-1}$) within the central 2-Mpc.  This mass-normalised SFR is an order of magnitude higher than seen in typical $z\sim$\,0.5--1 clusters and indicates a continued increase in the star formation activity in clusters out to the earliest epochs probed.  

\noindent$\bullet$ Comparing the spatial distribution of the far-infrared/submm bright members with less active galaxy populations in the structure, we find that the most active galaxies are not found in the densest regions, which are instead traced by luminous galaxies with colors compatible with passive stellar populations at $z\sim$\,1.6.  This is consistent with a scenario where  the activity we observe relates to infall of galaxies onto a pre-existing cluster core, which already contains a population of passive, but luminous galaxies.  We show that a toy model for the subsequent evolution of the passive and active populations in Cl\,0218.3$-$0510 matches the $H$-band luminosity distribution of elliptical galaxies at the present-day, with the most active galaxies at $z=$\,1.6 corresponding to the formation phase of some of the fainter elliptical galaxies seen at $z\sim$\,0.   If correct this places the earliest phase of the formation of (the most massive) cluster ellipticals at $z\gg$\,1.6, possibly extending out to $z\sim$\,5 (Simpson et al.\ 2014).

\acknowledgments 

We thank the Referee for their constructive and helpful comments on this work.
We also thank Bianca Poggianti, Cheng-Juin Ma, John Helly and Richard Bower for help and useful conversations.   IRS acknowledges support from  a Leverhulme Fellowship, the ERC Advanced Investigator programme DUSTYGAL 321334 and a Royal Society/Wolfson Merit Award.  AMS gratefully acknowledges an STFC Advanced Fellowship through grant number ST/H005234/1 and all of the Durham co-authors acknowledge STFC through grant number ST/I001573/1. JMS and ALRD acknowledge the support of STFC studentships (ST/J501013/1 and ST/F007299/1, respectively).  AK acknowledges support by the Collaborative Research Council 956, sub-project A1, funded by the Deutsche Forschungsgemeinschaft (DFG). This work is based on observations carried out with SCUBA-2 on the James Clerk Maxwell Telescope (JCMT). The JCMT is operated by the Joint Astronomy Centre (JAC) on behalf of the Science and Technology Facilities Council (STFC) of the United Kingdom, the National Research Council of Canada, and (until 31 March 2013) the Netherlands Organisation for Scientific Research. Additional funds for the construction of SCUBA-2 were provided by the Canada Foundation for Innovation. The United Kingdom Infrared Telescope is operated by the JAC on behalf of the STFC of the UK. This research also made use of data taken as part of the HerMES Key Programme from the SPIRE instrument team, ESAC scientists and a mission scientist. {\it Herschel} is an ESA space observatory with science instruments provided by European-led Principal Investigator consortia and with important participation from NASA.  In addition this paper makes use of the following ALMA data: ADS/JAO.ALMA\#2012.1.00090.S. ALMA is a partnership of ESO (representing its member states), NSF (USA) and NINS (Japan), together with NRC (Canada) and NSC and ASIAA (Taiwan), in cooperation with the Republic of Chile. The Joint ALMA Observatory is operated by ESO, AUI/NRAO, and NAOJ. The data used in this paper can be obtained from the JCMT, UKIRT, {\it Herschel}, {\it Spitzer}, VLA, ALMA  and Subaru data archives.

\setlength{\tabcolsep}{3pt} 

%
%
\begin{center}{
\begin{table}[ht]
\centerline{\sc Table 1}
\centerline{\sc Properties of SMGs}
\smallskip
\small
\begin{tabular}{lcccccccccccccc}
\hline
ID & R.A.\ & Dec.\ & $z_{\rm phot}$ & $S_{\rm 1.4\,GHz}$ & $S_{\rm 24\mu m}$ &  $z'$ & $H$ & 4.5\,$\mu$m & $S_{\rm 250\mu m}$ &  $S_{\rm 350\mu m}$ &  $S_{\rm 500\mu m}$ &  $S_{\rm 850\mu m}$ & $T_{\rm d}$ & $L_{\rm bol}$  \\
   & \multispan2{~(J2000)}     &  & ($\mu$Jy) & ($\mu$Jy) & (AB) & (AB) & (AB) & (mJy) & (mJy) & (mJy) & (mJy) & (K) &  (10$^{12}$\,L$_\odot$) \\
\hline\hline
 ~4 &  34.77112 &  $-$5.16960 &  1.63 &   ...   &   875.4 &   23.34 &   21.57 &   20.33 &    11$\pm$3 &    15$\pm$3 &   $<$\,12 &   $<$\,7 & 30$\pm$5 &  0.8$^{+1.4}_{-0.6}$ \\
 11 &  34.50192 &  $-$5.12132 &  1.67 &   ...   &   581.2 &   24.83 &   23.17 &   21.38 &    14$\pm$3 &    15$\pm$3 &   $<$\,12 &   $<$\,7 & 34$\pm$7 &  1.1$^{+2.2}_{-0.8}$ \\
 12 &  34.53913 &  $-$5.12828 &  1.68 &    99.8 &   ...   &   24.26 &   22.54 &   21.39 &    31$\pm$4 &    25$\pm$4 &    16$\pm$4 &   $<$\,7 & 34$\pm$4 &  2.1$^{+2.2}_{-1.2}$ \\
 14 &  34.33656 &  $-$5.14899 &  1.66 &   ...   &   586.4 &   23.76 &   21.88 &   21.06 &    15$\pm$3 &    16$\pm$3 &    13$\pm$4 &   $<$\,7 & 28$\pm$3 &  0.9$^{+1.0}_{-0.5}$ \\
 15 &  34.44959 &  $-$5.14940 &  1.69 &   147.5 &   347.1 &   24.17 &   22.27 &   20.60 &    42$\pm$5 &    30$\pm$4 &    12$\pm$3 &   $<$\,7 & 43$\pm$6 &  4$^{+5}_{-3}$ \\
 16 &  34.54811 &  $-$5.15113 &  1.70 &   275.8 &   ...   &   25.47 &   23.07 &   21.19 &    21$\pm$3 &    14$\pm$3 &   $<$\,12 &   $<$\,7 & 44$\pm$12 &  2$^{+7}_{-2}$ \\
 17 &  34.28927 &  $-$5.15375 &  1.66 &   ...   &   373.7 &   23.83 &   22.09 &   20.45 &    11$\pm$3 &    11$\pm$3 &    16$\pm$4 &   $<$\,7 & 25$\pm$3 &  0.7$^{+0.7}_{-0.4}$ \\
 26 &  34.50498 &  $-$5.24101 &  1.67 &   ...   &   483.2 &   24.61 &   22.67 &   20.91 &    19$\pm$3 &    22$\pm$4 &   $<$\,12 &   $<$\,7 & 35$\pm$6 &  1.5$^{+2.6}_{-1.1}$ \\
 28 &  34.69529 &  $-$5.24341 &  1.64 &    97.5 &   ...   &   24.25 &   21.92 &   20.35 &    14$\pm$3 &    14$\pm$3 &   $<$\,12 &   $<$\,7 & 34$\pm$7 &  1.0$^{+2.2}_{-0.8}$ \\
 29 &  34.40054 &  $-$5.25554 &  1.67 &   115.1 &   361.8 &   23.74 &   21.99 &   20.25 &    29$\pm$4 &    28$\pm$4 &    16$\pm$4 &     6$\pm$1 & 29$\pm$2 &  1.7$^{+1.2}_{-0.8}$ \\
 30 &  34.66084 &  $-$5.26395 &  1.67 &   ...   &   451.4 &   23.05 &   21.83 &   21.10 &    24$\pm$3 &    19$\pm$3 &    16$\pm$4 &   $<$\,7 & 32$\pm$4 &  1.6$^{+1.7}_{-0.9}$ \\
 33 &  34.26542 &  $-$5.27759 &  1.67 &   110.5 &   550.4 &   24.85 &   22.44 &   20.39 &    44$\pm$5 &    37$\pm$5 &    22$\pm$4 &   $<$\,7 & 35$\pm$4 &  3.2$^{+2.9}_{-1.7}$ \\
 37 &  34.70097 &  $-$5.30141 &  1.67 &   125.2 &   435.0 &   24.39 &   22.22 &   20.61 &    30$\pm$4 &    31$\pm$4 &    21$\pm$4 &     9$\pm$1 & 26$\pm$2 &  1.8$^{+1.0}_{-0.7}$ \\
 39 &  34.66438 &  $-$5.30628 &  1.60 &    86.1 &   ...   &   25.46 &   23.40 &   21.88 &    23$\pm$3 &    12$\pm$3 &   $<$\,12 &   $<$\,7 & 53$\pm$20 &  4$^{+18}_{-3}$ \\
 40 &  34.30795 &  $-$5.32148 &  1.66 &   106.5 &  1052.8 &   23.59 &   21.39 &   19.51 &    48$\pm$5 &    40$\pm$5 &    27$\pm$5 &    10$\pm$1 & 29$\pm$2 &  2.8$^{+1.5}_{-1.0}$ \\
 42 &  34.68115 &  $-$5.33646 &  1.66 &    61.7 &   ...   &   24.93 &   23.28 &   22.20 &    12$\pm$3 &    17$\pm$3 &   $<$\,12 &   $<$\,7 & 30$\pm$5 &  0.8$^{+1.4}_{-0.6}$ \\
 44 &  34.30778 &  $-$5.36366 &  1.69 &   ...   &   948.8 &   24.16 &   22.47 &   20.64 &    36$\pm$4 &    30$\pm$4 &    26$\pm$5 &   $<$\,7 & 32$\pm$3 &  2.4$^{+2.1}_{-1.3}$ \\
 48 &  34.64560 &  $-$5.41950 &  1.66 &   ...   &   573.6 &   24.21 &   22.06 &   20.19 &    25$\pm$4 &    26$\pm$4 &    12$\pm$4 &   $<$\,7 & 33$\pm$5 &  1.7$^{+1.7}_{-1.0}$ \\
 51 &  34.25920 &  $-$5.44522 &  1.69 &   ...   &   445.2 &   24.78 &   22.89 &   20.99 &    13$\pm$3 &    11$\pm$3 &   $<$\,12 &   $<$\,7 & 33$\pm$7 &  0.9$^{+2.2}_{-0.7}$ \\
 52 &  34.53305 &  $-$5.02929 &  1.66 &   301.1 &   392.8 &   25.05 &   22.92 &   20.85 &    11$\pm$3 &    19$\pm$3 &    17$\pm$4 &   $<$\,7 & 24$\pm$2 &  0.8$^{+0.7}_{-0.4}$ \\
 57 &  34.30774 &  $-$5.05717 &  1.70 &   ...   &   651.2 &   24.57 &   22.81 &   21.06 &    30$\pm$4 &    20$\pm$3 &   $<$\,12 &     6$\pm$1 & 36$\pm$6 &  2.1$^{+3.0}_{-1.5}$ \\
 58 &  34.41113 &  $-$5.06099 &  1.69 &   206.9 &   589.4 &   24.44 &   22.42 &   20.48 &    31$\pm$4 &    27$\pm$10 &   $<$\,12 &     7$\pm$1 & 37$\pm$6 &  2.4$^{+3.9}_{-1.7}$ \\
\noalign{\smallskip}\\
MIPS & ... & ... & ... & ... & ... & ... & ... & ... & 5.2$\pm$1.0 & 4.7$\pm$1.0 & 3.4$\pm$1.3 & 0.1$\pm$0.5 & 33$\pm$6 & 0.4$^{+0.6}_{-0.3}$ \\
$[{\rm OII}]$ & ... & ... & ... & ... & ... & ... & ... & ... & 0.7$\pm$0.2 & 0.7$\pm$0.2 & 1.1$\pm$0.4 & 0.10$\pm$0.06 & 29$\pm$5 & 0.04$^{+0.07}_{-0.03}$  \\ 
\hline\end{tabular}
\smallskip
\end{table}}
\end{center}
\vspace*{-0.5cm}

\end{document}